\documentclass[
        reprint,
        superscriptaddress,
        aps,prc,
        noeprint,floatfix,nofootinbib,
        altaffilletter
        ]{revtex4-2}

\usepackage{graphicx}
\usepackage{upgreek}
\usepackage{amsmath}
\usepackage[normalem]{ulem}
\usepackage{lineno}
\usepackage{hyperref}
\usepackage{xcolor}
\usepackage{tabularx}
\usepackage{enumitem} 
\usepackage{makecell}
\usepackage{pifont}
\usepackage{siunitx}
\usepackage{blindtext}

\usepackage[caption=false]{subfig}

\usepackage{mathrsfs}

\DeclareSIUnit\eV{eV}
\DeclareSIUnit\bar{bar}
\DeclareSIUnit\sps{\ensuremath{\mathrm{SPS}}}
\newcommand{\SIadj}[2]{\SI[number-unit-product={\text{-}}]{#1}{#2}}
\PassOptionsToPackage{unicode}{hyperref}
\PassOptionsToPackage{naturalnames}{hyperref}

\begin{document}

\title{Cyclotron Radiation Emission Spectroscopy of Electrons from Tritium Beta Decay and \texorpdfstring{$\mathrm{^{83m}}$Kr}{} Internal Conversion}

\newcommand{\Washington}{\affiliation{Center for Experimental Nuclear Physics and Astrophysics and Department of Physics, University of Washington, Seattle, WA 98195, USA}}
\newcommand{\Mainz}{\affiliation{Institute for Physics, Johannes Gutenberg University Mainz, 55128 Mainz, Germany}}
\newcommand{\MIT}{\affiliation{Laboratory for Nuclear Science, Massachusetts Institute of Technology, Cambridge, MA 02139, USA}}
\newcommand{\PennState}{\affiliation{Department of Physics, Pennsylvania State University, University Park, PA 16802, USA}}
\newcommand{\PNNL}{\affiliation{Pacific Northwest National Laboratory, Richland, WA 99354, USA}}
\newcommand{\Yale}{\affiliation{Wright Laboratory and Department of Physics, Yale University, New Haven, CT 06520, USA}}
\newcommand{\Livermore}{\affiliation{Lawrence Livermore National Laboratory, Livermore, CA 94550, USA}}
\newcommand{\Case}{\affiliation{Department of Physics, Case Western Reserve University, Cleveland, OH 44106, USA}}
\newcommand{\Indiana}{\affiliation{Center for Exploration of Energy and Matter and Department of Physics, Indiana University, Bloomington, IN, 47405, USA}}
\newcommand{\KIT}{\affiliation{Institute of Astroparticle Physics, Karlsruhe Institute of Technology, 76021 Karlsruhe, Germany}}
\newcommand{\Sorbonne}{\affiliation{Laboratoire de Physique Nucl\'eaire et de Hautes \'Energies, Sorbonne Universit\'e, Universit\'e Paris Cit\'e, CNRS/IN2P3, 75005 Paris, France}}
\newcommand{\CfA}{\affiliation{Center for Astrophysics $\mid$ Harvard $\&$ Smithsonian, Cambridge, MA 02138, USA}}

\author{A.~Ashtari~Esfahani} \altaffiliation{Present Address: Department of Physics, Sharif University of Technology, P.O. Box 11155-9161, Tehran, Iran} \Washington
\author{S.~B\"oser} \Mainz
\author{N.~Buzinsky} \altaffiliation{Present Address: Center for Experimental Nuclear Physics and Astrophysics and Department of Physics, University of Washington, Seattle, WA 98195, USA} \MIT
\author{M.~C.~Carmona-Benitez} \PennState
\author{C.~Claessens} \altaffiliation{claesc@uw.edu} \Washington \Mainz
\author{L.~de~Viveiros} \PennState
\author{P.~J.~Doe} \Washington
\author{M.~Fertl} \Mainz
\author{J.~A.~Formaggio} \MIT
\author{J.~K.~Gaison} \PNNL
\author{L.~Gladstone} \Case
\author{M.~Guigue} \Sorbonne
\author{J.~Hartse} \Washington
\author{K.~M.~Heeger} \Yale
\author{X.~Huyan} \altaffiliation{Present Address: LeoLabs, Menlo Park, CA 94025, USA} \PNNL
\author{A.~M.~Jones} \altaffiliation{Present Address: Ozen Engineering, Sunnyvale, CA 94085, USA} \PNNL
\author{K.~Kazkaz} \Livermore
\author{B.~H.~LaRoque} \PNNL
\author{M.~Li} \MIT
\author{A.~Lindman} \Mainz
\author{E.~Machado} \Washington
\author{A.~Marsteller} \Washington
\author{C.~Matth\'e} \Mainz
\author{R.~Mohiuddin} \Case
\author{B.~Monreal} \Case
\author{R.~Mueller} \PennState
\author{J.~A.~Nikkel} \Yale
\author{E.~Novitski} \Washington
\author{N.~S.~Oblath} \PNNL
\author{J.~I.~Pe\~na} \MIT
\author{W.~Pettus} \Indiana
\author{R.~Reimann} \Mainz
\author{R.~G.~H.~Robertson} \Washington
\author{D.~Rosa~De~Jes\'us} \PNNL
\author{G.~Rybka} \Washington
\author{L.~Salda\~na} \Yale
\author{M.~Schram} \altaffiliation{Present Address: Thomas Jefferson National Accelerator Facility, Newport News, VA 23606, USA} \PNNL
\author{P.~L.~Slocum} \Yale
\author{J.~Stachurska} \MIT
\author{Y.-H.~Sun} \Case
\author{P.~T.~Surukuchi} \Yale
\author{J.~R.~Tedeschi} \PNNL
\author{A.~B.~Telles} \Yale
\author{F.~Thomas} \Mainz
\author{M.~Thomas} \altaffiliation{Present Address: Booz Allen Hamilton, San Antonio, Texas, 78226, USA} \PNNL
\author{L.~A.~Thorne} \Mainz
\author{T.~Th\"ummler} \KIT
\author{L.~Tvrznikova} \altaffiliation{Present Address: Waymo, Mountain View, CA 94043} \Livermore
\author{W.~Van~De~Pontseele} \MIT
\author{B.~A.~VanDevender} \Washington \PNNL
\author{J.~Weintroub} \CfA
\author{T.~E.~Weiss} \altaffiliation{talia.weiss@yale.edu} \Yale
\author{T.~Wendler} \PennState
\author{A.~Young} \altaffiliation{Present Address: Department of Astrophysics/IMAPP, Radboud University, PO Box 9010, 6500 GL Nijmegen, The Netherlands} \CfA
\author{E.~Zayas} \MIT
\author{A.~Ziegler} \PennState

\collaboration{Project 8 Collaboration}

\date{\today}

\begin{abstract}

Project 8 has developed a novel technique, Cyclotron Radiation Emission Spectroscopy (CRES), for direct neutrino mass measurements.
A CRES-based experiment on the beta spectrum of tritium has been carried out in a small-volume apparatus.  We provide a detailed account of the experiment, focusing on systematic effects and analysis techniques. 
In a Bayesian (frequentist) analysis, we measure the tritium endpoint as $18553^{+18}_{-19}$ ($18548^{+19}_{-19}$) eV and set upper limits of $155$ ($152$) eV (90\% C.L.) on the neutrino mass. No background events are observed beyond the endpoint in 82 days of running.
We also demonstrate an energy resolution of $1.66\pm0.19$\,eV in a resolution-optimized magnetic trap configuration by measuring $^{83\rm m}$Kr 17.8-keV internal-conversion electrons.
These measurements establish CRES as a low-background, high-resolution technique with the potential to advance neutrino mass sensitivity.

\end{abstract}

\maketitle

\tableofcontents

\section{Introduction}

The neutrino comes in three flavors\textemdash electron, muon, and tau\textemdash associated with the charged leptons.  Super-Kamiokande \cite{Fukuda:1998mi} and the Sudbury Neutrino Observatory \cite{Ahmad:2001an} showed that these flavors mix and oscillate, explaining the anomalies in atmospheric and solar neutrino fluxes. Oscillation between neutrino flavors requires that  neutrinos have mass, in contradiction with the Standard Model of particle physics wherein neutrinos are massless. It is now clear that neutrino mass eigenstates $m_{i=1,2,3}$ exist, of which at least two have non-zero mass.   The flavor eigenstates are linear combinations of $m_i$,  with amplitudes given by elements of a unitary matrix $U$, the Pontecorvo–Maki–Nakagawa–Sakata (PMNS) matrix \cite{Pontecorvo:1957cp,Maki:1962mu}.

 Oscillation experiments can determine only the differences of the squares $m_i^2-m_j^2$, not the mass scale. It is known from solar neutrino oscillations~\cite{Ahmad:2001an} that $m_2>m_1$, but whether $m_3$ is the lightest mass eigenstate (``inverted ordering'') or the heaviest (``normal ordering'')  is presently unknown. Oscillation experiments also constrain  the sum  of the three neutrino masses, $\Sigma m_i$, to be at least \SI{0.05}{\eV} \cite{PDG:2020} because masses are positive definite.  The neutrino constitutes the first and, so far, the only identified dark matter in the cosmos.  Limits on $\Sigma m_i$ have been obtained from observations of the cosmic microwave background and large-scale cosmic structure (see, for example, the Planck collaboration \cite{Planck:2018vyg} who reported a limit of \SI{<0.12}{\eV} within the framework of the $\Lambda$CDM model).

The most sensitive technique for a direct and model-independent neutrino mass determination is the analysis of the endpoint region of the tritium beta decay spectrum~\cite{Formaggio:2021nfz}.
The signal for neutrino mass emerges as a phase-space modification of the spectral shape close to the beta endpoint. This signal is independent of whether neutrinos are Majorana or Dirac particles.  When the energy resolution in a beta spectroscopy experiment is larger than the neutrino mass splittings, the experiment measures the electron-weighted neutrino mass,
\begin{eqnarray}
    m_\beta &=&\sqrt{\sum_{i=1}^3 \left| U_{\rm{e}i} \right|^2 m_i^2}\, . \quad
\end{eqnarray}
Neutrino oscillation experiments impose an ultimate lower bound of $m_\beta$\SI{\geq 0.009}{\eV} ($m_\beta$\SI{\geq 0.05}{\eV}) for the case of normal (inverted) mass ordering~\cite{PDG:2020}. 

Recently, the KATRIN experiment has set a new upper limit on $m_\beta$ of \SI{0.8}{\eV} at \SI{90}{\percent} confidence level~\cite{KATRIN:2021uub}.    KATRIN's design goal is a 0.2-\SI{}{\eV} mass sensitivity limit~\cite{KATRIN:2021dfa}, if the mass is not larger. KATRIN is expected to either exhaust the quasidegenerate range of neutrino masses, or to measure the neutrino mass if it lies in that range.

The neutrino mass might turn out to be smaller than the state-of-the-art experiment KATRIN can discover.  While there may be ways to extend the reach of KATRIN somewhat~\cite{KATRINSnowmass:2022}, this type of experiment approaches a fundamental limit as a result of its sheer size and  use of molecular tritium (T$_2$).  With T$_2$, one relies on a theoretical prediction of the spectrum of excited molecular states produced in beta decay, which broadens the spectral response~\cite{Doss:2006zz,Saenz:2000dul}.  The neutrino mass effect is spread over a range of slightly differing endpoints for the excitations, exacting both a statistical and a systematic price. An independent direct measurement of this spectrum of excited states is not possible, although specific tests can be performed \cite{TRIMS:2020nsv} and agree with theory at the percent level. 
To advance significantly beyond KATRIN's design sensitivity\textemdash which is the goal of a next-generation tritium endpoint experiment\textemdash requires a different approach. An experiment that sets a limit of $m_{\beta}<0.04$\,eV would reach the full range of masses allowed for the inverted ordering, so this experiment would either measure the neutrino mass scale or exclude the inverted ordering~\cite{Esfahani:2017dmu, AshtariEsfahani:2021moh}.
 
The Project 8 Collaboration has devised a new method called Cyclotron Radiation Emission Spectroscopy (CRES) \cite{Monreal:2009za,Esfahani:2017dmu} to reach this goal.
In  CRES, the emitted beta electron's energy is measured by detecting its cyclotron radiation as it spirals in a magnetic field. In special relativity, the cyclotron frequency $f_c$ in a magnetic field $B$ is related to kinetic energy $E_{\rm{kin}}$ as follows:
\begin{equation}
f_c = \frac{1}{2\pi}\frac{eB}{m_{\rm e}+E_{\rm{kin}}/c^2},
\label{eq:energytofrequency}
\end{equation}
where $e$ is the magnitude of the electron charge, $m_{\rm e}$ is the mass of the electron, and $c$ is the speed of light in vacuum. Project~8 aims to combine CRES with the use of atomic tritium as the beta decay source, to remove uncertainty from molecular states.

CRES was first demonstrated with a precision measurement of the energies of single electrons from $^{83\mathrm{m}}$Kr decay by the Project 8 collaboration in 2014 \cite{Asner:2014cwa} (Phase~I). Subsequently, Project 8 has performed a CRES-based tritium experiment (Phase II), the subject of this paper. We report details of tritium endpoint and neutrino mass results obtained from the small-scale Phase II apparatus. These results rely on the first measurement of the continuous tritium spectrum using CRES and the first quantitative exploration of systematic effects in CRES. A companion Letter~\cite{Project8:2022hun} provides a high-level overview of Project 8's Phase II results.

This paper is organized as follows. 
In \autoref{sec:apparatus}, the apparatus and data-taking conditions are described in sufficient detail to give context to the analysis. A more comprehensive paper on the apparatus is in preparation. \autoref{sec:datafeatures} describes the general features of CRES data and the data sets to be analyzed.
In \autoref{sec:simCRES}, CRES signals are reproduced in simulation and compared to experimental data.
\autoref{sec:analysisoverview} provides an overview of the analysis approach, including a general CRES signal model and all its inputs. \autoref{sec:Kr_model} presents calibration measurements with $^{83 \rm m}$Kr that produce inputs to tritium data analysis. $^{83 \rm m}$Kr data are also used to validate the frequency-energy relation in \autoref{eq:energytofrequency} and to demonstrate the high-resolution capabilities of CRES. 
\autoref{sec:tritium_model} presents two models of tritium spectra: one for Monte Carlo data generation and one for tritium data analysis. These models build on the general CRES signal model. 
\autoref{sec:systematic_uncertainties} describes parameter inputs and systematic uncertainties for the tritium analysis. 
\autoref{sec:final-analysis} then describes procedures for analyzing tritium data using both Bayesian and frequentist approaches. In \autoref{sec:results}, we present endpoint and neutrino mass results. In addition, a search for events above the endpoint produces a stringent limit on the background rate. The sensitivity of this Phase II experiment to the neutrino mass is compared to an analytic prediction.

\section{Apparatus \label{sec:apparatus}}

\begin{figure}[tb]
  \centering
  \includegraphics[width=\columnwidth]{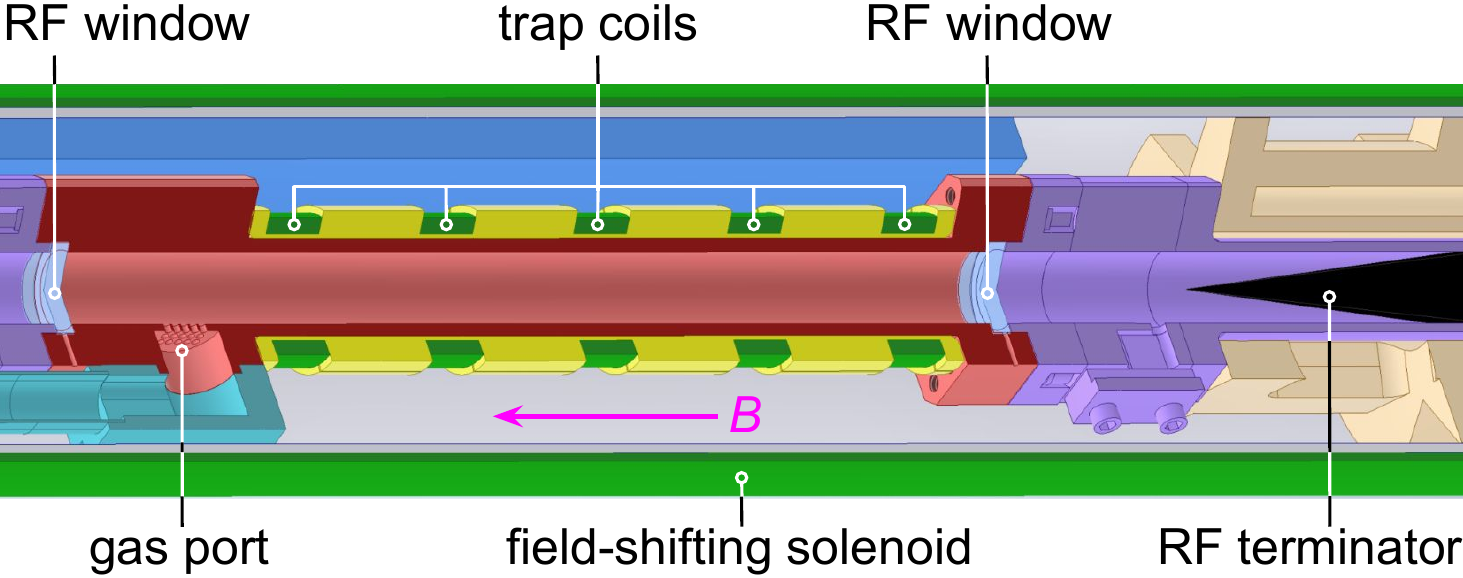}
  \caption{Cutaway of the cryogenic CRES cell, where electrons are produced in radioactive decay and magnetically trapped. The cell waveguide has a cold interior diameter of 10.03 mm and length of 132 mm (distance between RF windows). Cyclotron radiation travels axially up the waveguide (left in rotated view), toward the amplifiers and readout electronics.}
  \label{fig:apparatus}
\end{figure}

In the Project 8 Phase II apparatus,  molecular tritium or $^{83\mathrm{m}}$Kr is confined in a cryogenic gas cell (the ``CRES cell'') within the  field of a commercial warm-bore  superconducting magnet.  The cell, mechanically supported on an experimental insert, is positioned in the vertical  magnetic field of \SI{0.959}{\tesla}, which induces cyclotron motion and confines electrons radially.  The CRES cell is shown in \autoref{fig:apparatus}.  
Electrons emitted in radioactive decay are trapped axially in dips in the magnetic field created by coils wound around the gas cell. The trap geometry affects resolution, event rate, and event characteristics. For Phase II, electrons are trapped in multiple short traps where the electrons' axial motion is near-harmonic, rather than in a single longer trap---both because of non-uniformity in the background field, and because of a Doppler shift, as explained below.  Currents can be varied independently in the five coils to create traps of varying geometries.  Cyclotron radiation emitted by the electrons then propagates through a waveguide, is amplified in two low-noise cryogenic amplifiers in series, and passes on to room-temperature elements of the detection chain.

The 1.6-m insert which supports the CRES cell is cooled by a Cryomech AL-60 Gifford-McMahon cryocooler. 
The cell temperature is controlled by a heater on a PID loop.  The dominant noise source in the experiment is thermal, set by the temperature of a custom-made radio-frequency (RF)-absorbing terminator at the lower end of the cell.
To reduce noise, the cell is kept as cold as possible.  A lower limit on the cell temperature of \SI{85}{\kelvin} is set by the requirement that a sufficient density of $^{83\mathrm{m}}$Kr remains in the gas phase. This temperature is maintained during all data taking to avoid altering systematic effects between $^{83\mathrm{m}}$Kr and T$_2$.

The Phase II apparatus was previously described in~\cite{Esfahani:2017dmu}. Further details on the apparatus and tests thereof will be reported in a paper in preparation.

\subsection{Magnetic trap}\label{sec:electron_trap}\label{fss_procedure}

The criterion for adiabatically trapping a charged particle in a magnetic trap is~\cite{Jackson:1967zz}
\begin{eqnarray}
    \theta_0 &\geq&  \sin^{-1}\left( \sqrt{ \frac{ B_0}{B_{\rm max}}}\right),
    \label{eq:pitchanglerange}
\end{eqnarray}
where $\theta_0$ is the pitch angle at $B_0$, the lowest-field point, and $B_{\rm max}$ is the smaller of the maximum field values on either side of $B_0$.  Pitch angle is defined as the angle between the electron's momentum and the local field direction.

Cyclotron radiation propagates along the waveguide axis toward the detection system. 
Because the trapped electrons undergo axial motion, their cyclotron radiation is Doppler-shifted at twice the axial frequency $f_a$. The maximum Doppler shift is proportional to the electron's axial velocity as it passes through the trap minimum.  The resulting frequency modulation creates sidebands to the mean cyclotron frequency (termed the `carrier').  The modulation index $h$ is the ratio of the peak shift in carrier frequency to the modulation  (here, axial) frequency: 
\begin{eqnarray}
    h&=&\frac{\beta c f_c}{v_\phi f_a}\cos{\theta_0},
\end{eqnarray}
where $\beta c$ is the electron's speed and $v_\phi$ is the phase velocity in the waveguide.  There is also a frequency shift associated with the magnetic field variation along the electron's trajectory, but it is much smaller and neglected here.  For large $h$, sidebands proliferate and the carrier becomes too weak to be detected~\cite{Esfahani:2019mpr}.   Our noise threshold for detecting the carrier corresponds approximately to $h\le1$. In magnetic traps this condition translates directly to a lower limit on the axial frequency, which dictates short traps.  For a given value of $h$, the magnetic field at the turning point is:
\begin{eqnarray}
    B_{\rm m}=B_0\left[1-\left(\frac{hv_\phi f_a}{\beta c f_c}\right)^2\right]^{-1}.
\end{eqnarray}
To increase statistics, several short traps are spaced along the cell in the most uniform region of the background field.  

\begin{figure}[htb]
    \centering
    \includegraphics[width=0.48\textwidth]{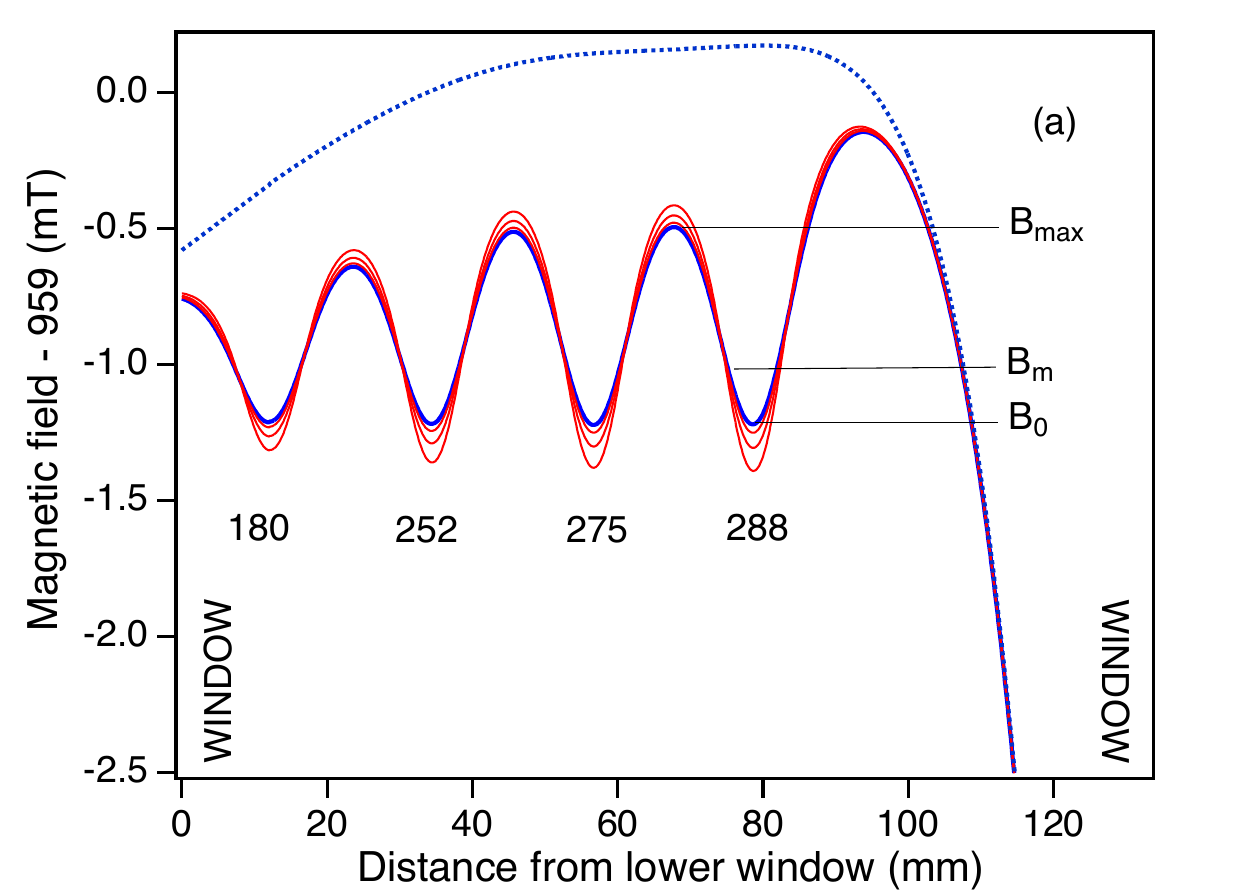}
    \includegraphics[width=0.48\textwidth]{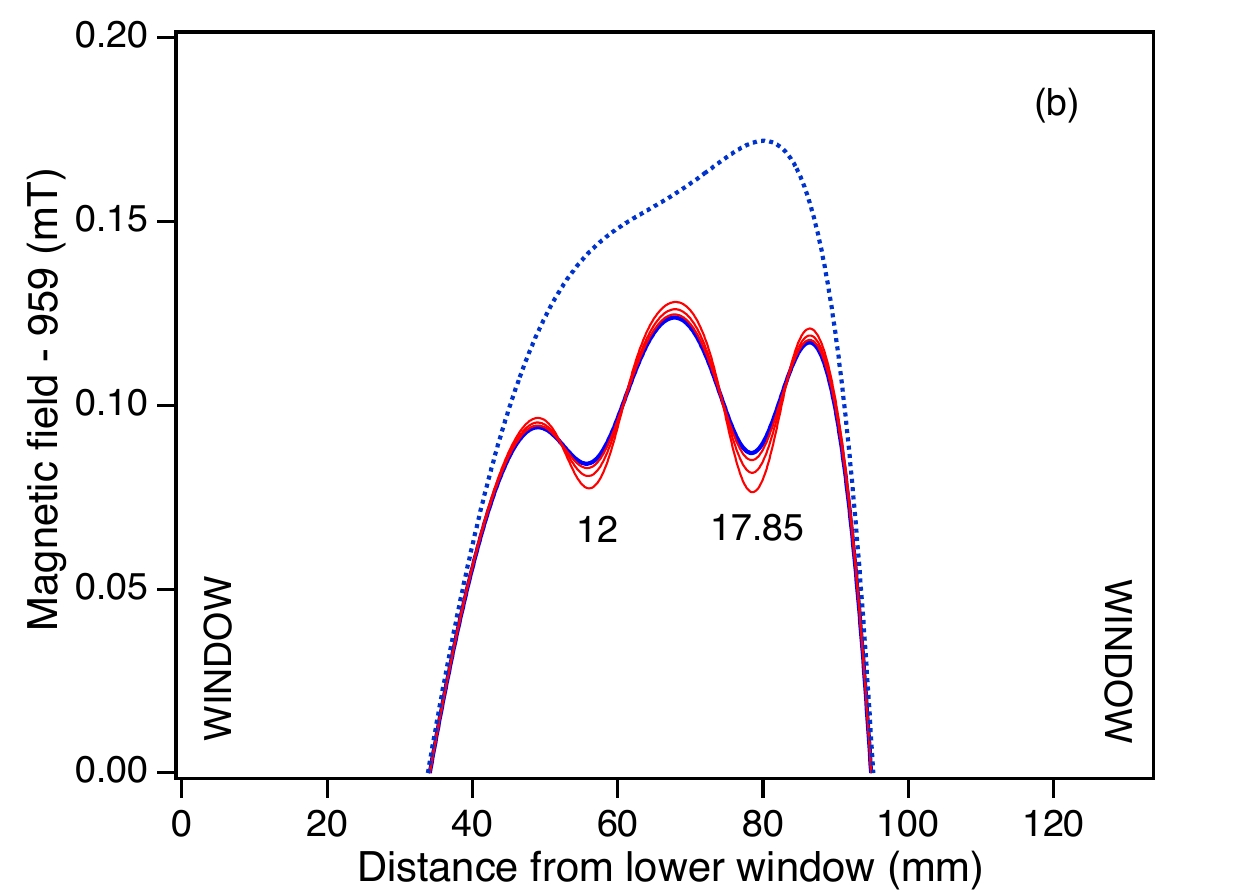}
    \caption{(a) Calculated deep quad trap coil field. (b) Calculated shallow double trap coil field. Note different scales on $y$-axes. The background field from NMR measurements is shown as a blue dotted line. The heavy solid blue lines are for a radius \SI{1}{\mm} from the axis, and the fine red lines are for 2, 3, and \SI{4}{\mm} in turn. The coil currents  are shown in \SI{}{\mA} under each trap. Trap depths were equalized for the 1-mm radius because the coupling to the waveguide's TE$_{11}$ mode maximizes on the axis.}
    \label{fig:quadtrapcoils}
\end{figure}

Data were taken in two trap configurations. A shallow double trap was used to demonstrate the high resolution capability of the CRES technique; this trap is shown in \autoref{fig:quadtrapcoils}(b).  A deep `quad' trap with the magnetic field shown in \autoref{fig:quadtrapcoils}(a) 
was used to maximize the number of trapped electrons and to increase effective volume $V_{\rm eff}$ in this small apparatus.  For a single trap,
\begin{eqnarray}
    V_{\rm eff}&=& 2\pi\int_0^{r_{\rm max}} \int_{z_{\rm min}-a_{\rm m}}^{z_{\rm min}+ a_{\rm m}}\sqrt{1-\frac{B(z,r)}{B_{\rm m}(r)}}rdrdz, \label{eq:veff}
\end{eqnarray}
where $r_{\rm max}$ is the waveguide radius, $z_{\rm min}$ is the location of the trap minimum,  and $a_{\rm m}$ is the axial distance between fields $B_0$ and $B_{\rm m}$. 
\autoref{fig:quadtrapcoils}(a) illustrates the defined fields for trap 4, with $B_{\rm m}$ shown for $h=1$ at 1-mm displacement from the axis.

Four of the trap coils shown in \autoref{fig:apparatus} were used; the fifth was not used because the background field varies too steeply there to form a trap.  The 0.959-\SI{}{\tesla} background field is also shown in \autoref{fig:quadtrapcoils}. 
It deviates from homogeneity at the \SI{5 e-4}{} level because of a non-functional trim coil of the superconducting magnet, which causes the slope and curvature seen in the figure.    

Table~\ref{tab:quadtrapcoils} lists parameters for the  traps. In the table, $z_{\rm min}$ is the location of the trap minimum relative to the lower CRES cell window.  Estimates of $B_0$ are also included; calibrations with $^{83\mathrm{m}}$Kr produced more precise field estimates, as discussed in \autoref{sec:deep-trap-data-and-fits} and \autoref{sec:Bfield_errors}. The angle $\theta_0(h=1)$ is the minimum detectable pitch angle at trap center. The effective volume $V_{\rm eff}$  of each trap is calculated numerically with the aid of \autoref{eq:veff} for $r_{\rm max}=$ \SI{5.02}{\mm} (the waveguide radius), and for $a_{\rm m}$, $B_{\rm m}$ corresponding to $h=1$. Other sources of inefficiency, such as the Larmor radius limitation, mode-coupling threshold,  and track and event reconstruction, are not included in $V_{\rm eff}$.  Those effects are treated as efficiency terms, described and tabulated in \autoref{sec:sensitivity}. Also shown in the table is the minimum trapped pitch angle $\theta_0$(min).
\begin{table}[tb]
    \centering
    \caption{Calculated quad and shallow trap configurations.}
    \begin{tabular}{lrrrrl}
    \hline\hline
         & Coil 1 & Coil 2 & Coil 3  & Coil 4  & Unit \\
    \hline
    Turns & 64 & 63 & 64 & 65  \\
    \hline
    \multicolumn{6}{l}{\bf Deep quad trap configuration parameters}\\
    Current & 180 & 252 & 275 & 288  & mA \\
    $z_{\rm min}$  & 11.9 & 34.4  & 56.7  &  78.6 & mm  \\
    $B_0-959$\phantom{aa}  & -1.21173 & -1.21768  & -1.22141  &  -1.21918 & mT \\
    Trap depth  & 0.45423 & 0.57720 & 0.70887 & 0.72418 & mT \\
    $\theta_0(h=1)$  & 89.47 & 89.37 & 89.33 &  89.30 & deg \\
    $\theta_0$(min)  & 88.75 & 88.59 & 88.44 &  88.42  & deg \\
    $V_{\rm eff}$  & 3.0 & 3.6 & 3.9 &  4.1 & mm$^3$ \\
    \hline
    \multicolumn{6}{l}{\bf Shallow double trap configuration parameters} \\
    Current & & & 12 & 17.85 & mA \\
    $z_{\rm min}$ & & & 55.6  &  78.4 & mm \\
    $B_0 - 959$ & &  & 0.084724  &  0.087937 & mT \\
    Trap depth &  &   & 0.009905  &  0.036838 & mT \\
    $\theta_0(h=1) $ &  &  & 89.89 & 89.84 & deg \\
    $\theta_0$(min)  &  &  &  89.82 & 89.64  & deg \\
    $V_{\rm eff}$ &  &  & 0.61 &  0.96 & mm$^3$\\
    \hline
    \hline
    \end{tabular}
    \label{tab:quadtrapcoils}
\end{table}

A long additional copper coil called the ``field-shifting solenoid'' (FSS) was inserted into the superconducting solenoid's warm bore that encompasses the CRES cell (\autoref{fig:apparatus}). The FSS was used to shift the homogeneous magnetic field for studies of detection efficiency as a function of frequency.
By running current through this additional coil, the field was shifted in steps of $\Delta B_{\rm FSS}=0.07$\,mT over a range of \SI{\pm 3}{mT}. 

A single-axis fluxgate magnetometer (Schonstedt Instrument Co.~DM2220) was used during the final $^{\mathrm{83m}}$Kr calibration phase to assess the role of environmental magnetic field changes in the laboratory.  Over a 3-day period, peak-to-peak variations of \SI{6}{\percent} in the 0.11-\SI{}{\milli\tesla} vertical laboratory background field were observed at a distance of \SI{3}{m} from the magnet. These variations correspond to a \SI{0.3}{\eV} difference in reconstructed electron energies, much smaller than the 50-eV line width of the quad trap. The variation within the magnet bore is expected to have been even lower due to the self-shielding provided by a superconducting magnet in persistent mode.  While the fluxgate magnetometer was not available during tritium running, the influence of environmental field variations is considered negligible at the relevant sensitivity level.

\subsection{Gas system}\label{sec:gas_system}

The radioactive gases were released into the apparatus from a custom gas manifold. This manifold connected to the CRES cell via a delivery line running along the insert and through a grid of  holes with diameter less than 0.1 wavelengths of the microwave radiation. The gas system could be run in two modes.  In neutrino mass data acquisition mode, molecular tritium gas was delivered, while in calibration mode, $^{83\mathrm{m}}$Kr gas was delivered. Tritium was stored in a 0.5-\SI{}{\g} non-evaporable getter (SAES ST 172/HI/7.5-7/150 C), the temperature of which was controlled by an ion gauge to maintain the desired operating pressure, usually \SIrange[range-phrase = --,range-units = single]{1.6}{2.6e-6}{\milli\bar} (calibrated for H$_2$). Pressure could be maintained to \SI{\pm3}{\percent}  run-to-run for the duration of data-taking.  In later data sets, we prevented accumulation of $^3$He (produced in tritium decay) by extracting gas continually through a leak valve to a getter-ion pump, which also lowered the concentration of traces of methane, Ar, CO, and CO$_2$ impurities; this is referred to as the ``pumped'' configuration. The initial \SIadj{2}{Ci} inventory of tritium was sufficient for a \SIadj{\sim 100}{\day} data-taking period and the gas was not recycled.  Gas composition could be monitored by two residual gas analyzers: an SRS-100 close to the SAES getter, and an Extorr XT100 at the getter-ion pump manifold.  

For calibration studies, the $^{83\mathrm{m}}$Kr gas emanated from $^{83}$Rb (\SI{6}{mCi} on July 19, 2019), adsorbed on zeolite  \cite{Venos:2005vn}. The Kr was mixed with H$_2$ to tune the mean time between electron-gas collisions to match that in tritium data. The H$_2$ was stored in a separate getter and pressure-controlled in the same way as the tritium.

\subsection{Radio-frequency system}\label{subsec:rf_system}

\begin{figure}[htb]
    \centering
    \includegraphics[width=\columnwidth]{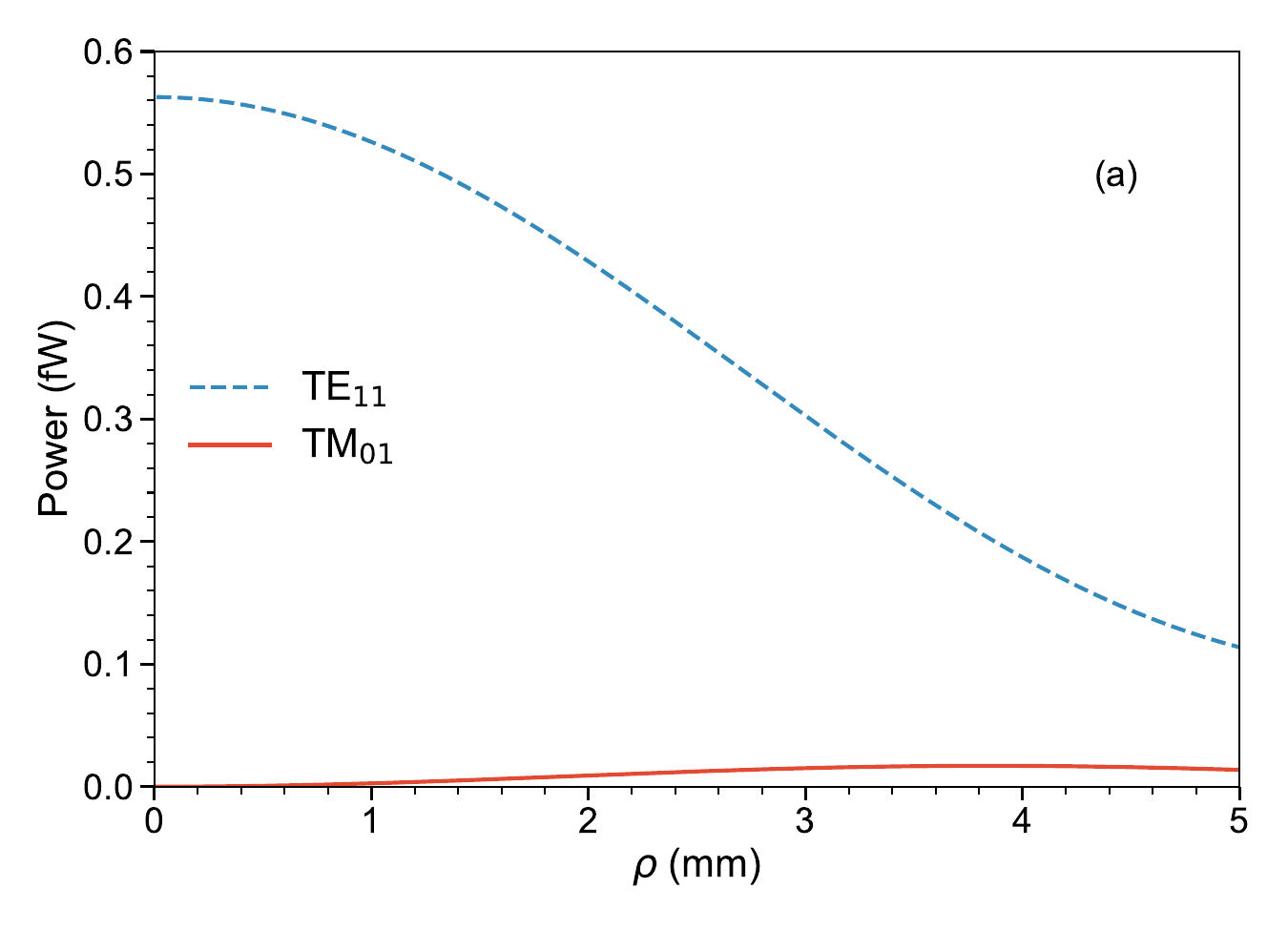}
    \includegraphics[width=\columnwidth]{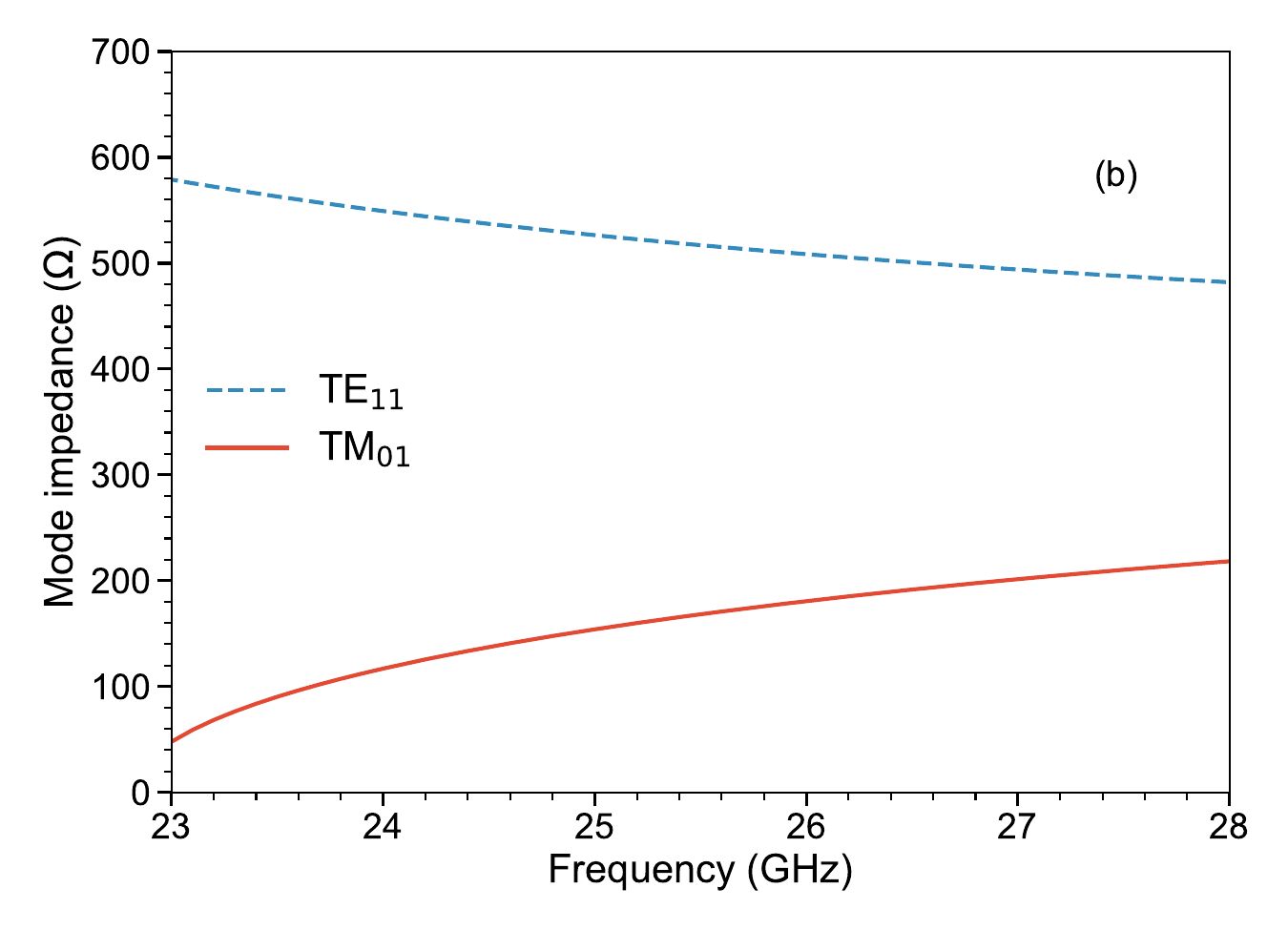}
    \caption{(a) Coupling of an electron to the two propagating modes as a function of the distance $\rho$ from the axis. The electron has energy 18.6\,keV ($f_c \approx 25.9\,\si{GHz}$) and  pitch angle $90^\circ$.  (b) Mode impedance for the two propagating modes.}
    \label{fig:modecoupling}
\end{figure}

The CRES cell forms the first section of the waveguide through which cyclotron radiation travels toward the amplifiers (Low Noise Factory LNF-LNC22\_40WA). For the Phase II decay cell used in this work, a circular waveguide was chosen over the rectangular WR-42 used previously~\cite{Asner:2014cwa}, to increase the volume and to accept circular polarization. In the frequency range \SIrange[range-units = single]{25.8}{27.0}{\giga\hertz}, waveguide of radius \SI{5.02}{\milli\meter} supports two propagating modes, TE$_{11}$ and TM$_{01}$. The electron couples well to the TE$_{11}$ mode and only weakly to the TM$_{01}$ mode (\autoref{fig:modecoupling}). 
The coupling is maximal on-axis and falls off to a small value near the wall.  

The cyclotron radiation emitted by a trapped electron propagates both downward and upward, passing through RF-transparent CaF$_2$ windows (United Crystals Inc.) that confine the radioactive gas (Fig.~\ref{fig:apparatus}). This crystalline material has a coefficient of thermal expansion that matches copper at low temperatures, and it is known to have low permeability to tritium~\cite{osti_4351657}. To minimize interference from reflections, the downward-propagating radiation is absorbed in a custom-made cryogenic RF terminator.  Beyond the uppermost window, the upward-propagating circularly polarized radiation is converted to linear polarization by a quarter-wave `plate.' The radiation is then
transmitted via a WR-42 single-mode rectangular waveguide, including a gold-coated stainless steel section for thermal insulation, to cryogenic amplifiers held at \SI{30}{\kelvin}. 
Residual reflections from the windows, joints, and transitions 
create weak resonant cavity modes within the gas cell, which enhance spontaneous emission at particular frequencies and locations within the trap.  These resonances significantly modify the response to signals from electrons as a function of trap position and frequency, presenting a difficult analysis challenge.

After cryogenic and room-temperature amplification, room-temperature RF electronics downmix the signal by \SI{24.5}{\giga\hertz}. The downshifted signal is digitized by a ROACH2 DAQ system~\cite{Hickish2016} at \SI{3.2} gigasamples per second, with an FPGA performing digital downconversion to \SI{200} megasamples per second and Fast Fourier Transforms (FFT)  for three separate frequency windows with independently-set center frequencies.
When two 40.96-\SI{}{\micro\second} bins within any 0.5-\SI{}{\milli\second} window exceed a signal-to-noise ratio (SNR) threshold, a compute node writes time-series data to disk. We calculate SNR as the ratio of the power in 24.4\,kHz wide frequency bins to the average power of all bins in a spectrogram. This SNR is used instead of absolute power in all stages of data analysis (triggering, event reconstruction, spectrum analysis) to avoid the effects of gain variation of the amplifiers and filters.

\section{CRES data features \label{sec:datafeatures}}

\subsection{Electron event properties}
\label{subsec:electron_data}

\begin{figure}[htb]
\centering
\includegraphics[width=0.45\textwidth]{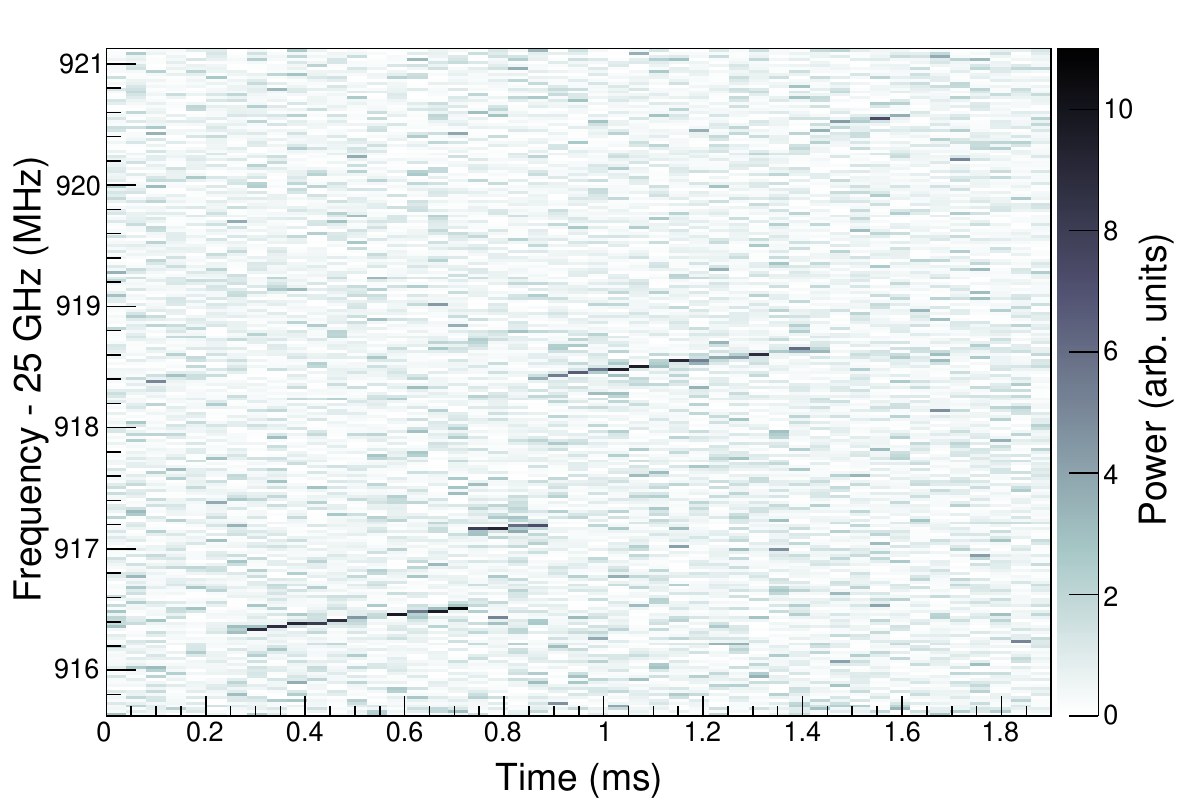}
\caption{Spectrogram of chirped electron signals of a single radiating electron. The bin size is 24.4\,kHz by 40.96 $\mu$s. }
\label{fig:spectrogram}
\end{figure}

Electron events may be displayed in a spectrogram (or ``waterfall plot'') of frequency vs.~time, with pixels indicating signal power by color or intensity. Figure~\ref{fig:spectrogram} shows a typical event, which is continuous in time but discontinuous in frequency. Events are composed of ``tracks,'' with jumps in frequency between tracks. The frequency jumps are due to both energy loss and pitch angle changes caused by collisions with gas molecules. 
Pitch-angle changes cause the amplitude of the electron's axial motion to increase or decrease, so the average magnetic field experienced by the electron may increase or decrease.  Between collisions, signals continuously chirp upward in frequency as the electrons radiate energy \cite{Esfahani:2019mpr}.

We use a point-clustering algorithm to identify high-SNR bins occurring close together in time and frequency as belonging to the same track~\cite{furse2015techniques}.  A reconstruction algorithm extracts the initial frequency of the first track of each event, identifying it as the ``start frequency'' of the event. It is this start frequency that is used to calculate the kinetic energy at the time of decay via \autoref{eq:energytofrequency}.

One of CRES's promising features is its immunity to  background. 
Charged particles originating on the wall are returned to the wall by the magnetic field within one cyclotron orbit or one axial cycle, before they can be observed. The interactions of cosmic rays and energetic beta and gamma backgrounds with the gas can in principle lead to the production and trapping of an electron in the right energy range, but this is a rare occurrence because of the low density of the gas.  
The dominant background is expected to be from RF noise fluctuations that form false tracks.
We distinguish electron events from RF noise by searching for upward-sloping tracks and by performing cuts as a function of three characteristics: the number of tracks in a candidate event, the duration of the first track, and the SNR of the first track. These characteristics are distributed differently for electron events and RF noise fluctuations.
We set the threshold for this cut before tritium data acquisition at a level expected to allow less than one RF-noise-induced background event beyond the endpoint per 100 days of run time at \SI{90}{\percent} confidence level.

Even with the successful elimination of false tracks, the first track {\em visible} in an event is not necessarily the first track following beta decay or internal conversion.  Tracks can be too short or too low in power to be detected. The detector response function described later accounts for such missed tracks.

\begin{table*}[htb]
\centering
\footnotesize
\caption{Characteristics of key data sets. $N_{\mathrm{events}}$ is the number of reconstructed events in a data set after all cuts. $N_{\mathrm{tracks}}^{\mathrm{true}}$ is the estimated mean number of electron tracks per event (including unreconstructed tracks). $\tau$ is the mean time between electron-gas collisions. Tritium data were acquired in an energy range of 16.2-19.8\,keV.}
\begingroup
\setlength{\tabcolsep}{6pt} 
\begin{tabular}{llllrrrr}
\hline\hline
Data Set & Purpose & \begin{tabular}[c]{@{}l@{}}Magnetic field\\ configuration\end{tabular} & \begin{tabular}[c]{@{}l@{}}Gas system\\ configuration\end{tabular} & $N_{\mathrm{events}}$ & $N_{\mathrm{tracks}}^{\mathrm{true}}$ & $\tau$ (ms) & Section(s) \\ \hline 
\begin{tabular}[c]{@{}l@{}}$^{83\mathrm{m}}$Kr\\ shallow\end{tabular} & \begin{tabular}[c]{@{}l@{}}\\Demonstrate best \\ resolution and probe\\ detector response\end{tabular} & \begin{tabular}[c]{@{}l@{}}Shallow \\ double trap\end{tabular} & Not pumped & 6831 & $1.24(12)$ & $0.342(5)$ & \begin{tabular}[c]{@{}l@{}} \autoref{sec:frequency-energy_relation}, \\ \autoref{sec:shallow_trap}
\end{tabular}\\
\begin{tabular}[c]{@{}l@{}}$^{83\mathrm{m}}$Kr \\ field-shifted\end{tabular} & \begin{tabular}[c]{@{}l@{}}\\  Measure frequency \\ dependence of \\ efficiency and \\ other characteristics\end{tabular} & \begin{tabular}[c]{@{}l@{}}\phantom{a}\\ Deep quad trap \&\\ deep single traps;\\ background B \\ field varied\end{tabular} & Not pumped & 9350 & $2.73(24)$ & $0.173(3)$ & \autoref{subsec:frequency_dependence} \\ 
\begin{tabular}[c]{@{}l@{}}$^{83\mathrm{m}}$Kr \\ pre-tritium\end{tabular} & \begin{tabular}[c]{@{}l@{}}\\Calibrate magnetic \\ field, scattering \\ environment\end{tabular} & Deep quad trap & Not pumped & 87634 & $2.40(27)$ & $0.162(1)$ & \autoref{sec:deep_trap} \\ 
Tritium & \begin{tabular}[c]{@{}l@{}}\\Spectroscopy of \\ tritium endpoint\end{tabular} & Deep quad trap & Pumped & 3770 & $2.36(34)$ & $0.146(3)$ &    
\begin{tabular}[c]{@{}l@{}}\autoref{sec:systematic_uncertainties}, \\ \autoref{sec:final-analysis}\end{tabular} \\
\begin{tabular}[c]{@{}l@{}}$^{83\mathrm{m}}$Kr\\ post-tritium\end{tabular} & \begin{tabular}[c]{@{}l@{}}\\Calibrate magnetic \\ field, scattering \\ environment\end{tabular} & Deep quad trap & Pumped & 47426 & $3.37(21)$ & $0.188(1)$ &  \autoref{sec:deep_trap} \\
\hline
\hline
\end{tabular}
\endgroup
\label{tab:mml_track_length_fit_results}
\label{tab:data_set_table}
\end{table*}

\subsection{Data set features}

Studies of the efficiency, energy response, and magnetic field were carried out with $^{83\mathrm{m}}$Kr during the second half of 2019.  Tritium data taking began in mid-December and extended into March 2020, with several days of downtime in February for laboratory maintenance. A final week of $^{\mathrm{83m}}$Kr measurements concluded the data-taking campaign in March as planned, just before a global pandemic precluded further data-taking and laboratory work. 

Properties of key data sets are summarized in \autoref{tab:mml_track_length_fit_results} in chronological order.\footnote{In the table and throughout the paper, 3.4(12) signifies $3.4\pm1.2$.}
The mean number of electron tracks per event, $N_{\mathrm{tracks}}^{\mathrm{true}}$, is determined from the post-reconstruction tracks per event in data combined with simulation studies of the relationship between true and reconstructed tracks per event. Section~\ref{sec:scatter_peak_errors} describes how $N_{\mathrm{tracks}}^{\mathrm{true}}$ is used in tritium data analysis. Section~\ref{subsec:track_length} describes how $\tau$, the mean time between electron-gas collisions, is extracted from the data and used in tritium analysis. For tritium data, and for all data sets that provide direct calibration input to the tritium analysis ($^{83\mathrm{m}}$Kr field-shifted, $^{83\mathrm{m}}$Kr pre-tritium, and $^{83\mathrm{m}}$Kr post-tritium), the gas composition and electron-gas scattering rate were kept as similar and stable as possible, despite the absence of krypton and the presence of helium during tritium data-taking.

$^{83\mathrm{m}}$Kr data  were acquired in periods lasting a few hours each for deep quad trap data sets. Shallow-trap data sets took several days to acquire adequate statistics. With maximum event durations of \SI{<10}{\milli\second} and rates of \SI{\sim 1}{cps} (counts per second), pileup effects were negligible.\footnote{Were pileup present, any two simultaneous electron events would typically have distinct cyclotron frequencies and would therefore be distinguishable in the data. The Project 8 collaboration is studying potential pileup effects in future phases.} A single \SIadj{100}{\mega\hertz}-bandwidth DAQ channel was used to acquire data on the \SIadj{17.8}{\kilo\eV} K internal-conversion line, and sometimes the second channel or both the second and third channels simultaneously took data on the L lines ($30.4\,\si{keV}$) or the M and N lines ($31.9\,\si{keV}$ and $32.1\,\si{keV}$).

Tritium data were taken over 82 days in the deep quad trap configuration, with a mean event rate of \SI{0.5 e-3}{cps}.
The analysis window spanned \SIrange[range-phrase = --,range-units = single]{25.81}{25.99}{\giga\hertz}, or \SIrange[range-phrase = --,range-units = single]{16.2}{19.8}{\kilo\eV} in electron energy.
The three DAQ windows overlapped to minimize efficiency variation with frequency due to windowing. Only the highest-efficiency channel is analyzed in the overlap regions. 

\subsection{Scattering}
\label{sec:gas_composition}
\label{gamma_i_determination}

Here we describe our assessment of the relative probability $\gamma_i$ for an electron to inelastically scatter with gas species $i$. These probabilities are needed to model the distribution of energy losses from missed tracks, since the energy loss between two tracks depends on which gas the electron scatters with.

We neglect elastic scattering as an energy-loss mechanism between tracks within an event because it tends to produce pitch angle changes of ${\geq} 5^\circ$~\cite{DavidJoy:ElectronScattering}, ejecting the electron from the trap and terminating the event. By contrast, inelastic scatters produce ${\sim}0.1 ^\circ$ changes \cite{DavidJoy:ElectronScattering}---generally small enough that the electron remains trapped.
In addition, elastic scattering cross sections are an order of magnitude smaller than inelastic cross sections for the most prevalent gas species. 
   
For key data sets, $\gamma_i$ are derived from the mass composition measured with the quadrupole mass analyzer, with
\begin{eqnarray}
    \gamma_i &=& \frac{\sigma_{i,E}\, C_{\mathrm{temp},i}\, p_{\mathrm{raw,}i}/s_i}{\sum_n(\sigma_{n,E}\, C_{\mathrm{temp,}n}\, p_{\mathrm{raw,}n}/s_n)},
    \label{eq:gammai}
\end{eqnarray}
where $\sigma_{i,E}$ is the total inelastic scattering cross section of an electron of energy $E$ with gas $i$; $C_{\mathrm{temp},i}$ is a factor accounting for gas freezing to CRES cell walls; $p_{\mathrm{raw,}i}$ is the uncorrected partial-pressure reading of the quadrupole mass analyzer; and $s_i$ is the sensitivity factor of the quadrupole mass analyzer to gas species $i$. The uncertainty on each of these quantities is propagated through to the uncertainty on $\gamma_i$ in the standard way. 

The inelastic scattering cross sections $\sigma_{i,E}$ are derived from literature values. Contributions to uncertainties on $\sigma_{i,E}$ include both uncertainties within and differences between published data sets. Values of $\sigma_{i,E}$ for H$_2$ and T$_2$ come from a measurement at 18.6 keV \cite{Aseev2000} and are scaled according to \cite{Liu:1987ka} for 17.8\,keV. Cross sections on $^3$He, Kr, and Ar are taken from \cite{Cullen:1989ig}, with uncertainties from \cite{nagy1980absolute, Brusa:1996hw, zecca2000electron, Cartwright:1992he}. For CO, the cross section is evaluated using the expression from \cite{Hwang:1996hf}, with uncertainties from \cite{Rapp:1965kf, Kanik:1993ge}. 

\begin{table*}[htbp]
  \centering
  \renewcommand{\arraystretch}{1.15}
    \begingroup
    \setlength{\tabcolsep}{6pt} 
  \begin{tabular}{ l  c  c  c  c c}
              \hline \hline
Data set        & Tritium        & $^{83\mathrm{m}}$Kr shallow           & $^{83\mathrm{m}}$Kr field-shifted          & $^{83\mathrm{m}}$Kr pre-tritium  & $^{83\mathrm{m}}$Kr post-tritium    \\ \hline \hline
Hydrogen isotopes & 91(5)   & 9-98    & 38-98      & 23-91    & 41-99 \\ 
Helium-3          & 8(4)    & 0-99    & 0-59       & 0-67     & 0-58 \\ 
Argon             &         & ${<}$1  & ${<}$1     & 5-10     & ${<}$1 \\ 
Krypton           &         & 1-3     & 2-3        & 2-5      & ${<}$1 \\ 
Carbon monoxide   & 1(1)    &         &            &          & ${<}$1 \\ \hline \hline
\end{tabular}
\caption{Possible ranges of percentages of inelastic scatters due to each of the main gas species present during the tritium data set and the primary $^{\mathrm{83m}}$Kr data sets. Scattering from all non-listed gases for a given data set is negligible. Values are given as ranges with near-uniform probability (with sigmoid edges) for $^{83\mathrm{m}}$Kr data, where deuterium contamination interfered with distinguishing $^3$He from hydrogen isotopes.}
\label{tab:scattering_fraction_results}
\endgroup
\end{table*}

Because Kr is the only relevant gas species for which significant adsorption to cold walls is expected at 85 K, we take $C_{\mathrm{temp},i}$=1 for all other species. The C$_{\mathrm{temp},\mathrm{Kr}}$ value of 0.90(5) is measured from temperature-varying $^{\mathrm{83m}}$Kr CRES data, taking advantage of the direct dependence of event rate on Kr density in the cell.

The manufacturer of the quadrupole mass analyzer used does not publish sensitivity factors for its product, so we adopt sensitivity factors $s_i$ from other quadrupole mass analyzer manuals, with uncertainties estimated from differences between different manufacturers'  values.

We measured the raw partial pressures $p_{\mathrm{raw,}i}$ using the quadrupole mass analyzer. Since $^{\mathrm{83m}}$Kr data sets were completed in hours or at most a few days, gas conditions were stable. Therefore, each data set's $p_{\mathrm{raw,}i}$ measurements are determined from a single representative quadrupole mass analyzer scan. In contrast, tritium data were taken over months, with some variation in conditions, especially initially as pumping speed settings were optimized. Gas composition for tritium data is therefore an average: the sum of measurements taken in each state weighted by the accumulated counts in that state.

One complication in interpreting $p_{\mathrm{raw,}i}$ values comes from the inability to distinguish species with identical charge-to-mass ratio using the available quadrupole mass analyzers. This creates a challenge because deuterium gas was used in the initial testing during commissioning of the gas system, and was mistakenly allowed to contaminate the reservoir of H$_2$ used for $^{\mathrm{83m}}$Kr data. For $^{\mathrm{83m}}$Kr data sets, therefore, mass-3 signals due to $^3$He and HD cannot be distinguished. This modest-quality quadrupole mass analyzer also suffered from zero-blast, making mass-1 and mass-2 measurements insufficiently reliable to measure the relative partial pressures of H$^+$ and D$^+$. This mass-3 $^3$He/HD uncertainty is reflected in the larger error bars on gas composition in the $^{\mathrm{83m}}$Kr data sets. In contrast, the tritium gas supply was not deuterium-contaminated, so the tritium data do not suffer from this uncertainty.

Table~\ref{tab:scattering_fraction_results} shows the inelastic scattering fraction results, as derived from the mass composition measured with the quadrupole mass analyzer and \autoref{eq:gammai}. 
For fits to the $^{\mathrm{83m}}$Kr pre-tritium and post-tritium data sets, to propagate uncertainties, we sample $\gamma_i$ from near-uniform distributions defined according to the ranges in this table, requiring that the sampled fractions sum to 1. This near-uniform shape, flat with sigmoid ends, reflects the inability to distinguish $^3$He from HD in $^{\mathrm{83m}}$Kr data. 

In tritium and $^{\mathrm{83m}}$Kr data, scattering from molecular hydrogen isotopes is dominant, and $^3$He scattering is the next-largest contributor. The $^3$He gas is produced by the decay of tritium in the storage getter and adsorbed to the gas system walls. Tritium data include a small contribution from a mass-28 gas species, likely CO. $^{\mathrm{83m}}$Kr data include contributions from krypton gas and, in some cases, argon (the latter in data sets taken before the pumped gas system configuration was set up, which then enabled lower impurity levels). 

\subsection{Mean track duration \texorpdfstring{$\tau$}{}}
\label{subsec:track_length}

Track duration is the time between successive scatters of an electron with gas molecules.
The efficiency and detector response function are affected by the interaction of the track reconstruction process with the distribution of track durations  (See Appendix \ref{appendix:track_distribution_response_function_etc} for details), making it necessary to assess this distribution for each data set. 

The track duration distribution is modeled as follows. Since track duration is determined by random scattering, which is a Poisson process, the underlying probability density function (PDF) is exponential. However, short tracks are less efficiently detected, so the number $N$ of tracks detected with a duration $t$ is given by
\begin{eqnarray}\label{eqn:track_length_distribution}
 \ N(t) &\propto& P_d(t)e^{-t/\tau},
\end{eqnarray}
where $\tau$ is the mean time between collisions with gas molecules and $P_d(t)$ is the relative probability of detection as a function of $t$. 
We use the following empirical model for $P_d(t)$, which provides good fits to the data and accounts for the roll-off in detection efficiency at low $t$:

\begin{eqnarray}\label{eqn:detection_probability}
 \ P_d(t) &=& \mathrm{erfc}\left(\psi - \sqrt{\frac{\zeta}{t}}\right).
\end{eqnarray}
Here, $\mathrm{erfc}$ is the complementary error function and $\psi$ and $\zeta$ are determined solely by conditions held constant for all data sets (except the field-shifted data), such as SNR and the event-reconstruction algorithm's success rate at reconstructing short-duration tracks. 

To extract $\tau$ in the core tritium and $\mathrm{^{83m}}$Kr calibration data sets, we first determine $\psi$ and $\zeta$ independently by analyzing reconstructed first-track-duration distributions of $\mathrm{^{83m}}$Kr data sets at five different pressures, and therefore five values of mean track duration. We use the Stan software package \cite{Carpenter2017} to perform a Markov chain Monte Carlo (MCMC) Bayesian analysis in which $\psi$ and $\zeta$ are shared for all data sets and $\tau$ is allowed to vary between data sets, with weakly informative priors. The best-fit values and uncertainties for $\psi$ and $\zeta$ are determined from the resulting posterior distributions. 

\begin{figure}[htb]
  \centering
  \includegraphics[width=1.0\columnwidth]{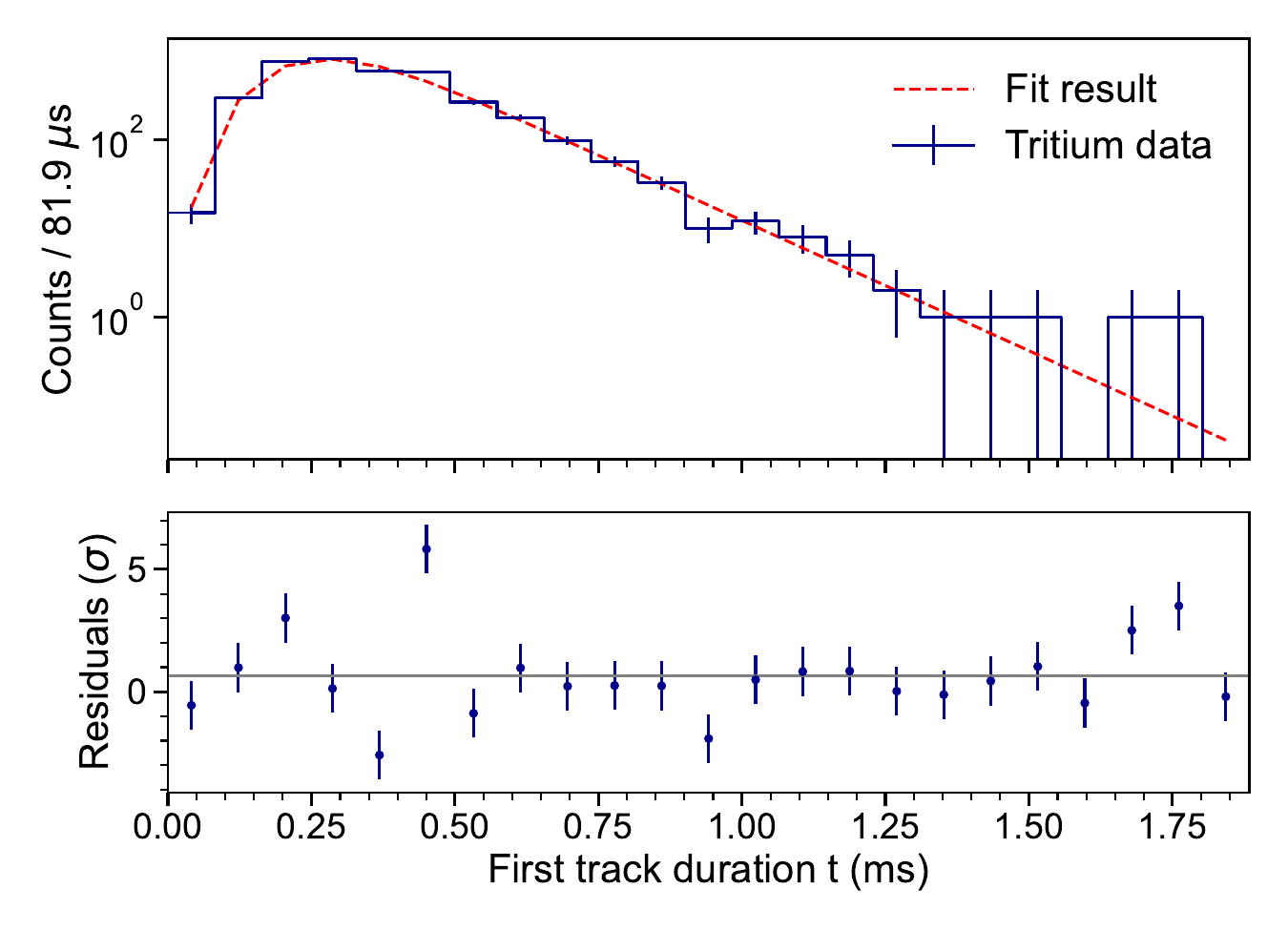}
  \caption{    Tritium data track duration fit using \autoref{eqn:track_length_distribution} and \autoref{eqn:detection_probability}. At low track durations ($< 0.5\,\si{ms}$), the combination of binning and duration-specific cut thresholds in event reconstruction lead to variations in the track counts per bin in this histogram that are not captured by the fitted model. For the analysis in this paper, only the underlying mean track duration is relevant, which is determined by the exponential behavior of the data at long track durations.}
  \label{fig:mll_track_length_fits}
\end{figure}

This information about $\psi$ and $\zeta$ is used in track-duration distribution fits, in which we extract $\tau$ for the core tritium and $\mathrm{^{83m}}$Kr calibration data sets using negative log-likelihood minimization.
\autoref{fig:mll_track_length_fits} shows the fit result for the tritium data set, and \autoref{tab:mml_track_length_fit_results} lists the extracted mean track durations for all data sets.

\section{Simulated CRES data \label{sec:simCRES}}

\label{sec:sim-events}

Simulations are used to generate inputs to the analysis model and to evaluate the performance of event detection and reconstruction methods. It is therefore crucial that simulated events accurately reproduce the features of real data, especially in the main properties relevant for reconstruction: number of tracks per event, track duration, and SNR.

\subsection{CRES signal generation with Locust}
The Locust software package~\cite{AshtariEsfahani:2019mwv} simulates the detection of RF signals by modeling the response of an antenna and receiver to time-varying electromagnetic fields. Locust can independently generate a custom signal to use as input to its receiver chain algorithm, which processes the signal prior to digitization and recording. We developed a Locust signal generator module that simulates chirped data with typical electron event properties. 
The Phase~II waveguide and trap geometries are implemented in this generator to create realistic Phase~II-like event signals. 
Event starting conditions are sampled from probability density functions. 
The generated CRES signals are added to Gaussian white noise. The relative amplitudes of event signals and noise are set to reproduce the SNR observed in experimental data.  
Simulated and experimental data are processed with the same event reconstruction methods.

We generate a set of simulated events to compare to experimental data. 
For this purpose, electrons are sampled at the $\mathrm{^{83m}Kr}$ K-line energy at different radial and axial positions in the waveguide, and with different pitch angles, generating trajectories in response to the magnetic field map of a single-coil trap in the apparatus.
To reproduce multi-trap effects (e.g. SNR and field variation differences between traps), the events from simulations with different trapping fields are combined. 

\subsection{Signal frequency and power}
The average CRES signal frequency and power are calculated from the magnetic field along the electron's trajectory and the power coupled to the $\mathrm{TE}_{11}$ mode. This calculation accounts for the frequency modulation associated with an electron's pitch angle (\autoref{sec:electron_trap}).
Smaller pitch angles lead to reduced power in the carrier and more power in sidebands.  Only the carrier is detected in this experiment.
Field-shifting studies measured SNR differences between the single traps (\autoref{subsec:efficiency}). These differences are accounted for by multiplying the signal power in each trap with a relative SNR factor. For generating a simulated data set that can directly be compared to recorded data, the overall SNR scale is determined by iteratively adjusting the maximum coupled power until the first-track SNR distribution after reconstruction matches that of real data. The maximum coupled power corresponds to SNR$_{\rm max}$, the SNR of a $90^{\circ}$ electron at $r=0\,\si{mm}$ in trap 3. This is the trap in which power is coupled most effectively into the transporting waveguide mode at the CRES frequency of the $\rm{^{83m}Kr}$ K-line. Later, this procedure for setting the SNR scale is replaced by the method described in \autoref{sec:max-SNR-optimization}.

\begin{figure}[tb]
    \includegraphics[width=0.95\columnwidth]{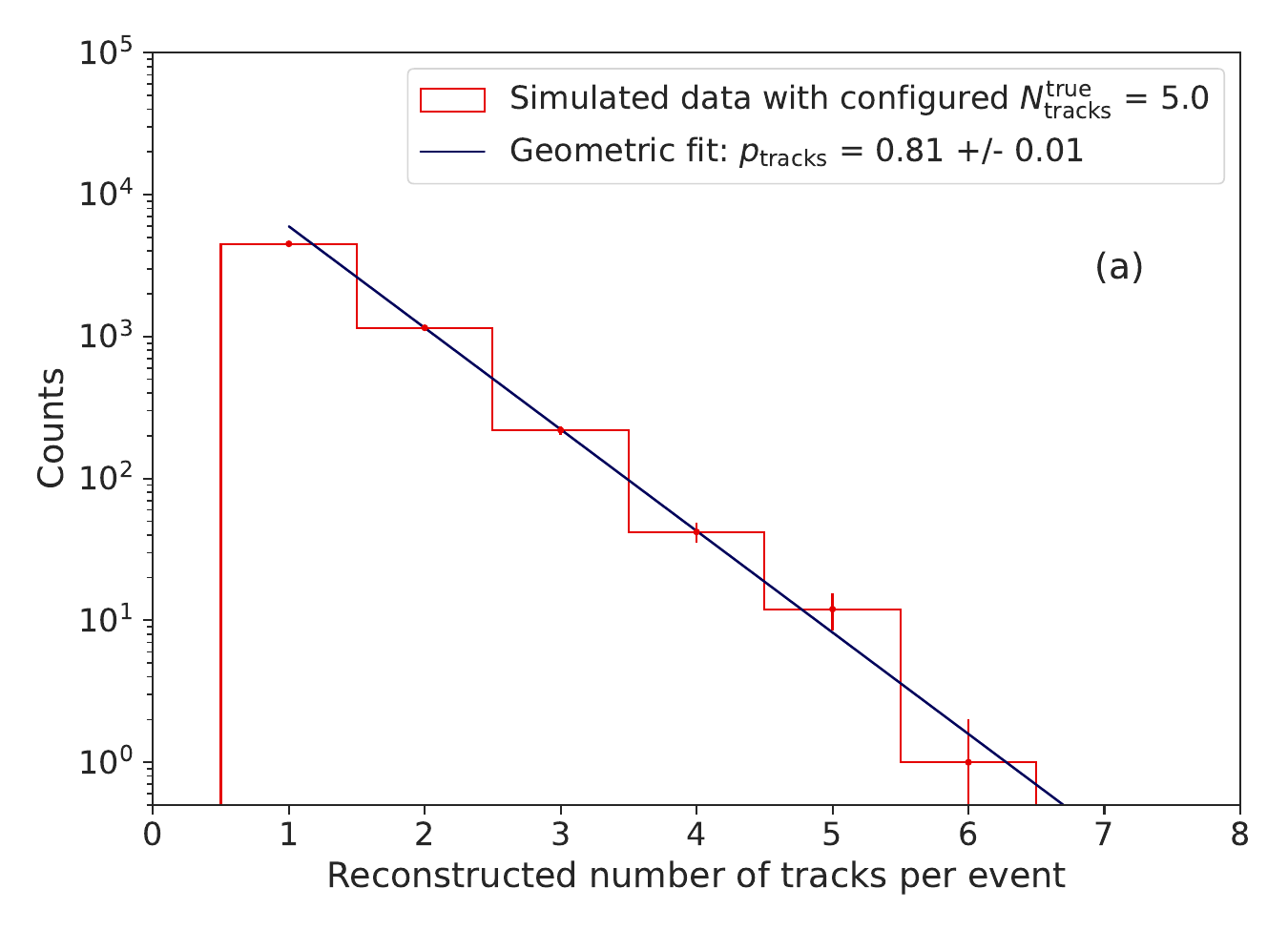}
    \includegraphics[width=0.95\columnwidth]{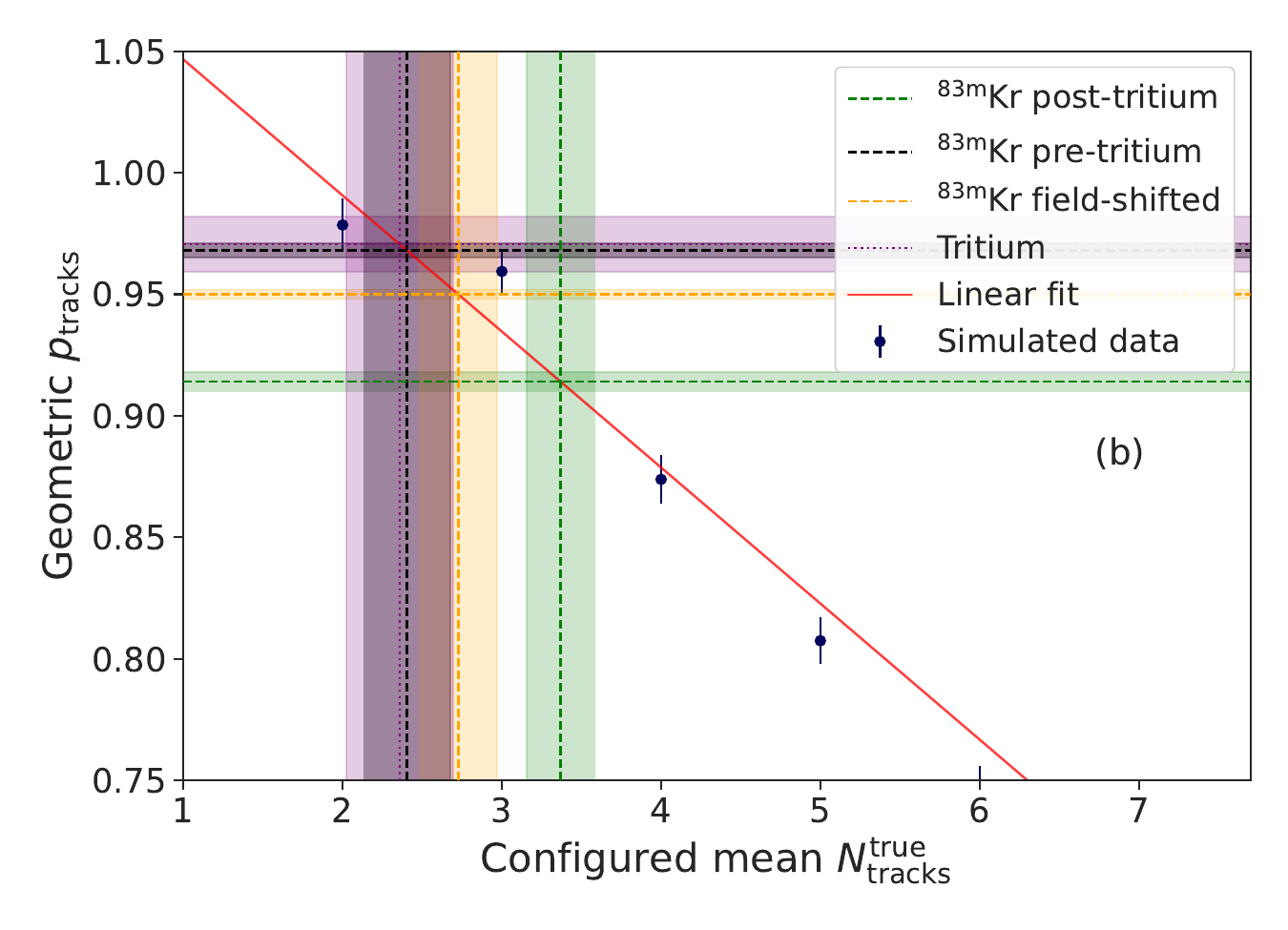}
    \caption{Procedure for extracting the underlying mean number of tracks per event from simulations: (a) For each configured value of $N_{\mathrm{tracks}}^{\mathrm{true}}$, the distribution of reconstructed number of tracks per event is fitted with a geometric distribution of parameter $p_\textnormal{tracks}$. The fit excludes the first bin since the counts are reduced in this bin by event cuts. (b) The intercepts of a linear fit to the fitted $p_\textnormal{tracks}$ for all tested $N_{\mathrm{tracks}}^{\mathrm{true}}$ allows us to find the optimum configuration for each data set. The uncertainty on $p_\textnormal{tracks}$ (horizontal bands) and the uncertainty of the linear fit parameters are propagated to an uncertainty on the extracted mean $N_{\mathrm{tracks}}^{\mathrm{true}}$ (vertical bands).}
    \label{fig:number_of_tracks_extraction}
\end{figure}

\subsection{Simulated event properties}
To simulate multi-track events, a sequence of chirped signals is generated with start frequency and power calculated as described above.
Track slopes are sampled from a Gaussian distribution with a mean ($352.3\,\si{MHz/s}$) and standard deviation ($54.5\,\si{MHz/s}$) corresponding to the mean and standard deviation observed in the deep quad trap $^{\mathrm{83m}}$Kr data. The Gaussian assumption is only approximately valid, but since reconstruction efficiency is relatively insensitive to track slopes as long as they are within several $100 \,\si{MHz/s}$ of the mean, achieving a better agreement of the slope distribution is unnecessary.

The track durations are drawn from an exponential distribution. For each event, the mean track duration $\tau$ that defines this distribution is drawn from a Gaussian with mean and width according to the fit results listed in \autoref{tab:mml_track_length_fit_results}. This way, the uncertainty on the mean track duration is propagated to the simulated data.

\begin{figure}[tbp]
\centering
\includegraphics[width=0.9\columnwidth]{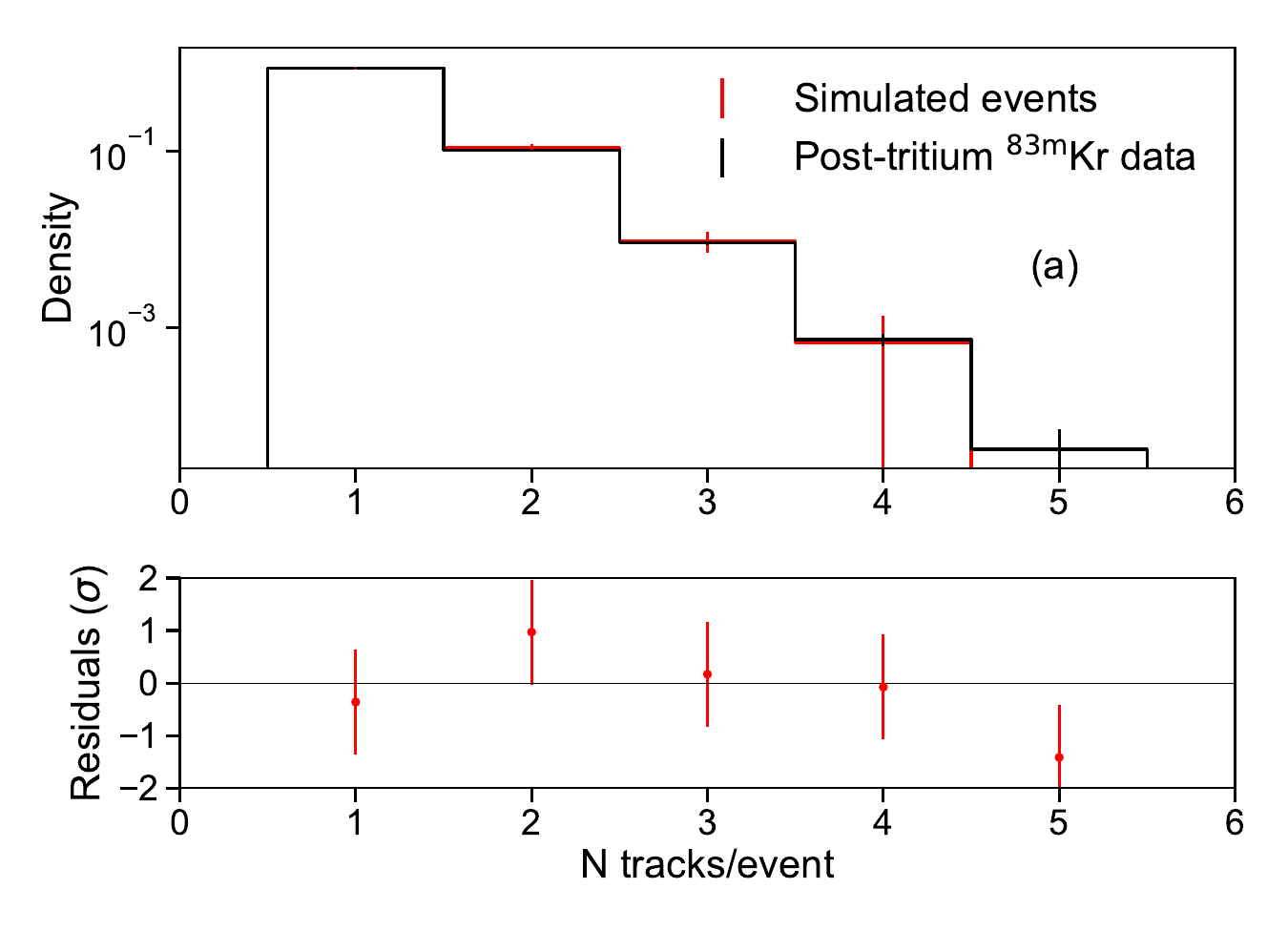}
\includegraphics[width=0.9\columnwidth]{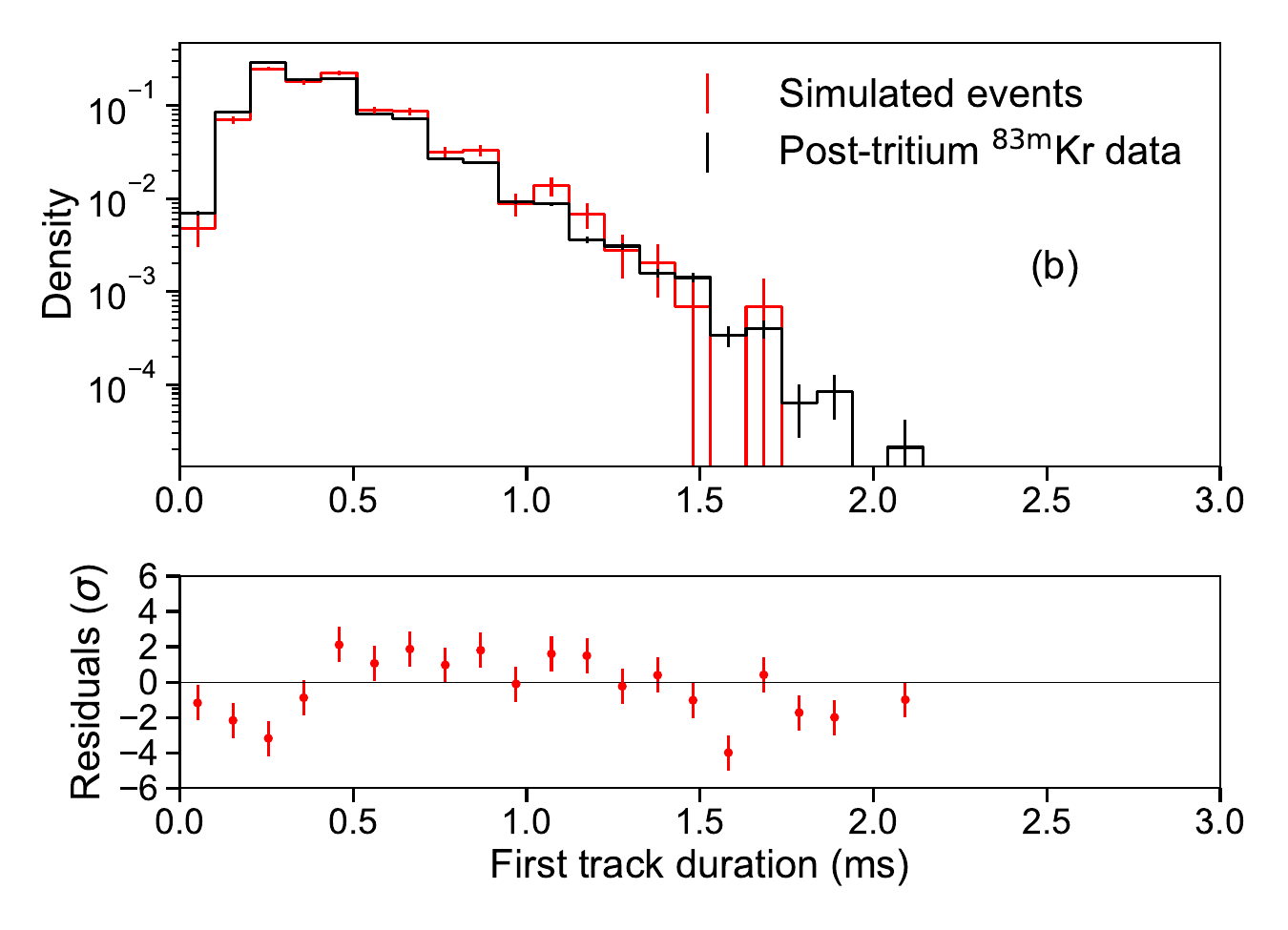}
\includegraphics[width=0.9\columnwidth]{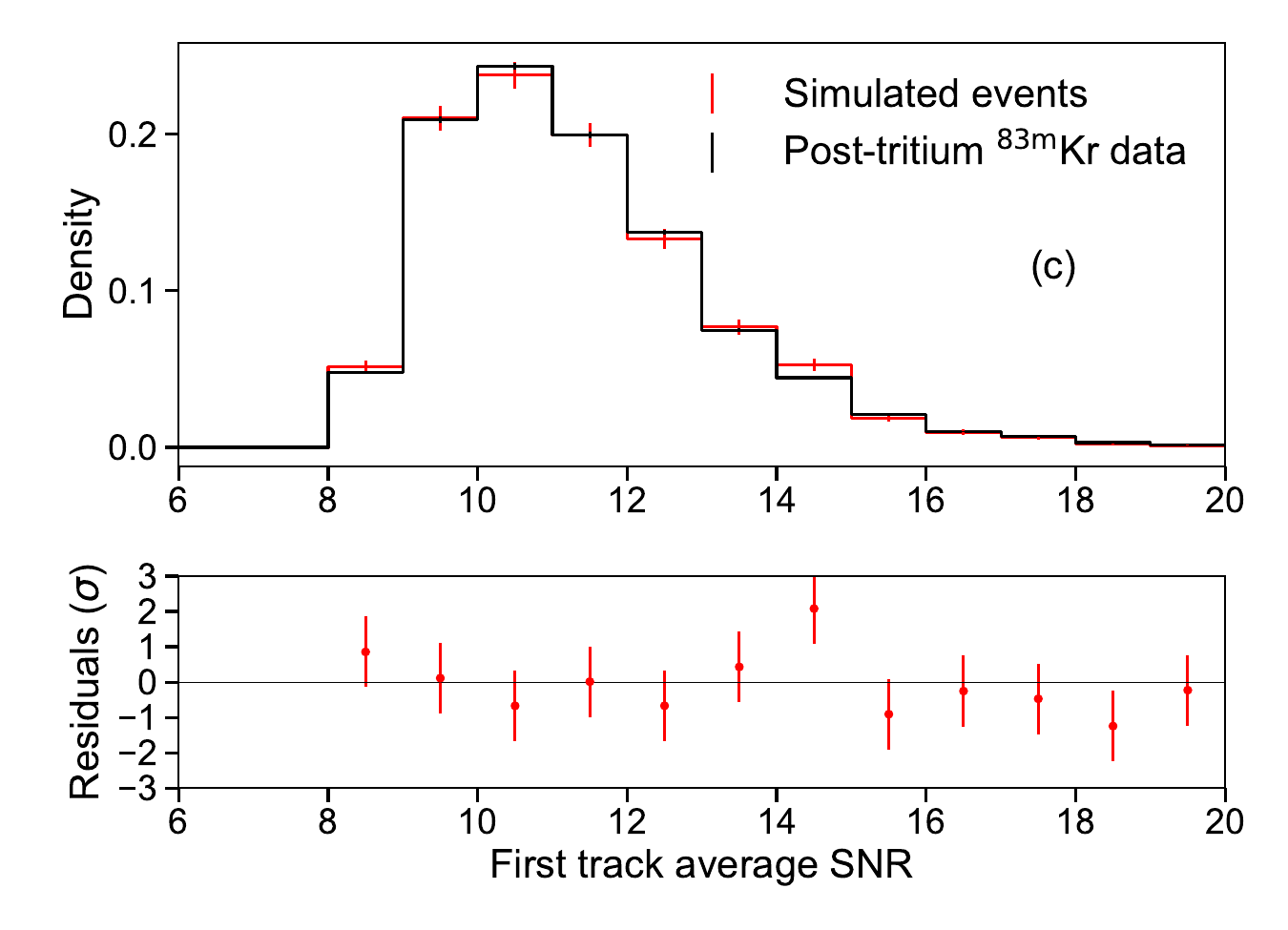}
\caption{Comparison of simulated data to the post-tritium $\mathrm{^{83m}Kr}$ data set after triggering and event reconstruction: (a) number of tracks per event,  (b) first track duration, and (c) first track SNR. The simulations were optimized to reproduce these data.
The good agreement validates the simulations and enables their use to generate input to the analysis. For this purpose simulated data sets were generated to match each data set listed in \autoref{tab:data_set_table}.
}
\label{fig:simulation_validation_track_length_ntracks_snr}
\end{figure}

The number of tracks per event is drawn from a geometric distribution with a configurable expectation value.
The observed mean number of tracks after event reconstruction does not correspond to the underlying truth ($N_{\mathrm{tracks}}^{\mathrm{true}}$), because sometimes tracks are missed or two tracks are combined into one during reconstruction. 
To find the right $N_{\mathrm{tracks}}^{\mathrm{true}}$ for all data sets, events of different $N_{\mathrm{tracks}}^{\mathrm{true}}$ are simulated and reconstructed. 
The reconstructed number of tracks is then fitted with a geometric distribution, as shown in \autoref{fig:number_of_tracks_extraction}(a). 
The distribution is characterized by its success probability $p_{\rm{tracks}}$, which corresponds to the probability of a detected track to not be followed by another track. 
The relation between the reconstructed and the true mean number of tracks per event was found to be linear in a separate study in preparation for publication. 
Figure~\ref{fig:number_of_tracks_extraction}(b) shows $p_{\mathrm{tracks}}$ vs. inputted $N\mathrm{_{tracks}^{true}}$. 
The underlying $N\mathrm{_{tracks}^{true}}$ for each data set can be read off from the intersection of the linear fit with the data set's $p_\mathrm{tracks}$. For each data set, vertical bands indicate the uncertainty on $N\mathrm{_{tracks}^{true}}$ from two contributions: the linear fit uncertainty and the $p_\mathrm{tracks}$ uncertainty. 

The sizes of frequency jumps between tracks are drawn from the energy loss function for electron-hydrogen scattering, converted to frequency via \autoref{eq:energytofrequency}. 
Since the distribution of first tracks is the only information from simulations that we use as analysis input (see \autoref{sec:gen_resolution}), there is little sensitivity to the loss function and hydrogen serves for all gases.  It is only required that the jump size be large enough to prevent the reconstruction algorithm from joining tracks that are in fact separate.
Pitch angle changes during inelastic scatters are assumed to be small and are ignored, and the power of consecutive tracks in an event is kept constant.  Despite these approximations, after processing with the Phase~II trigger and reconstruction methods, real data sets are well reproduced by simulated events in the main properties relevant for reconstruction  (\autoref{fig:simulation_validation_track_length_ntracks_snr}).

\section{CRES spectrum analysis \label{sec:analysisoverview}}

This section describes the signal model of CRES spectra used in the analysis, with an overview flow chart in \autoref{fig:simple_flowchart}. \autoref{fig:flowchart} of Appendix~\ref{sec:flowchart} contains a detailed flowchart that reflects all analysis steps and the interdependence of the $\mathrm{^{83m}Kr}$ and $\mathrm{T_2}$ analyses. 
\begin{figure*}[tb]
\centering
\includegraphics[width=0.9\textwidth]{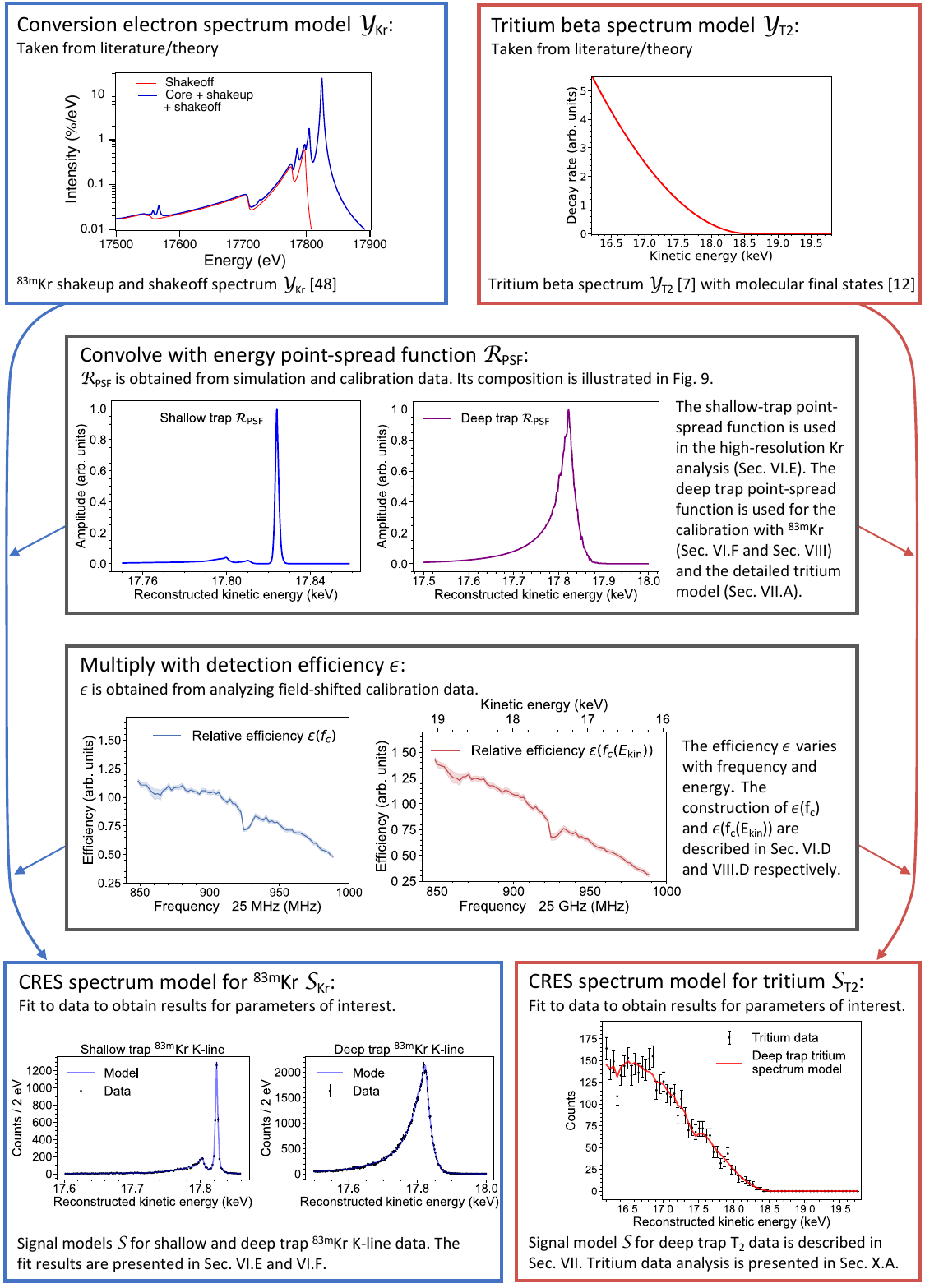}
\caption{Flow chart of construction of Phase II CRES analysis models (\autoref{eq:FullModel1}). To obtain a full model spectrum $\mathcal{S}$, the underlying electron energy spectrum $\mathcal{Y}$ is convolved with the point spread function $\mathcal{R}_{\mathrm{PSF}}$ before being multiplied with the detection efficiency $\mathcal{\epsilon}$. }
\label{fig:simple_flowchart}
\end{figure*}

\subsection{CRES energy spectrum model}\label{sec:Kr_detector_response}

\begin{figure}[tb]
  \centering
  \includegraphics[width=1.0\columnwidth]{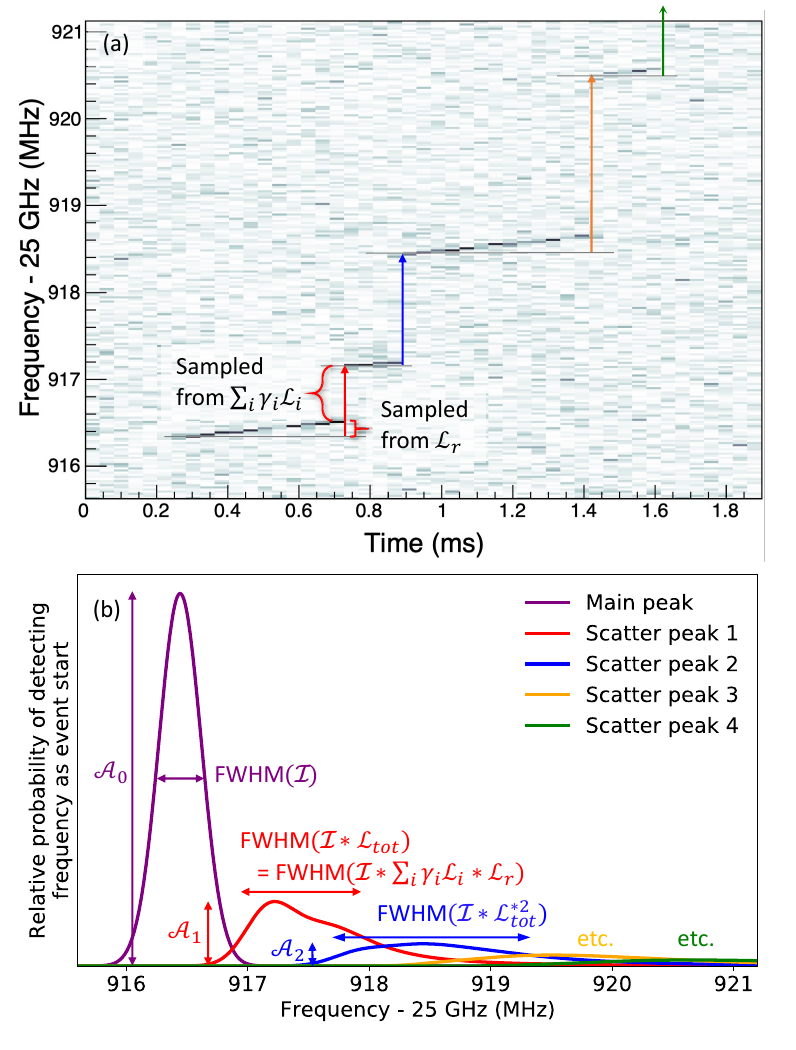}
  \caption{(a) A CRES event with frequency jumps corresponding to inelastic scattering  $\sum_i(\gamma_i \mathcal{L}_i)$ and energy loss due to cyclotron radiation $\mathcal{L}_r$ labelled. These energy losses only affect the spectrum when early tracks in the event are missed. (b) A modeled spectrum broken down into its constituent scatter peaks and labelled to show the roles of model parameters: scatter peak amplitudes $\mathcal{A}_j$, instrumental resolution $\mathcal{I}$, and energy loss spectra $\mathcal{L}_{\mathrm{tot}}^{*j}$. A Gaussian $\mathcal{I}$ is used here for illustration purposes only; a finer-grained instrumental resolution is used in the analysis, as described later.}
  \label{fig:scattering_illustration}
\end{figure}

We model a generic detected CRES signal spectrum $\mathcal{S}$ as
\begin{eqnarray}
    \mathcal{S} &=& \epsilon\left(\mathcal{Y}*\mathcal{R}_{\mathrm{PSF}}\right), 
  \label{eq:FullModel1} \\
    \mathcal{R}_{\mathrm{PSF}} &=& \sum_{j=0}^{j_{\mathrm{max}}}\mathcal{A}_j\left(\mathcal{I}*\mathcal{L}_{\mathrm{tot}}^{*j}\right).
    \label{eq:FullModel2}
\end{eqnarray}
Diagrams of these two equations are shown in \autoref{fig:simple_flowchart} and \autoref{fig:scattering_illustration}, respectively. In both equations, all variables are functions of $E_\mathrm{kin}$, as denoted by script lettering. The symbol $*$ represents convolution and $^{*j}$ represents self-convolution $j$ times. The efficiency function $\epsilon$ encodes the probability of detecting electron events. The underlying true energy spectrum of the electrons is $\mathcal{Y}$.
$\mathcal{R}_{\mathrm{PSF}}$ is the point-spread function, which represents the energy response for mono-energetic electrons---in other words,  how reconstructed energies are shifted and broadened relative to true energies. 

Equation~\ref{eq:FullModel2} shows that the energy point-spread function $\mathcal{R}_{\mathrm{PSF}}$ is comprised of a sum of scatter peaks $\mathcal{I}*\mathcal{L}_{\mathrm{tot}}^{*j}$ weighted by amplitudes $\mathcal{A}_j$, as illustrated in \autoref{fig:scattering_illustration}.  These amplitudes describe the relative likelihood that an electron will first be detected after $j$ scattering events. As an example, for tritium data, the best estimates for the first few $A_j$ values are $A_0=1$ (by definition), $A_1=0.41$, $A_2=0.24$, and $A_3=0.15$. We account for the possibility of up to $j_{\mathrm{max}}$ scatters before the first detection. We use $j_{\mathrm{max}}=20$ in both $^\mathrm{83m}$Kr and tritium fits, as increasing $j_{\mathrm{max}}$ further has no observable effect on results for Phase II conditions.  The instrumental resolution $\mathcal{I}$  is the spectrum that a source of mono-energetic electrons would have if they were all detected before scattering. The distribution of electrons' energy losses between scatters $\mathcal{L}_{\mathrm{tot}}$ depends on the gas composition, the differential cross section on each gas component, and the loss to cyclotron radiation.  The elements in $\mathcal{R}_{\mathrm{PSF}}$ are described further in the remainder of this section.

\subsection{Instrumental resolution \texorpdfstring{$\mathcal{I}$}{}}\label{sec:gen_resolution}
In Phase~II, the instrumental resolution $\mathcal{I}$ accounts for broadening from the differences in mean magnetic fields sampled by detected electrons with different pitch angles and radial positions. These mean field distributions vary with trapping geometry. 
To obtain $\mathcal{I}$ for each data set, mono-energetic events in all constituent single traps are simulated as described in \autoref{sec:sim-events}.
In future Project 8 phases, $\mathcal{I}$ will also account for uncertainties on the mean cyclotron frequencies (\emph{e.g.}, due to frequency binning and noise). In Phase II, this effect was small (${\sim}$0.2\,eV) compared to magnetic field variation.

The shape of $\mathcal{I}$ depends on the range of track SNR values accepted in analysis. This is because SNR thresholds limit the range of  axial excursions in the trap, which in turn limits the range of mean fields experienced by detected electrons.  
Locust computes relative signal powers to reflect all physical effects in the waveguide. Hence, only the absolute power scale must be set by configuring the power corresponding to the measured SNR$_{\rm max}$. The value of SNR$_{\rm max}$ is specific to each data set and its optimization is described in \autoref{sec:max-SNR-optimization}.

The simulated events are filtered by an efficiency matrix, which is a binned look-up table of the probability for an event to be accepted as a function of SNR, first-track duration, and number of tracks in an event. 
The efficiency matrix is produced by simulating 100,000  events covering the full parameter space and analyzing the event detection probability  with respect to those three event properties.
We use this matrix to avoid processing all simulated data with the trigger and reconstruction methods, thereby greatly reducing the processing time. This allows us to iteratively optimize, for example, the event SNR in a data set, with a quick turn-around time. 

\begin{figure}[htb]
\centering   
\includegraphics[width=\columnwidth]{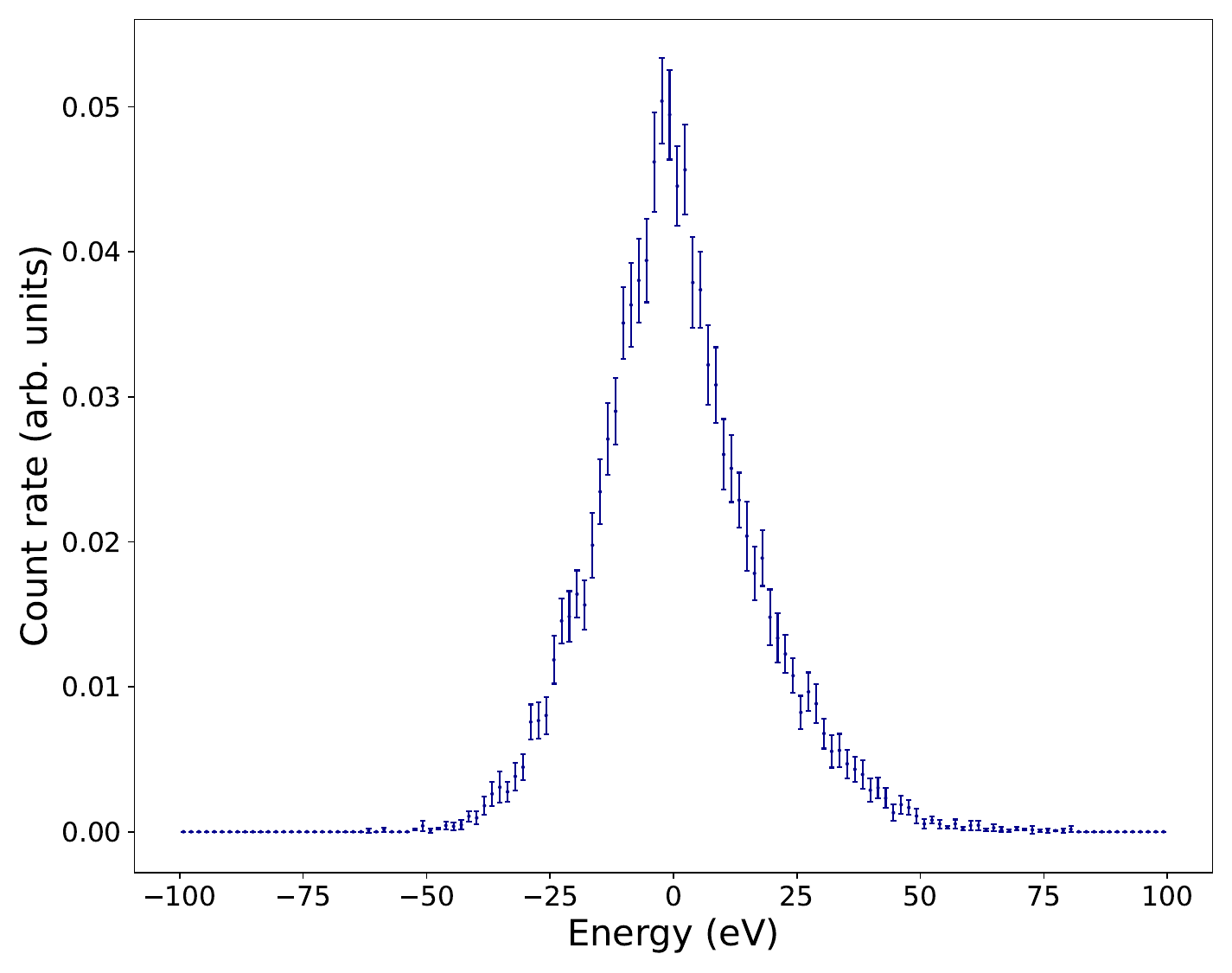}
\caption{Instrumental resolution $\mathcal{I}$ of the pre-tritium $^{\mathrm{83m}}$Kr calibration data set. The error bars include uncertainties from Poisson counting, the efficiency matrix, and the trap weights.}
\label{fig:kr_ftc_res}
\end{figure}

The density histogram of simulated events that survive the efficiency filter is a frequency resolution distribution, which is converted to an energy resolution via \autoref{eq:energytofrequency}. 
We center the energy resolution distribution from each trap on $0\,\si{eV}$ to align the distributions before combining them (in the experiment the trapping field strengths are aligned to minimize the resolution width of recorded data). The total $\mathcal{I}$ is a weighted average of the resolutions of individual traps.   The relative SNR scales of individual traps are determined using the mapping from SNR to counts, requiring that the fraction of events in each trap's resolution matches the fraction collected in the trap in real data.  An example of a simulated $\mathcal{I}$ is shown in \autoref{fig:kr_ftc_res}. 
Because the resolution is centered on $0\,\si{eV}$, a fit of the full spectrum model to data will find the overall energy scale, set by the mean magnetic field experienced by the detected electrons.

\subsection{Energy loss spectra \texorpdfstring{$\mathcal{L}_{\mathrm{tot}}^{*j}$}{}}
\label{L_model}
Electrons lose energy primarily by inelastic scattering with gas molecules, causing the jumps in \autoref{fig:spectrogram}.
Cyclotron radiation is a smaller, continuous source of electron energy loss, causing the upward track slopes in \autoref{fig:spectrogram}. 
$\mathcal{L}_{\mathrm{tot}}^{*j}$ comprises the distribution of possible energy losses an electron has experienced \emph{before} the first detected track due to both of these effects, with the self-convolution $j$ times accounting for $j$ scatters. The electron energy loss spectrum for a single scatter is
\begin{equation}
\begin{aligned}
    \mathcal{L}_{\mathrm{tot}} = (\gamma_1 \mathcal{L}_1 + \gamma_2 \mathcal{L}_2 + \cdots + \gamma_n \mathcal{L}_n)*\mathcal{L}_r\,,\label{eq:combine_peaks_from_different_gases}
\end{aligned}
\end{equation}
where $\mathcal{L}_i$ is the electron inelastic energy loss spectrum for the $i^{\rm th}$ gas species, each $\gamma_i$ is the fraction of inelastic scatters that are due to the specific gas species $i$, and $\mathcal{L}_r$ is the energy loss spectrum due to cyclotron radiation during the missed track.

We determine bounds on $\gamma_i$ from quadrupole mass analyzer data as described in \autoref{gamma_i_determination}. For the shallow trap data, the high resolution allows for a more precise estimate of gas scattering fractions for H$_2$ and He to be determined from the fit to $^\mathrm{83m}$Kr data.  We neglect the energy dependence of the scatter fractions $\gamma_i$ over the small range of energy change.

Each $\mathcal{L}_i$ is calculated in the Bethe theory of electron inelastic scattering (as in~\cite{Inokuti:InelasticScattering1971}), given by
\begin{eqnarray}
\hspace{-0.1in}\frac{d\sigma}{dE} &=& \frac{4 \pi a_0^2R}{E_{\rm kin}} \left[\frac{R}{E}\frac{df}{dE}\ln\left(\frac{4c_EE_{\rm kin}}{R}\right) + \mathcal{O}\left(\frac{R}{E_{\rm kin}}\right)\right], \quad
\label{eq:inelastic scatter energy loss }
\end{eqnarray}
where $R$ is the Rydberg energy, $a_0$ is the Bohr radius, $E_{\rm kin}$ is the incident energy of the electron, $E$ is the energy loss of the electron, $c_E \approx (R/E)^2$, and $df/dE$ is the optical oscillator strength of the gas molecules.
The optical oscillator strength \cite{Chan:AbsoluteOscillatorStrength1991,chan1992absolute,olney1997absolute} data for the relevant gas species are from the LXCat database~\cite{LXCat_carbone2021data,pitchford2017lxcat,pancheshnyi2012lxcat,LXCat:database}. 

\begin{figure}[tb]
  \centering
  \includegraphics[width=1.0\columnwidth]{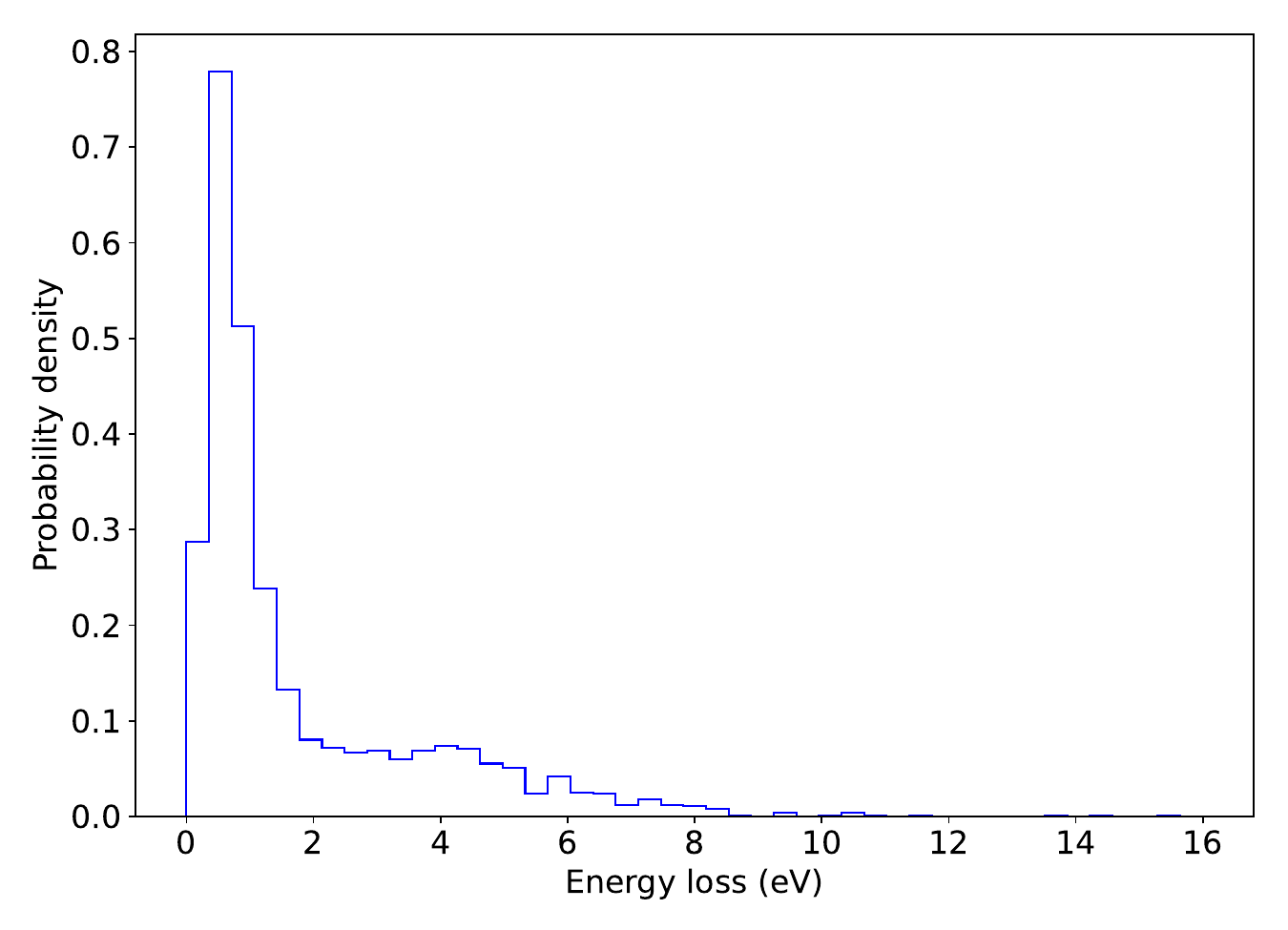}
  \caption{Simulated $\mathcal{L}_r$, the energy loss distribution due to cyclotron radiation during a missed track, for conditions (\emph{e.g.}, track duration) corresponding to $^{\mathrm{83m}}$Kr pre-tritium data. The width of this distribution scales with the average track durations and slopes (slopes correspond to radiated power).}
  \label{fig:radiation_loss}
\end{figure}

We determine the loss due to cyclotron radiation $\mathcal{L}_r$ using the simulated data described in \autoref{sec:sim-events}.
In these simulated data, missed tracks associated with detected events are identified.
The distribution of differences between track end frequency and track start frequency among these tracks is converted to energy and taken as the radiative energy loss spectrum $\mathcal{L}_r$ (\autoref{fig:radiation_loss}).
With most of its weight in a peak between 0 and 3 eV\textemdash reflecting the low likelihood of missing long tracks\textemdash and a modest tail out to ${\sim}$10 eV, $\mathcal{L}_r$ has a much smaller impact than inelastic scatters. 
 
\begin{figure}[htb]
\centering
\includegraphics[width=\linewidth]{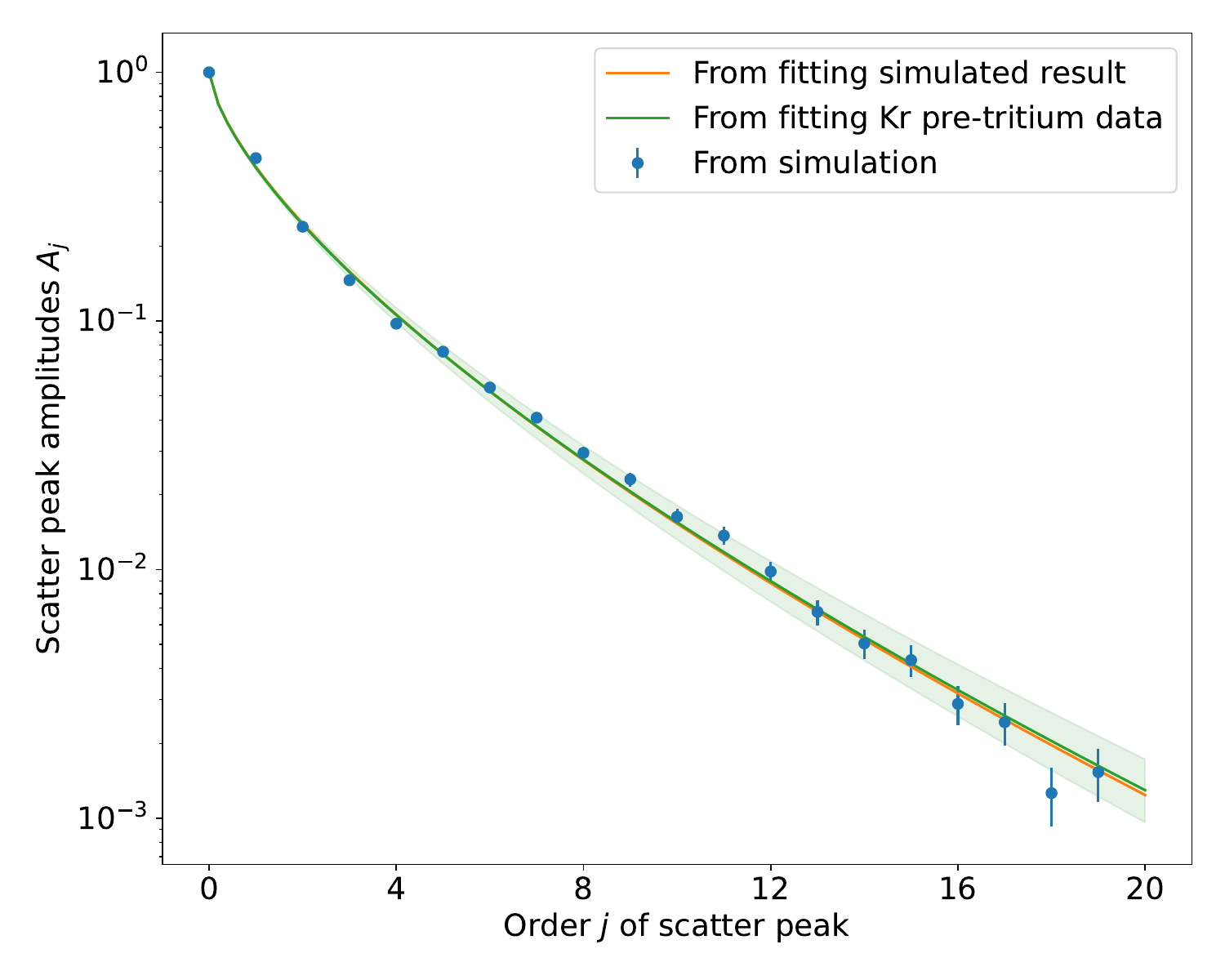}
\caption{From simulation, amplitudes $\mathcal{A}_j$ of scatter peaks in the CRES response function (blue points), caused by missing $j$ tracks in an event. $\mathcal{A}_j$ are simulated accounting for electron pitch angle changes from scattering (modeled by \autoref{eq:inelastic scatter angle}). $\mathcal{A}_j$ are modeled by a modified exponential function parameterized by $p$ and $q$ (\autoref{eq:scatter_peak_amplitude}). Fitting $p$ and $q$ for pre-tritium $\mathrm{^{83m}Kr}$ data results in a curve (green) in good agreement with simulation (orange).}
\label{fig:scatter-peak_amplitude_simulation}
\end{figure}

\subsection{Scatter peak amplitudes\label{sec:scatter-peak-amplitudes} \texorpdfstring{$\mathcal{A}_j$}{}} 
Each amplitude  $\mathcal{A}_j$ is the relative likelihood of missing the first $j$ tracks in an event and detecting track $j+1$. The function $\mathcal{A}_j(j)$ is nearly exponential. It deviates from an exponential due to pitch-angle changes from scattering (which alter the probability of an electron being trapped and detectable) and event reconstruction thresholds (which here depend on first track duration, first track SNR, and number of tracks in an event).

To determine the functional form of $\mathcal{A}_j(j)$, including the deviation from exponentiality, we perform a toy model simulation and reconstruction of events, then count events in each scatter peak. It is assumed that inelastic scattering leads to energy loss and small pitch angle changes, while elastic scattering removes electrons from the trap before the next inelastic scatter~\cite{DavidJoy:ElectronScattering}. 
The simulated inelastic scattering angle $\theta_s$ follows the distribution~\cite{Rudd:1991differential}
\begin{align}
    P(\theta_s) \propto \left(1 + \frac{\cos^2\theta_s}{\alpha^2}\right)^{-1}\,,
\label{eq:inelastic scatter angle}
\end{align}
where $\alpha$ is a constant that depends on gas composition. A fraction $\kappa$ of electrons leave the trap between inelastic scatters due to elastic scatters. As an example, we find $\alpha=0.0018$ (corresponding to an average sampled scattering angle of 0.48$^\circ$) and $\kappa=0.19$ for pre-tritium $^{\mathrm{83m}}$Kr data (similar to tritium data). This simulation and reconstruction procedure is described in more detail in Appendix~\ref{appendix:scatter_peak_amplitude_simulation}.

\autoref{fig:scatter-peak_amplitude_simulation} shows the dependence of $\mathcal{A}_j$ on $j$ from the simulations. The curve may be parameterized by
\begin{eqnarray}\label{eq:scatter_peak_amplitude}
    \mathcal{A}_j &=&\exp\Big[-pj^{(-dp+q)}\Big],
\end{eqnarray} 
with free parameters $p$ and $q$. The constant $d=0.4955$ is chosen to minimize the correlation between $p$ and $q$ and held fixed. For a specific $^{\mathrm{83m}}$Kr data set, $\mathcal{A}_j(j)$ is determined by fitting the CRES spectrum while using  a response function model that includes \autoref{eq:scatter_peak_amplitude}. This produces estimates of  $p$ and $q$. These parameters must be fitted because $\alpha$ and $\kappa$ are not known externally; they are instead tuned to match CRES data. In future experiments, $\alpha$ and $\kappa$ could be predicted using more precise calibration of gas composition and scattering effects. 

The $p$ and $q$ values for tritium analysis are extrapolated from $^{\mathrm{83m}}$Kr $p$ and $q$ values as a function of the average number of tracks per event ($N_{\mathrm{tracks}}^{\mathrm{true}}$), as described in \autoref{sec:scatter_peak_errors}. The result for tritium data is $p=0.89\pm 0.11$, $q=1.12 \pm 0.05$. The variation in $N_{\mathrm{tracks}}^{\mathrm{true}}$ among data sets stems mainly from differences in gas composition, which change $\alpha$ and $\kappa$, thus changing $\mathcal{A}_j(j)$.

\section{Calibration with \texorpdfstring{$^{\mathrm{83m}}\mathrm{Kr}$}{}\label{sec:Kr_model}}

Fits to $^{\mathrm{83m}}$Kr electron lines are used to both characterize the apparatus and estimate parameters for tritium data analysis. \autoref{sec:frequency-energy_relation} details how we performed  Voigt fits to  $^{\mathrm{83m}}$Kr lines to verify the relation between energy and cyclotron frequency. 
The remaining subsections describe fits to the 17.8-keV $^{\mathrm{83m}}$Kr line using the full CRES model in \autoref{eq:FullModel1}, producing estimates of the mean field $B$, 
scattering parameters $p$ and $q$, and detection efficiency $\epsilon$ as a function of frequency.

\begin{figure}[htb]
    \centering
    \includegraphics[width=1\columnwidth]{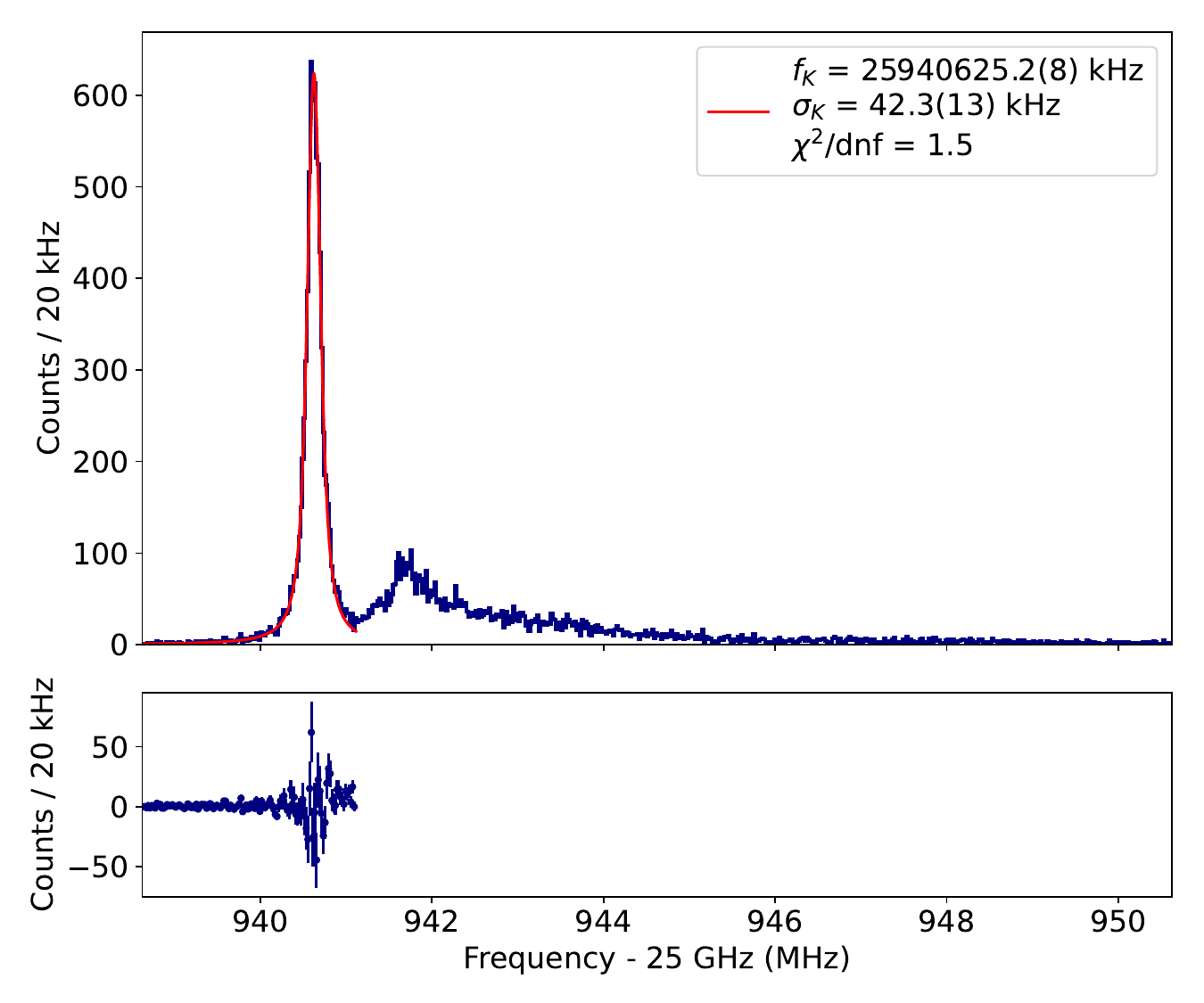}
    \caption{Top: Fit of a Voigt profile (red) to $^{\mathrm{83m}}$Kr K conversion electron events (blue) recorded with the shallow trap. 
    The center frequency $f_K$ and the standard deviation $\sigma_K$ of the instrumental resolution are extracted from the fit. Bottom: Residuals in the fitted frequency range. }
    \label{fig:linearity_Kr_K_freq_line}
\end{figure}

\begin{figure}[htb]
    \centering
    \includegraphics[width = 1\columnwidth]{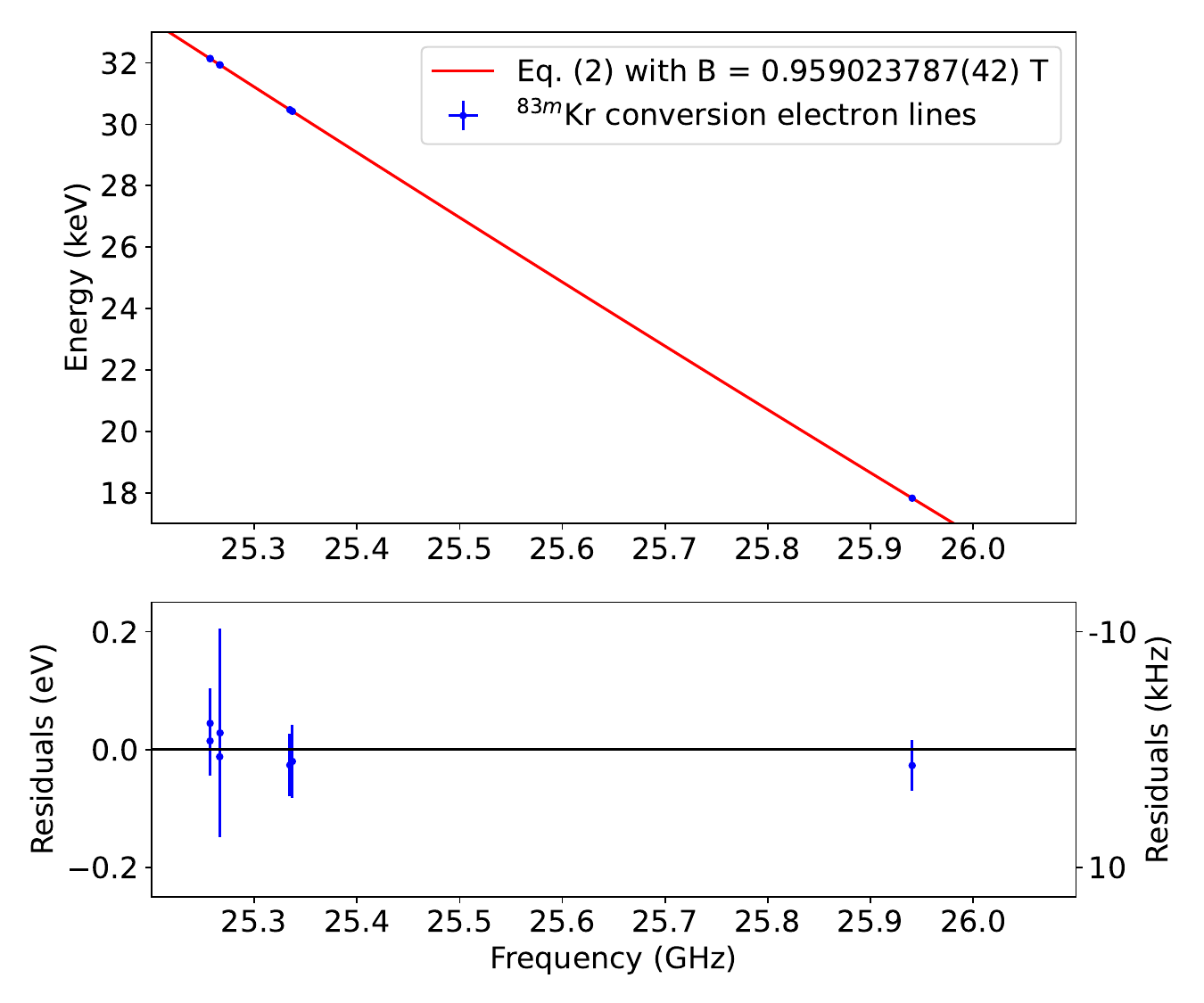}
    \caption{Fit via \autoref{eq:energytofrequency} of the measured frequencies of conversion lines to their kinetic energies as given in \autoref{table:Kr energy lines Venos}.  From right to left the lines are K, L2, L3, M2, M3, N2 and N3.  The magnetic field is the fit parameter.  The error bars do not include a 0.5-eV energy scale uncertainty from the gamma energy. In the residuals, the uncertainties in the frequencies are projected and added in quadrature to those in the binding energies. The frequency scale shown for the residual is not exact but shows the overall correspondence between the magnitudes of the energy difference and the frequency difference for convenience.}
    \label{fig:linearity}
\end{figure}

\begin{table*}[htb]
\caption{The frequencies of the conversion electron lines recorded in the shallow trap configuration. The N2 and N3 lines are not resolved but their frequencies are fitted separately by fixing the intensity ratio and separation between the two lines. The conversion electron line energies are calculated by fixing the gamma energy at the literature value of 32151.6 eV, and using the binding energies and recoil energies from \cite{V_nos_2018}. The 0.5-eV energy scale uncertainty from the gamma energy is not included.} 
\label{table:Kr energy lines Venos}
  \centering
  \renewcommand{\arraystretch}{1.15}
    \begingroup
    \setlength{\tabcolsep}{6pt} 
  \begin{tabular}{ l  c  c  c  c}
 \hline\hline
 Line & Conversion & Binding & Recoil &  \phantom{aa}Shallow trap frequency \\  
  & electron &  energy (eV) &  energy (eV) & (kHz) \\ & energy (eV) \\
 \hline\hline
 K & 17\,824.23(4)\phantom{a} & 14\,327.26(4) & 0.120 & 25\,940\,625.2(8)\phantom{a}\\ 
 \hline
 L2 & 30\,419.49(6)\phantom{a} & 1\,731.91(6) & 0.207 & 25\,337\,157.0(6)\phantom{a} \\
 
 L3 & 30\,472.19(5)\phantom{a} & 1\,679.21(5) & 0.207 & 25\,334\,690.7(8)\phantom{a} \\
 \hline
 M2 & 31\,929.26(17) & 222.12(17)  & 0.218 & 25\,266\,701.5(21) \\
 
 M3 & 31\,936.85(11) & 214.54(11) & 0.218 & 25\,266\,348.0(11) \\  
 \hline
 N2 & 32\,136.72(1)\phantom{a} & 14.67(1) & 0.219 & 25\,257\,051.7(27) \\
 N3 & 32137.39(1)\phantom{a} & 14.00(1) & 0.219 & 25\,257\,019.2(27) \\
 \hline\hline
\end{tabular}
\endgroup
\end{table*}

\subsection{Test of the frequency-energy relation} \label{sec:frequency-energy_relation}
To verify the predicted CRES energy-frequency relationship (\autoref{eq:energytofrequency}) across a 14.3-keV range, the $^{\mathrm{83m}}$Kr shallow trap data included measurements of the K, L2, L3, M2, M3, N2 and N3 internal-conversion lines of the 32-keV transition. For each line, the main peak is well separated from the scattering tail and from a $^{\mathrm{83m}}$Kr shakeup/shakeoff structure~\cite{Robertson:2020boa} in this high-resolution trap. This makes it possible to extract the central frequency of the main peak in each $^{\mathrm{83m}}$Kr spectrum by fitting it with a Voigt profile, which has a fixed Lorentzian width as tabulated in~\cite{V_nos_2018}. A constant background is added as a fit parameter when events from the tail of a different $^{\mathrm{83m}}$Kr line are present within the fit range. The frequencies extracted are given in \autoref{table:Kr energy lines Venos} and the fit to the K line is shown in \autoref{fig:linearity_Kr_K_freq_line}.

The energy of each conversion line is calculated in \cite{V_nos_2018} using the 32-keV gamma energy, as well as a binding energy and recoil energy specific to that line (shown in \autoref{table:Kr energy lines Venos}).  
\autoref{fig:linearity} shows the fitted frequency-energy relation with the mean magnetic field $B$ as free parameter.
The magnetic field found in the fit is $B = 0.959023787(42)$\,T.  Note that this does not include the uncertainty from the gamma energy scale of 0.5\,eV. The points in the residual plot below the figure illustrate the good internal agreement of the data with the  equation. The conversion line energies are calculated by fixing the gamma energy at the literature value of  32151.6 eV provided in \cite{V_nos_2018}. An improvement in the gamma energy measurement is planned by the KATRIN collaboration~\cite{Rodenbeck:2022fxc} and could also be made via CRES with a precise independent determination of the magnetic field by NMR.

\subsection{\texorpdfstring{$^{\mathrm{83m}}$Kr}{} fit procedure with CRES spectrum model}
\label{subsec:Kr Line model}
For the remainder of this paper, all $^{\mathrm{83m}}$Kr fits use the CRES spectrum model in \autoref{eq:FullModel1} to fit the 17.8-keV conversion-electron line. Since the structure of this CRES spectrum model is common between $^{\mathrm{83m}}$Kr and tritium data, we can use $^{\mathrm{83m}}$Kr fits to calibrate the tritium energy point-spread function $\mathcal{R}_{\mathrm{PSF}}$ and detection efficiency curve $\epsilon$. The 17.8-keV $^{\mathrm{83m}}$Kr line is a powerful tool, given its narrow (2.774-eV) natural line width~\cite{Altenmuller:2019ddl}, well understood shape, and closeness to the tritium endpoint at 18.6 keV. 
The underlying spectrum $\mathcal{Y}_{\mathrm{Kr}}$ includes the 17.8-keV $^{\mathrm{83m}}$Kr  main peak with its natural line width, as well as a lower-energy satellite structure from shakeup and shakeoff \cite{Robertson:2020boa}.

In these fits, the magnetic field $B$ and scattering parameters $p$ and $q$ are left free. For the fit to the deep quad trap data, the scatter fractions $\gamma_i$ are inputted, while for the shallow trap data, the scatter fractions for  H$_2$ and He are extracted from the fit, as motivated in \autoref{L_model}.
In the final fits, the detection efficiency variation with frequency $\epsilon$ (as determined in \autoref{subsec:efficiency}) is included in the model.
No background component is included in the fits due to the short run durations and negligible expected background rate. Section~\ref{subsec:electron_data} explains the reasons for expecting negligible background, and \autoref{subsec:background_limit} confirms this assumption.

Numerical scatter peaks serve as fixed inputs to the fitting function. These scatter peaks are produced by convolving data-set-specific simulated instrumental resolutions $\mathcal{I}$ (see \autoref{subsec:ins_res}) with electron energy loss spectra $\mathcal{L}_{\mathrm{tot}}$. To determine $\mathcal{L}_{\mathrm{tot}}$, loss spectra are combined according to \autoref{eq:combine_peaks_from_different_gases}, accounting for $\mathcal{L}_r$ and scattering from gases present in $^{\mathrm{83m}}$Kr data: Kr, $^3$He, Ar, and H$_2$ and its isotopologues.

Fits to $^{\mathrm{83m}}$Kr data are performed by minimizing a Poisson likelihood chi-squared \cite{Baker:ChiSquare1984},
\begin{align}
    \chi^2_{\lambda,p} = 2 \sum \big[y_i - n_i + n_i\ln{(n_i/y_i)}\big]\,,
\end{align}
where $y_i$ is the expected number of events in bin $i$ according to Eqs.~\ref{eq:FullModel1} and \ref{eq:FullModel2}, and $n_i$ is the measured  number of events in that bin. Because the spectra contain many bins with zero or few counts, $\chi^2$/DOF is not an optimal figure of merit. Instead, goodness-of-fit testing is performed  as suggested in~\cite{Baker:ChiSquare1984}. Treating the fitted spectrum as the truth, the Poisson $\chi^2_{\lambda,p}$ is sampled by Monte-Carlo, and the distribution of Poisson $\chi^2_{\lambda,p}$ is compared to the $\chi^2_{\lambda,p}$ for the data.

\begin{figure}[htb]
    \centering
    \includegraphics[width=1\columnwidth]{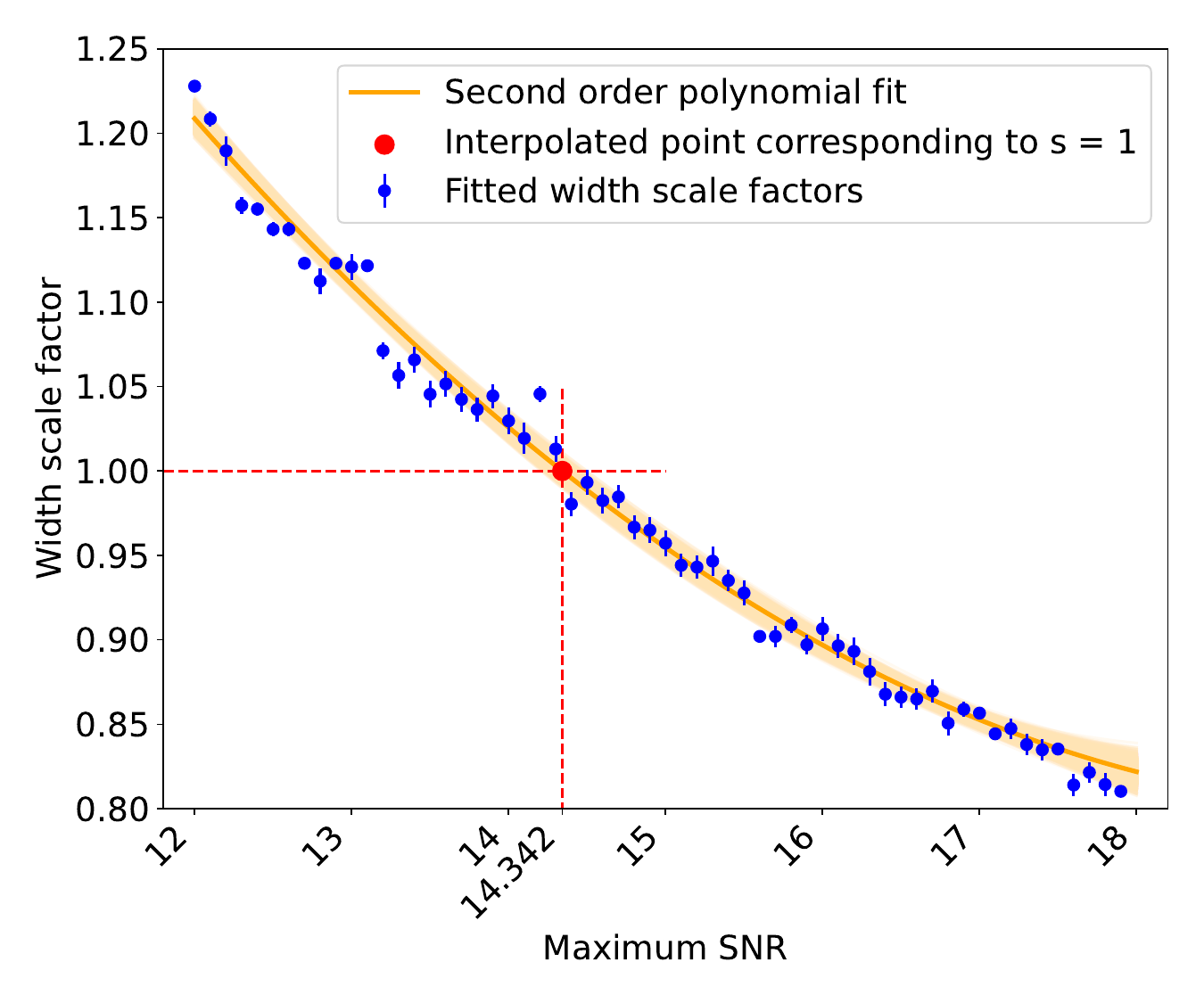}
    \caption{SNR$_{\rm max}$ optimization for $\mathrm{^{83m}Kr}$ pre-tritium data. The vertical axis is the scale factor $s$ that adjusts the simulated width of $\mathcal{I}$ to match the experimental one for each choice of SNR$_{\rm max}$ in the simulation.}
    \label{fig:max_snr_optimization_oct_data}
\end{figure}

\subsection{Instrumental resolution \texorpdfstring{$\mathcal{I}$}{}}\label{subsec:ins_res}
The instrumental resolution $\mathcal{I}$ is an input to $^{\mathrm{83m}}$Kr fits. 
$\mathcal{I}$  is determined for each trap configuration by simulation 
as described in \autoref{sec:gen_resolution}.

\subsubsection{SNR scaling optimization}\label{sec:max-SNR-optimization}
Each $\mathcal{I}$ distribution has an associated value of SNR$_{\rm max}$, the SNR of a $90^{\circ}$ electron in trap 3 at $r=0\,\si{mm}$. SNR$_{\rm max}$ mostly affects the width of $\mathcal{I}$ while maintaining the distribution's overall shape. In particular, a higher SNR$_{\rm max}$ corresponds to a wider $\mathcal{I}$ distribution because the overall SNR in track bins is higher, making electrons with smaller pitch angles more detectable. These small-pitch-angle electrons explore a larger range of magnetic fields, broadening the detected frequency spectrum.

The total system gain and noise temperature are not known well enough for each $^{\mathrm{83m}}$Kr data set to determine SNR$_{\rm max}$.
Instead, to estimate SNR$_{\rm max}$, 60 $^{\mathrm{83m}}$Kr fits are performed using inputted $\mathcal{I}$ distributions corresponding to SNR$_{\rm max}$ values ranging from 12 to 18.  
We add a fit parameter to the $^{\mathrm{83m}}$Kr model: a scale factor $s$, which widens or compresses $\mathcal{I}$ during the fit. When $s=1$, this indicates that $\mathcal{I}$ has the best width to describe the data, and thus the best SNR scale. We fit the 60 (SNR$_{\rm max}$, $s$) points to a quadratic function and predict the SNR$_{\rm max}$ for $s=1$. This procedure anchors SNR$_{\rm max}$ to experimental data. It also produces a best-estimate for the standard deviation of $\mathcal{I}$ for each $^{\mathrm{83m}}$Kr data set.

For pre-tritium $^{\mathrm{83m}}$Kr data, the outcome of the SNR scaling procedure (SNR$_{\rm max}$ = 14.3) is shown in \autoref{fig:max_snr_optimization_oct_data}, and the resulting  $\mathcal{I}$ is displayed in \autoref{fig:kr_ftc_res}. To simulate $\mathcal{I}$ for the tritium analysis, we use the same SNR$_{\rm max}$ value as in the pre-tritium 
$^{\mathrm{83m}}$Kr data, since its properties most closely resemble those of tritium data (see \autoref{tab:mml_track_length_fit_results}). The two data sets are only distinguishable in track duration, which has a sub-dominant effect on the width of $\mathcal{I}$.

\subsubsection{Uncertainties on \texorpdfstring{$\mathcal{I}$}{} propagated to \texorpdfstring{$^{\mathrm{83m}}$Kr}{} fit results}\label{sec:res-errors-to-Kr}
\label{sec:instrumental_resolution_uncertainty_propagation}

A simulated, fixed $\mathcal{I}$ distribution is inputted to each $\mathrm{^{83m}Kr}$ K-line fit listed in \autoref{tab:mml_track_length_fit_results}. As a result, uncertainties on $\mathcal{I}$ propagate to the fit parameter results ($p$, $q$ and $B$), which in turn feed into the tritium analysis.
Thus, we estimate the uncertainties in $p$, $q$ and $B$ due to both $\mathcal{I}$ simulation uncertainties and SNR$_{\rm max}$ uncertainties.

For each data set, $\mathcal{I}$  simulation uncertainties are obtained from 100 bootstrapped resolution shapes, which are produced by repeatedly sampling counts in all bins from Gaussian distributions. Each Gaussian's standard deviation equals the bin simulation uncertainty, which includes uncertainties from Poisson counting, the efficiency matrix, and the trap weights. $\mathrm{^{83m}Kr}$ K-line fits are then repeated 100 times, once with each bootstrapped $\mathcal{I}$ as input, to obtain uncertainty distributions for fit parameters. 
Separately, we estimate the SNR$_{\rm max}$ contribution to $\mathrm{^{83m}Kr}$ fit parameter uncertainties. To do so, we fit the data 100 times, each time using an inputted resolution simulated with a different SNR$_{\rm max}$ value sampled from a normal distribution (with a mean from the procedure in \autoref{sec:max-SNR-optimization} and an uncertainty calculated as described in \autoref{sec:sigma_errors}). 
Simulation and SNR$_{\rm max}$ uncertainties are added in quadrature.

\begin{figure}[b]
  \centering
  \includegraphics[width=1.0\columnwidth]{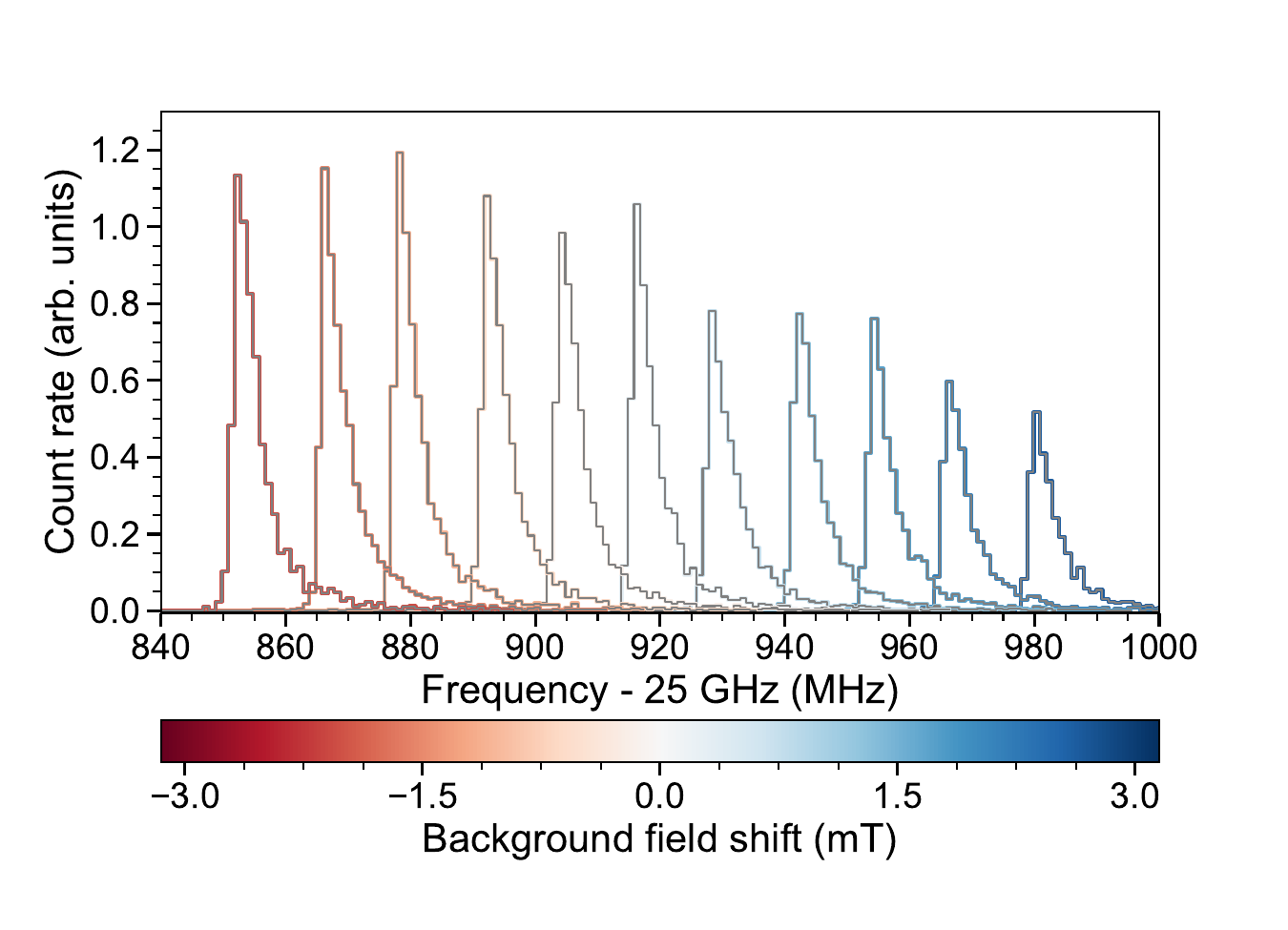}
  \caption{
    The 17.8-keV $\mathrm{^{83m}Kr}$ conversion electron line recorded in the deep quad trap at different magnetic background fields (red / blue).
  }
  \label{fig:q300_fss}
\end{figure}

\begin{figure}
\includegraphics[width=1.0\columnwidth]{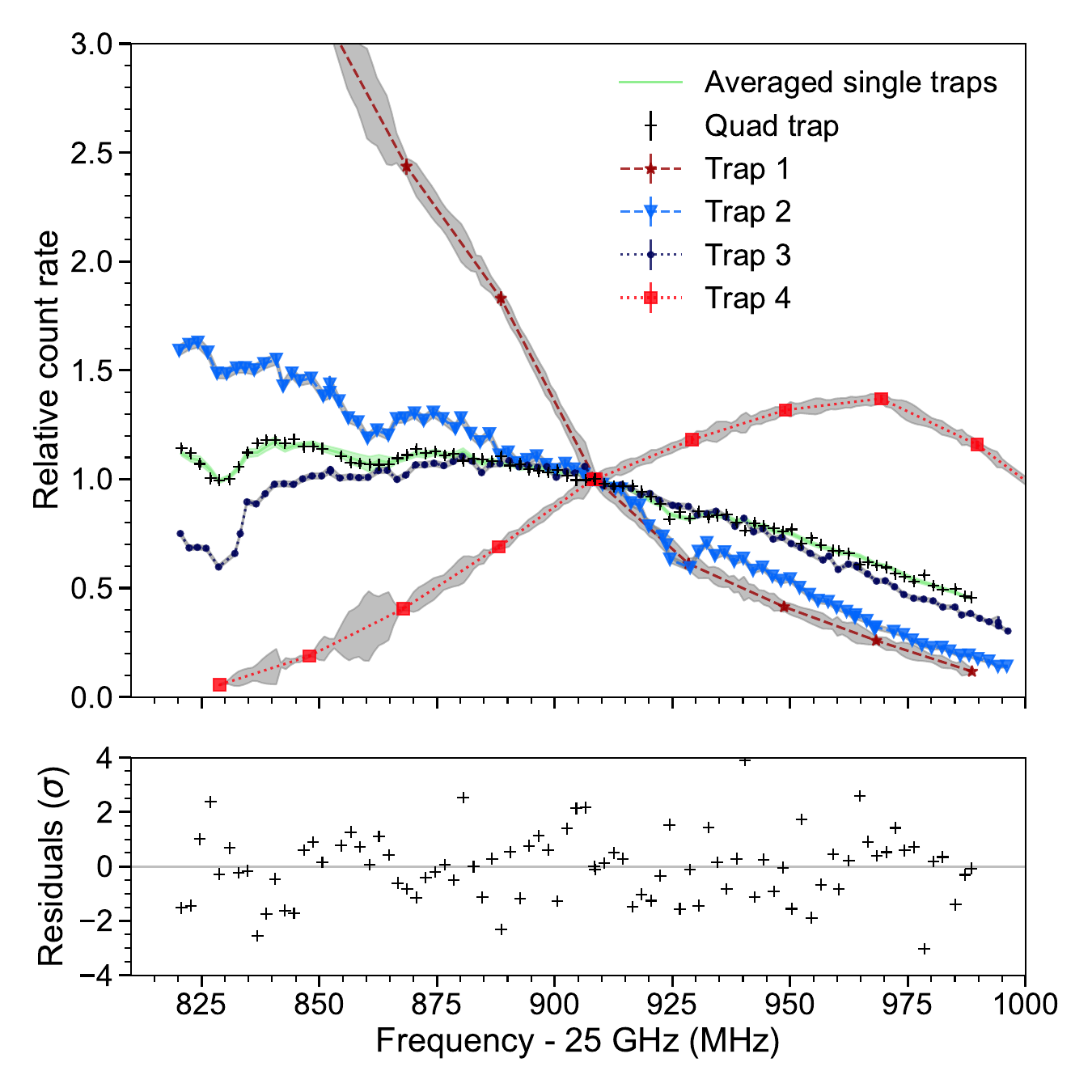}
\caption{Event detection rates from $^{\mathrm{83m}}$Kr K-line data recorded with different single-coil traps (red and blue) and the quad trap (black) in field-shifted data relative to the respective count rate at $f_c \approx 25.91\,\si{GHz}$ (where $B=B_0$). The uncertainties from interpolation are shown in grey. The relative count rates can be summed with weights (green) to match the quad-trap count rate curve (black). The residuals show the differences of the summed single-trap rates and the quad-trap rates divided by the quad-trap count rate uncertainties. The standard deviation of the residuals is larger than 1 and the uncertainties on the tritium efficiency $\epsilon$ (\autoref{sec:energy_correction}) are inflated to account for this.}
\label{fig:fss_event_rates}
\end{figure}

\subsection{Field-shifted \texorpdfstring{$\mathrm{^{83m}Kr}$}{} data analysis}
\label{subsec:frequency_dependence}
 
\subsubsection{Measurement of detection efficiency vs. frequency}
\label{subsec:efficiency}
Detection efficiency as a function of frequency is an input to the CRES spectrum model. To study the frequency response, we recorded $\mathrm{^{83m}Kr}$ data at a range of background magnetic field values, as described in \autoref{fss_procedure} and in the ``$\mathrm{^{83m}Kr}$ field-shifted'' row in \autoref{tab:mml_track_length_fit_results}. Data were taken in the full quad trap configuration as well as in each individual trapping coil in isolation.
A subset of the $\mathrm{^{83m}Kr}$ K-line data recorded in the quad trap configuration is shown in \autoref{fig:q300_fss}. 
To measure the detection efficiency vs.~frequency, we extracted $\epsilon(f_c)$ at the frequency center of each recorded peak by fitting the data with a reduced version of the full CRES spectrum model that does not include $\epsilon(f_c)$. The number of reconstructed events within $\pm$1~$\si{MHz}$ of the fitted peak's frequency location is compared to the number of events for the data at the unshifted background field ($B=B_0$).
The motivation for the start-frequency cut of $\pm1\,\si{MHz}$ around the peak center is to not average the detection efficiency over a larger frequency range while maintaining a sufficiently high statistical precision for the efficiency analysis.
The obtained relative count rate vs.~frequency in a given trap (shown in \autoref{fig:fss_event_rates}) is equivalent to the relative $\epsilon(f_c)$ in this trap for (quasi) mono-energetic data like the $\mathrm{^{83m}Kr}$ K-line (the energy spread of K-line electrons is small compared to the resolution width $\mathcal{I}$). For tritium data analysis in the quad trap, $\epsilon$ is summed from the single-trap count rates after a correction for the dependence of SNR on kinetic energy. We motivate and describe this correction in \autoref{sec:efficiency_for_tritium}.

\subsubsection{Extraction of statistical trap weights}
\label{subsec:trap_weights}
The statistical trap weights $w_i$ correspond to the relative number of detected events in each trap. These weights are used for two purposes: to sum the simulated instrumental resolutions of the 4 traps that compose the quad trap, and to correct the measured efficiency variation with frequency for the tritium analysis as will be discussed in \autoref{sec:energy_correction}.
We extract the $w_i$ at $B=B_0$ from the field-shifted data by minimizing the summed squared differences between the quad trap count rates vs.~frequency and the weighted sum of the single-trap count rates vs.~frequency, with $w_i$ being free parameters.
The resulting weights are $w_1 = 0.076(3)$, $w_2 = 0.341(13)$, $w_3 = 0.381(14)$, and $w_4 = 0.203(20)$, which are in good agreement with the observed count rate differences at $B=B_0$.

Note that the field step sizes are \SI{0.07}{\milli\tesla} in traps 2, 3, and the quad trap. We chose the step sizes in trap 1 and trap 4 to be \SI{0.7}{\milli\tesla} to reduce the total duration of these field-shifting scans for the traps with the  lowest count rate at the nominal frequency position of the K-line ($\approx 25.91\,\si{GHz}$).
For the summation, the count rates from trap 1 and 4 are interpolated linearly.
The uncertainties in the interpolated frequency ranges are taken to be equal to the largest deviation from a linear interpolation over the same range in trap 2 or 3 (shown in grey in \autoref{fig:fss_event_rates}).

\subsection{\texorpdfstring{$^{\mathrm{83m}}$Kr}{} shallow trap data and fits \label{sec:shallow_trap}}

\begin{figure}[htb]
  \centering
\includegraphics[width=1.0\columnwidth]{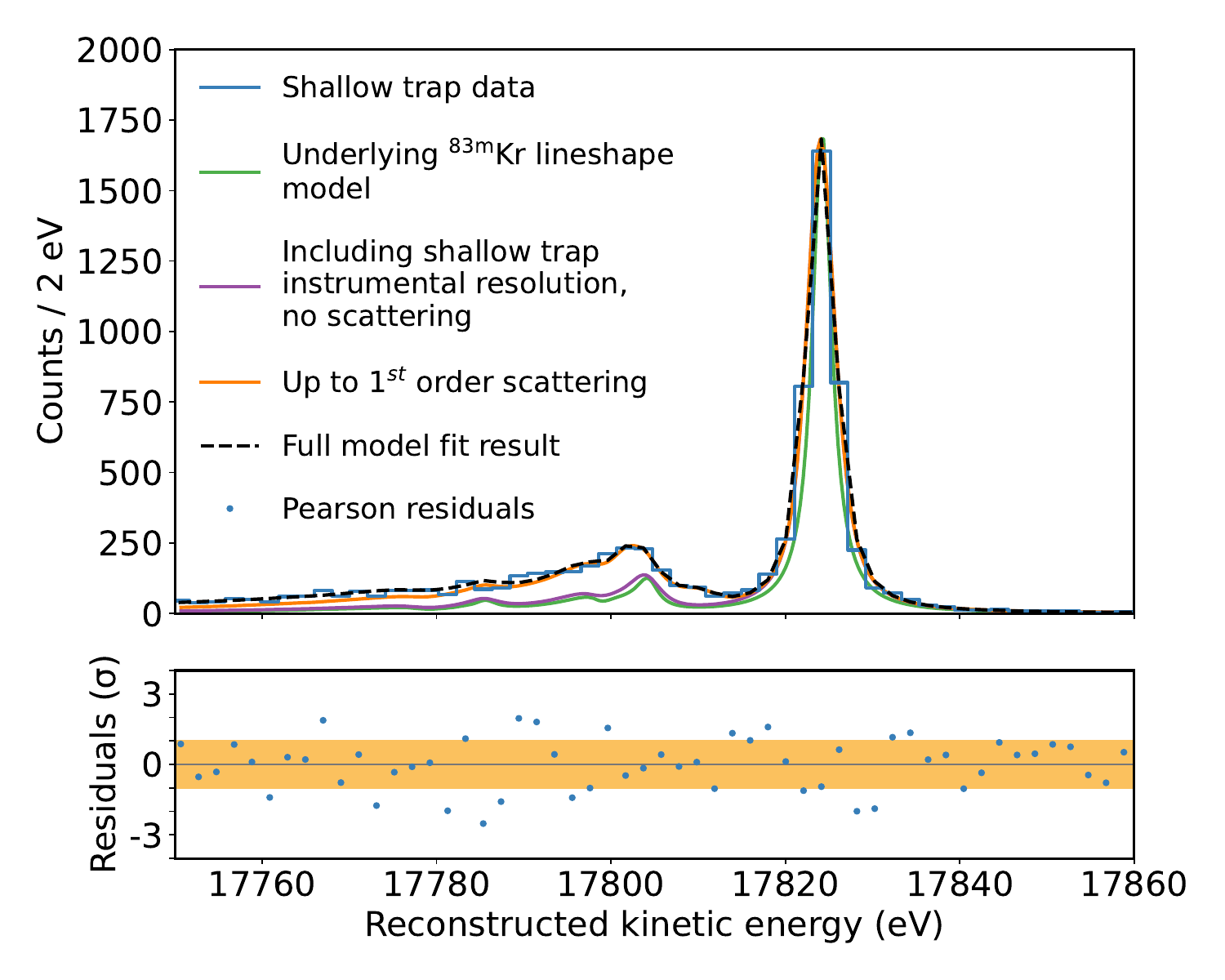} \caption{The 17.8-keV $^{\mathrm{83m}}$Kr K-conversion electron line, as measured with CRES in the shallow (high-resolution) electron trapping configuration, with FWHM of 4.0\,eV. The data are the $^{83\mathrm{m}}$Kr shallow data set (Table \ref{tab:mml_track_length_fit_results}).
  }
  
  \label{fig:krypton shallow trap}
\end{figure}

To explore the best resolution achievable in Phase II, and to test the CRES spectrum model (\autoref{eq:FullModel1}), we took $^{\mathrm{83m}}$Kr data with the trap coil currents set to the shallow trap configuration in \autoref{fig:quadtrapcoils}.
Figure~\ref{fig:krypton shallow trap} shows the fit to these data.
Also shown is the underlying $^{\mathrm{83m}}$Kr lineshape model $\mathcal{Y}_{\mathrm{Kr}}$, which includes both the main peak and the shakeup/shakeoff satellites. 
The figure displays intermediate lineshapes in which contributions to the model are included one by one, to exhibit the effects of magnetic field inhomogeneity (treated as equivalent to instrumental resolution $\mathcal{I}$) and scattering. 
In the shallow trap, there are only small differences between the average magnetic fields experienced by trapped electrons with different pitch angles. Accordingly, the broadening from $\mathcal{I}$ (included in the purple curve) is $1.66(19)$\,eV FWHM.
This combines with the natural linewidth of 2.774\,eV FWHM~\cite{Altenmuller:2019ddl} to produce a main peak with a FWHM of 4.0\,eV.  
Out of all events, 69\% are detected before scattering. Additional curves in \autoref{fig:krypton shallow trap} show events detected after a single scatter and after up to 20 scatters.  In the low-energy tail (below 17.814 eV), scattering events comprise 61\% of counts.

The summed $\chi^2_{\lambda,p}$ of the binned data (631) falls within 1$\sigma$ of the mean of the distribution of summed $\chi^2_{\lambda,p}$ values from MC simulations (607$\pm$ 40), verifying goodness of fit. This demonstrates the high-resolution capabilities of CRES and validates the $^{\mathrm{83m}}$Kr model.

\subsection{\texorpdfstring{$^{\mathrm{83m}}$Kr}{} pre-tritium and post-tritium quad trap data and fits\label{sec:deep_trap}}\label{sec:deep-trap-data-and-fits}

\begin{figure}[t]
  \centering
  \includegraphics[width=1.0\columnwidth]{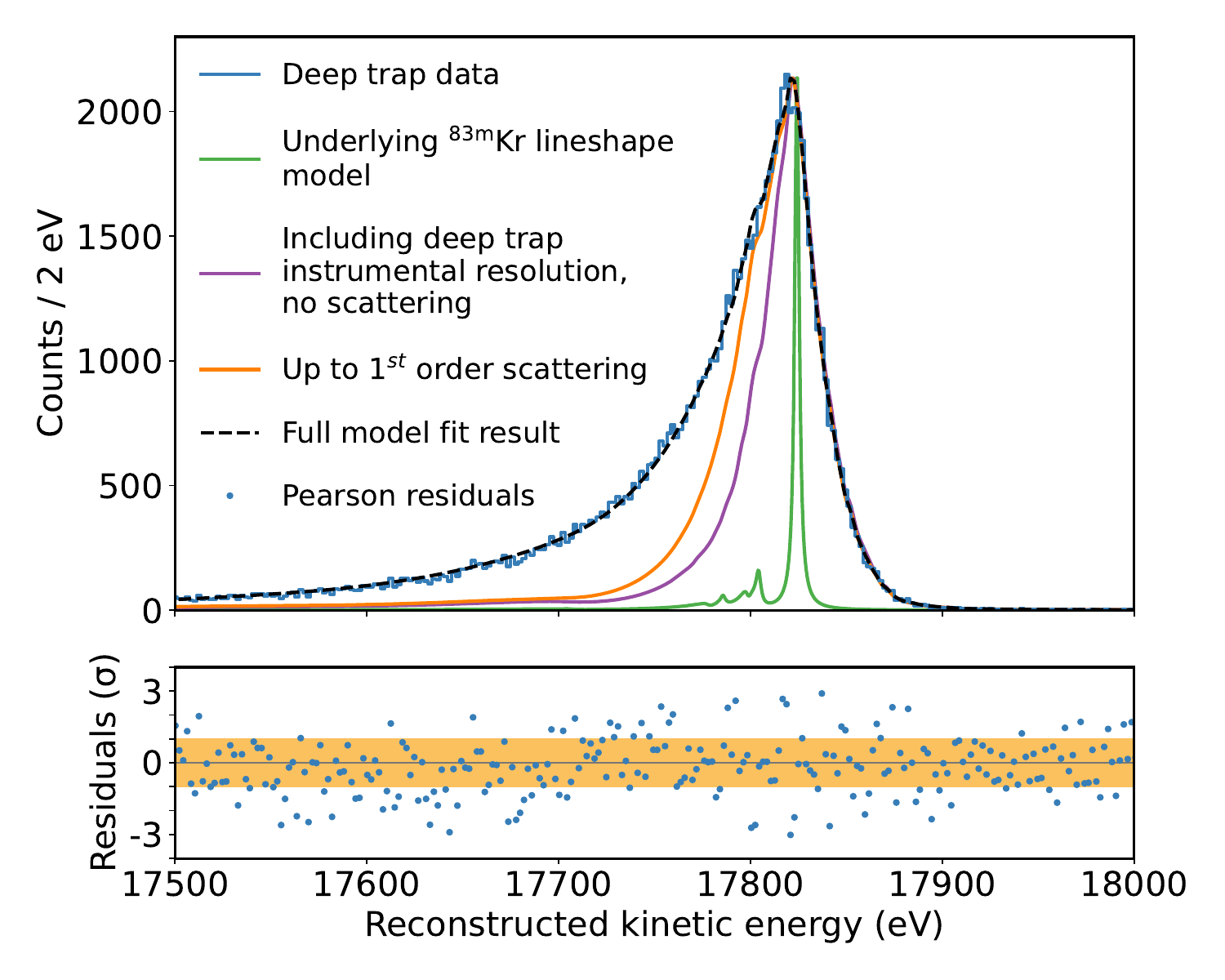}
  \caption{
    The 17.8-keV $^{\mathrm{83m}}$Kr K-conversion electron line, as measured with CRES in the deep (high-statistics) electron trapping configuration, with FWHM of 54.3\,eV. The data are the $^{83\mathrm{m}}$Kr pre-tritium data set (Table \ref{tab:mml_track_length_fit_results}).
  }
  \label{fig:krypton deep trap}
\end{figure}

The $^{\mathrm{83m}}$Kr ``pre-tritium" and ``post-tritium" data sets (see \autoref{tab:mml_track_length_fit_results}) were taken in the same deep quad trap as tritium data, to calibrate the mean  field $B$ and scattering parameters $p$ and $q$ for the tritium analysis. \autoref{fig:krypton deep trap} shows the $^{\mathrm{83m}}$Kr pre-tritium data and fit. The $^{\mathrm{83m}}$Kr line shape is significantly broadened by the 35.6\,eV FWHM instrumental resolution $\mathcal{I}$ (\autoref{fig:kr_ftc_res}), due to the large range of average magnetic fields experienced by electrons. 

Compared with the shallow trap, the larger pitch angle acceptance in the deep trap causes more electrons to remain trapped after scattering, leading to a higher average number of tracks per event. This also leads to a smaller proportion (53\%) of events being detected before scattering, since events that begin in non-detectable pitch angles have a larger phase space of detectable pitch angles to scatter into.
This gives rise to the enhanced low-energy tail and brings the FWHM to 54.3\,eV.
Note that here the scatter peaks merge with the instrumental resolution into a single broad peak. In the deep trap, the FWHM therefore contains both events detected before scattering and events first detected after scattering.

For the $^{\mathrm{83m}}$Kr pre-tritium and post-tritium data sets, the comparison of the summed $\chi^2_{\lambda,p}$ for binned data (1211 and 1112, respectively) with the distributions of MC-simulated summed $\chi^2_{\lambda,p}$ values (884$\pm$38, 830$\pm$59) indicated underfitting. This tension likely stems from small imperfections in the simulated instrumental resolution, relative to data. 
To account for the uncertainty associated with this tension, the uncertainties for $B$, $p$ and $q$ from the maximum likelihood fit are inflated by 17\% and 5\% for the pre-tritium and post-tritium data sets, respectively.
These fit uncertainties are combined with the larger uncertainty contributions from $\mathcal{I}$ and gas composition to produce the total uncertainties on $B$, $p$ and $q$.\footnote{When fit uncertainties are inflated to account for underfitting, this increases the total uncertainties on $B$, $p$ and $q$ by only 1.6\%, 0.3\% and 0.1\%, respectively.} Uncertainties on $\mathcal{I}$ are propagated using the sampling-and-refitting method described in \autoref{sec:res-errors-to-Kr}. The uncertainty on $\mathcal{I}$ due to SNR$_{\rm max}$ is not propagated to $B$, since those variables are independent. SNR$_{\rm max}$ primarily affects the width of $\mathcal{I}$, while $B$ controls the location of the distribution's center; these are two separate moments of $\mathcal{I}$.
To propagate the uncertainty from gas composition, the $^{\mathrm{83m}}$Kr fits are repeated 300 times while sampling the inputted gas scattering contributions from the distributions defined in \autoref{tab:scattering_fraction_results}. The gas composition uncertainties on $B$, $p$ and $q$ are the standard deviations of results from these 300 fits. 

With fit, $\mathcal{I}$, and gas composition uncertainties included, the best estimates of $B$ from $^{\mathrm{83m}}$Kr pre- and post-tritium data differ by 1.6\,$\sigma$. Estimates for $p$ and $q$ are not expected to be consistent between the two quad trap data sets, due to a difference in the mean number of tracks per event (see \autoref{sec:scatter_peak_errors}).

\section{Tritium models \label{sec:tritium_model}}

In this paper, we employ two models of tritium CRES data: a highly detailed model for data generation in Monte Carlo (MC) studies, and a simplified, analytic model for analysis.
In the detailed generation model, the beta spectrum function is numerically convolved with the energy point-spread function $\mathcal{R}_{\mathrm{PSF}}$, and no approximations are made to either function. In the analysis model, several approximations are made for computational efficiency. MC studies demonstrate that each of these approximations do not affect endpoint ($E_0$) and neutrino mass ($m_{\mathrm{\beta}}$) results, as discussed in \autoref{sec:MCstudies}.  The remainder of this section describes the tritium data generation and analysis models.

\subsection{Detailed tritium model for MC data generation}\label{sec:detailed_gen_model}
For tritium data, the underlying spectrum in \autoref{eq:FullModel1} is the beta spectrum $\mathcal{Y}_{\mathrm{tritium}}$, given by the product of neutrino and electron phase space density factors $D_{\nu}$ and $D_e$. 
When experimental sensitivity is insufficient to resolve individual mass eigenstates, the beta spectrum for molecular tritium is given by~\cite{Formaggio:2021nfz} 
\begin{equation}
\begin{split}
    &\mathcal{Y}_{\mathrm{tritium}} = D_{\nu} \cdot D_e, \quad \mathrm{where} \\
    &D_{\nu} \propto  \epsilon_\nu\big[\epsilon_\nu^2 -  m_\beta^2\big]^{1/2} \Theta(\epsilon_\nu-m_\beta) \\
    &D_e \propto F(Z, p_\mathrm{e})p_\mathrm{e} E_\mathrm{e}.
    \label{eq:betaspectrumfull}
\end{split}
\end{equation}
In this equation, $\epsilon_\nu = E_0 - V_k - E_\mathrm{kin}$, where $V_k$ is the energy supplied to rotational, vibrational, and electronic excitations of $^3$HeT$^+$ during the decay \cite{Bodine:2015sma}. $\Theta$ is the Heaviside step function,  $F(Z, p_\mathrm{e})$ is the relativistic Fermi function for charge $Z=2$ of the daughter nucleus, and $p_\mathrm{e}$ and $E_\mathrm{e}$ are the electron momentum and total energy, respectively. The MC data generation model uses \autoref{eq:betaspectrumfull}, including all atomic physics corrections to the Fermi function from~\cite{Kleesiek:2018mel}. We numerically convolve \autoref{eq:betaspectrumfull} with the final state distribution for $^3$HeT$^+$, which is the probability distribution of $V_k$ values. We use the final state distribution calculated by Saenz et al.~\cite{Saenz:2000dul} down to a binding energy of $-2288\,$eV.

The $\mathcal{R}_{\mathrm{PSF}}$ model includes a simulated, tritium-specific instrumental resolution $\mathcal{I}$,
which is numerically convolved (according to \autoref{eq:FullModel2}-\ref{eq:combine_peaks_from_different_gases}) with inelastic scatter spectra calculated from~\cite{LXCat_carbone2021data,pitchford2017lxcat,pancheshnyi2012lxcat,LXCat:database}.
We account for rare scatters with CO during generation but not analysis. Twenty scatter peaks are generated ($j_{\mathrm{max}}=20$). An MC study shows that including higher-order peaks in the generation and/or analysis models does not alter results. $\mathcal{R}_{\mathrm{PSF}}$ is numerically convolved with $\mathcal{Y}_{\mathrm{tritium}}$.

\subsection{Approximate tritium model for analysis}
\label{subsec:tritium_analysis_approximations}
The tritium analysis model is used to fit both the tritium spectrum obtained from the apparatus and Monte Carlo spectra. The analysis model includes approximate, analytic expressions for both  $\mathcal{Y}_{\mathrm{tritium}}$ and $\mathcal{R}_{\mathrm{PSF}}$, enabling computationally efficient inference. This is crucial for the Bayesian analysis, since algorithms that perform Bayesian inference in many dimensions---corresponding to many nuisance parameters---tend to be slow at numerical integration. The analytic model of $\mathcal{R}_{\mathrm{PSF}}$ also substantially speeds up the calculation of the response in the frequentist analysis.

For each model approximation described below, a Monte Carlo test is performed by generating an ensemble of spectra with the detailed tritium model, then analyzing those spectra using a model which includes the approximation. These studies show that $E_0$ and $m_\beta$ fit results are unaffected by each approximation, for Phase II data. Even for the resolution and statistics expected in Project 8's final planned phase, the simplified model of $\mathcal{Y}_{\mathrm{tritium}}$ was shown to yield accurate results~\cite{AshtariEsfahani:2021moh}. Some simplifications of $\mathcal{R}_{\mathrm{PSF}}$  may not hold in future experiments. We will refine the $\mathcal{R}_{\mathrm{PSF}}$ model and incorporate numerical components, as needed (which may be practical for Bayesian inference with state-of-the-art tools). 

\subsubsection{Tritium beta decay analysis model\label{sec:T2-beta-model}}
The frequentist analysis model uses \autoref{eq:betaspectrumfull} for the underlying beta spectrum $\mathcal{Y}_{\mathrm{tritium}}$. In the Bayesian analysis, $D_\nu$ is approximated according to the formalism in~\cite{AshtariEsfahani:2021moh}. This involves Taylor expanding in $m_\beta^2$ to produce the expression
\begin{equation}
    D_{\nu} \approx  \big[\epsilon_\nu^2 -  m_\beta^2/2\big] \Theta(\epsilon_\nu-m_\beta).
    \label{eq:betaspectrum}
\end{equation}
In addition, $D_e$ is Taylor expanded to first order around the energy at the center of the  analysis region of interest (ROI), neglecting atomic physics factors that correct the Fermi function. 
The resulting model for $\mathcal{Y}_{\mathrm{tritium}}$ may be analytically convolved with a normal distribution. Thus, for any model of $\mathcal{R}_{\mathrm{PSF}}$ that is expressed as a weighted sum of Gaussians, the full tritium model is analytic.
Here, unlike in Ref.~\cite{AshtariEsfahani:2021moh}, the low-energy edge of the spectrum is not smeared out by magnetic field broadening, since a hard maximum-frequency cut is performed before analysis.\footnote{A low-energy smearing of 0.001\,eV is included in the model for computational stability, negligibly affecting results. In Bayesian MCMC inference, infinitely steep drops in probability density can cause Markov chains to behave pathologically~\cite{Stanual2022, Betancourt2015}.}   

If the final state distribution of $^3$HeT$^+$ were neglected in the analysis model, this would bias the $E_0$ result by \SI{-8.1 \pm 0.8}{eV}, as computed from a Bayesian MC study. 
To speed up computation, the frequentist and Bayesian models use a sparse approximation of the  final state distribution down to 240 eV below the endpoint. The sparse distribution uses only every 4th excitation energy and associated probability from~\cite{Saenz:2000dul}.
While some electrons are produced by HT, which decays to $^3$HeH$^+$, this molecule's final state distribution is similar to that of $^3$HeT$^+$, compared with our resolution. These simplifications introduce no biases in the results.

\subsubsection{Tritium energy response function analysis model}\label{sec:T2-det-response}
For tritium analysis, the energy response point-spread function $\mathcal{R}_{\mathrm{PSF}}$ is modeled as a sum of Gaussians.
This allows $\mathcal{R}_{\mathrm{PSF}}$ to be analytically convolved with the beta spectrum, simplifying computation.

Within $\mathcal{R}_{\mathrm{PSF}}$, the energy loss function $\mathcal{L}_{\mathrm{tot}}$ accounts for scattering with H$_2$ and $^3$He, as these have the largest inelastic scatter fractions $\gamma_i$ in tritium data: \SI{0.911 \pm 0.045} and \SI{0.075 \pm 0.040}, respectively. The scatter fraction \SI{0.014 \pm 0.009} of CO is omitted from the tritium fit model. We further simplify by modeling each scatter peak as a weighted sum of H$_2$ and $^3$He peaks. This is akin to assuming that a given electron scatters with the same gas type after all missed tracks. The radiative loss $\mathcal{L}_r$ is omitted from the model, since it is small relative to scattering losses. MC studies validate these simplifications.

Each scatter peak $\mathcal{I}*\mathcal{L}_{\mathrm{tot}}^{*j}$
is expressed as a function of $\sigma$, the calculated standard deviation of the simulated resolution $\mathcal{I}$. This enables us to propagate uncertainty on the resolution width to the endpoint and neutrino mass, via $\sigma$. The $j=0$ peak is modeled separately from $j\geq1$ peaks (up to $j_{\mathrm{max}}=20$), as described below.

The $j=0$ term reduces to $\mathcal{I}$.  Because this term is the dominant contribution to $\mathcal{R}_{\mathrm{PSF}}$, it is important to model the peak with a closely-fitting distribution. A simple Gaussian would underestimate the tails of $\mathcal{I}$ and fail to account for its small asymmetry. Instead, the simulated $\mathcal{I}$ is fitted with a sum of two normal distributions $\mathcal{N}$ with means $\mu_0^{[i]}$ and standard deviations $\sigma_0^{[i]}$ ($i=1,2$), weighted by a parameter $0\le\eta\le 1$:
 
\begin{equation}
\begin{split}
\mathcal{I}*\mathcal{L}_{\mathrm{tot}}^{*0} \approx \eta\mathcal{N}\Big(\mu_0^{[1]}, \sigma_0^{[1]}(\sigma)\Big) +(1-\eta)\mathcal{N}\Big(\mu_0^{[2]}, \sigma_0^{[2]}(\sigma)\Big).
\end{split}
\label{eq:T2_ins_res_model}
\end{equation}

To find how $\sigma_0^{[1]}$ and $\sigma_0^{[2]}$ depend on $\sigma$, we perform fits to simulated resolutions with a range of $\sigma$ values (see Appendix~\ref{sec:simplified_res_sims}).  
The result of this procedure is $\sigma_0^{[1]}(\sigma) = 1.1\sigma + 1.9\,$eV and $\sigma_0^{[2]}(\sigma) = 0.8\sigma - 3.7\,$eV.
This procedure also demonstrates that $\eta=0.66$ is constant as $\sigma$ varies. By fixing $\eta$ and plugging the expressions for $\sigma_0^{[1]}(\sigma)$ and $\sigma_0^{[2]}(\sigma)$ into \autoref{eq:T2_ins_res_model}, we obtain a model for $\mathcal{I}$ with only three free parameters: $\sigma$, $\mu_0^{[1]}$ and $\mu_0^{[2]}$.
This ``reduced model" 
can scale in width based on $\sigma$ and captures the slight asymmetry in $\mathcal{I}$ via $\mu_0^{[1]}$ and $\mu_0^{[2]}$.
We confirm that the fitted value of $\sigma$ matches the standard deviation $\sigma$ calculated directly from $\mathcal{I}$. 

\begin{figure}[t]
  \includegraphics[width=1.0\columnwidth]{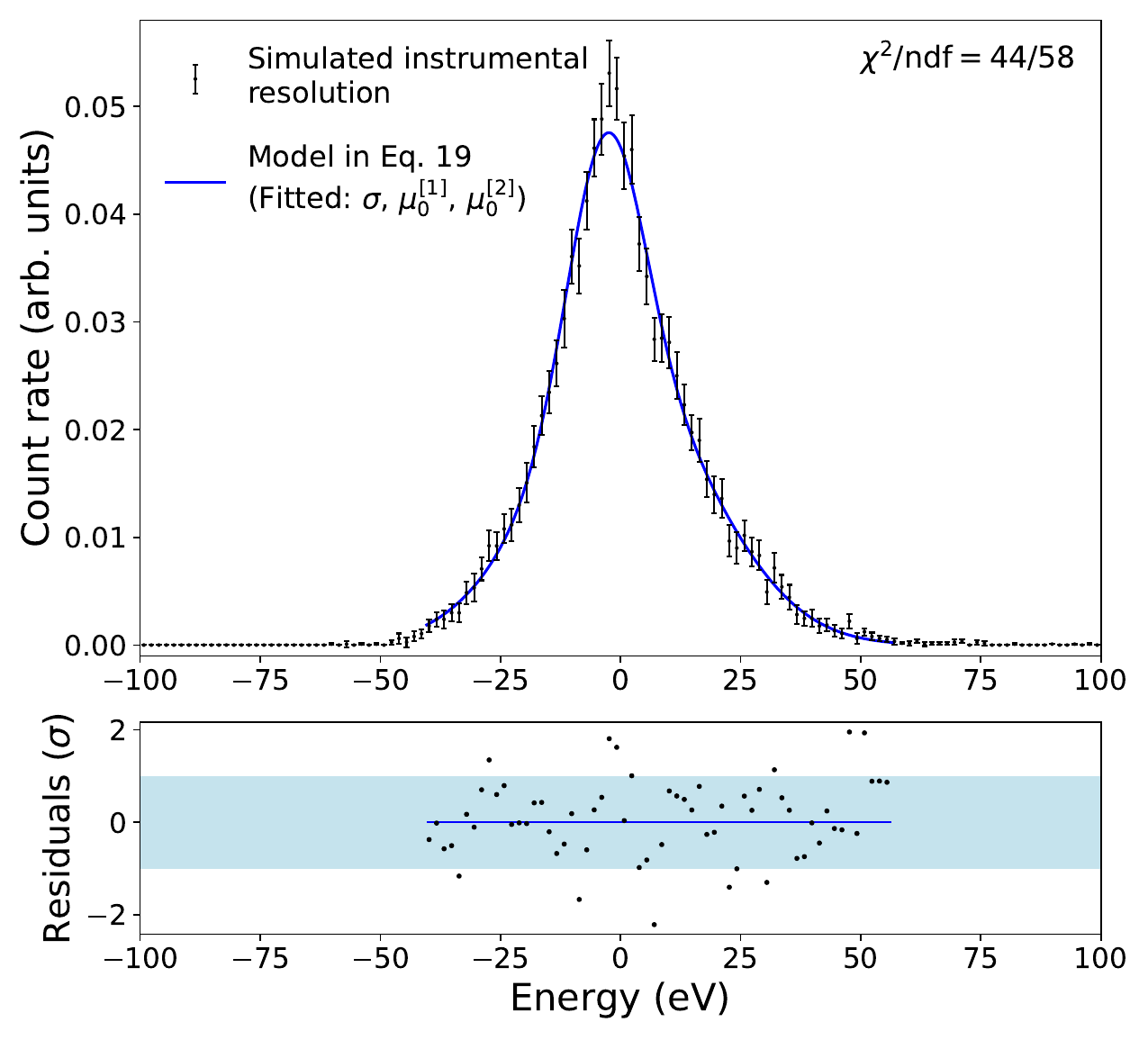}
  \caption{
   Fit of \autoref{eq:T2_ins_res_model} to the simulated resolution $\mathcal{I}$ for tritium data. Bin errors are approximately Gaussian, from sources described in \autoref{subsec:ins_res}. Fit parameters are means $\mu_0^{[1]}$, $\mu_0^{[2]}$ of the two normal distributions and standard deviation $\sigma$ of $\mathcal{I}$. The fraction of counts $\eta$ in the first normal distribution is inputted, determined by computing the average $\eta$ from fits to 100 resolutions with different $\sigma$ values.}
  \label{fig:T2_res_fit}
\end{figure}

We find the best estimates of $\sigma$, $\mu_0^{[1]}$ and $\mu_0^{[2]}$ by fitting the reduced model to the simulated $\mathcal{I}$ distribution that was generated with the best-estimate SNR$_{\rm max}$ value, as shown in \autoref{fig:T2_res_fit}. The $\chi^2/\mathrm{ndf}$ is $44/58$, and the fit result for $\sigma$ is consistent with the standard deviation calculated directly from $\mathcal{I}$. 
The fit energy range is limited to produce a good fit to the central region of $\mathcal{I}$, which has the largest impact on tritium data fits. During tritium analysis, uncertainties are propagated for $\sigma$ (see \autoref{sec:sigma_errors}) but not for $\mu_0^{[1, 2]}$, since an MC study shows that neglecting uncertainties on $\mu_0^{[1, 2]}$ negligibly affects results.

When $j\geq1$, each scattering term $\mathcal{I}*\mathcal{L}_{(\mathrm{H}_2)}^{*j}$ or $\mathcal{I}*\mathcal{L}_{(\mathrm{He})}^{*j}$
can be modeled by a normal distribution, with a mean and standard deviation that depend only on $\sigma$. This approximation holds despite the asymmetry in $\mathcal{L}_{(\mathrm{H}_2, \mathrm{He})}^{*j}$ for two reasons:  the $j\geq1$ peaks each contribute sub-dominantly to $\mathcal{R}_{\mathrm{PSF}}$, and they are broadened by $\mathcal{I}$, making them more Gaussian.
Scatter peaks are modeled as

\begin{equation}
\begin{split}
&\mathcal{I}*\mathcal{L}_{\mathrm{tot}}^{*j} (j\geq1)  \approx \gamma_{\mathrm{H}_2}\Big[\mathcal{I}*\mathcal{L}_{(\mathrm{H}_2)}^{*j}\Big] + (1-\gamma_{\mathrm{H}_2})\Big[\mathcal{I}*\mathcal{L}_{(\mathrm{He})}^{*j}\Big] \\ &\!\!\!\!\!\rightarrow  \gamma_{\mathrm{H}_2}\mathcal{N}\Big(\mu_j^{\mathrm{H}_2}(\sigma), \sigma_j^{\mathrm{H}_2}(\sigma)\Big)  +(1\!-\!\gamma_{\mathrm{H}_2})\mathcal{N}\Big(\mu_j^{\mathrm{He}}(\sigma), \sigma_j^{\mathrm{He}}(\sigma)\Big),
\end{split}
\label{eq:simplified_scattering}
\end{equation}

\noindent where $\gamma_{\mathrm{H}_2}$ is the hydrogen inelastic scatter fraction and $1-\gamma_{\mathrm{H}_2}$ is the helium inelastic scatter fraction.  The Gaussian means $\mu_j$ depend on $\sigma$ because $\mathcal{L}_{(\mathrm{H}_2, \mathrm{He})}^{*j}$ is asymmetric, so the convolution with $\mathcal{I}$ can shift the center of each scatter peak when $\sigma$ is large enough (as is the case for deep quad trap data).

The slopes and $y$-intercepts of $\mu_j(\sigma)$ and $\sigma_j(\sigma)$ are fixed during tritium data analysis, so the scatter tail shape depends only on $\sigma$, $\gamma_{\mathrm{H}_2}$ and scatter peak amplitudes. For each gas, the slopes and intercepts for $\mu_j$ and $\sigma_j$ are determined through the following procedure. For a given $j$, we fit Gaussians to 20 sets of scatter peaks broadened by 20 different resolution widths, ranging from $0.5 \sigma$ to $1.5 \sigma$. 
This procedure produces $\mu_j$ and $\sigma_j$ for a range of $\sigma$ values. We then observe and fit the linear dependence of $\mu_j$ and $\sigma_j$ on $\sigma$. 

Each scatter peak is multiplied by the corresponding amplitude $\mathcal{A}_j(p, q)$.
Tritium-specific $p$ and $q$ values are estimated in \autoref{sec:scatter_peak_errors} by slightly shifting $p$ and $q$ from the fit to $^{\mathrm{83m}}$Kr pre-tritium data, to account for a difference in the mean number of tracks per event ($N_{\mathrm{tracks}}^{\mathrm{true}}$) between $^{\mathrm{83m}}$Kr and tritium data.
Combining the $j=0$ and $j\geq1$ peaks, the full $\mathcal{R}_{\mathrm{PSF}}$  model for tritium analysis includes a limited set of free parameters with propagated uncertainties: $\gamma_{\mathrm{H}_2}$, $\sigma$, $p$, and $q$.

\subsection{Event rate model}\label{sec:event_rate}
For both tritium data generation and analysis models, the signal probability density function $\mathcal{S}(E_\mathrm{kin})$ is given by \autoref{eq:FullModel1}. A false event probability density function $\mathcal{F}(E_\mathrm{kin})$ is also introduced.  $\mathcal{F}(E_\mathrm{kin})$ is assumed to be flat in energy because the probability to measure RF noise (the only expected significant background source) is uniform as a function of cyclotron frequency, and energy is approximately linearly related to frequency over a limited range. Combining signal and background, the expected tritium event rate is  
\begin{equation}
  \frac{dN}{dE_\mathrm{kin}}(E_\mathrm{kin}) = r_s \mathcal{S}(E_\mathrm{kin})+r_f\mathcal{F}(E_\mathrm{kin}),
  \label{eq:fullT2model}
\end{equation}
\noindent where $r_s$ is the signal rate and $r_f$ is the false event rate. Binning of data is handled differently in Bayesian and frequentist analyses, as  discussed in \autoref{sec:final-analysis}.

\section{Tritium parameter estimates and uncertainties}
\label{sec:systematic_uncertainties}

We study and quantify the following systematic effects for the tritium data analysis: 
\begin{enumerate}
\item The mean magnetic field $B$ that converts cyclotron frequencies to energies;
\item The tritium-specific simulated instrumental resolution $\mathcal{I}$, which determines $\sigma$;
\item Scatter peak amplitudes $\mathcal{A}_j$ (parameterized by $p$ and $q$);
\item The energy-dependent event detection efficiency $\epsilon$;
\item The frequency dependence of the energy point-spread function $\mathcal{R}_{\mathrm{PSF}}$ (specifically, of $\sigma$, $p$, and $q$); and
\item Gas composition, which determines the hydrogen inelastic scattering fraction $f_{\mathrm{H}_2}$ in tritium data.
\end{enumerate}
This section covers items 1-5. Item 6\textemdash gas composition and associated uncertainties\textemdash was discussed in \autoref{sec:gas_composition} and \autoref{sec:T2-det-response}.

Systematic factors affect the tritium data analysis via three pathways. First, $^{\mathrm{83m}}$Kr-specific estimates of some uncertainties (on simulated resolutions and gas composition) are incorporated into the $^{\mathrm{83m}}$Kr quad trap analysis and propagated to $B$, $p$ and $q$---the three tritium model parameters from $^{\mathrm{83m}}$Kr fits. Second, tritium-specific estimates of some uncertainties (on $\sigma$ and $f_{\mathrm{H}_2}$) are directly employed in tritium data analysis and propagated to the endpoint $E_0$ and neutrino mass $m_\beta$ as described in \autoref{sec:final-analysis}. Third, $^{\mathrm{83m}}$Kr fits are used to estimate the variation of parameters ($\epsilon$, $\sigma$, $p$ and $q$) across the tritium ROI, as well as the uncertainty on this variation.

Below, we describe procedures for estimating  parameters in the tritium model, their systematic uncertainties, and where relevant, their correlations. At the end of this section, \autoref{sec:summaryofpriors} summarizes the probability distributions for all tritium model parameters, which account for uncertainties on these parameters.

\subsection{Mean magnetic field \texorpdfstring{$B$}{} \label{sec:Bfield_errors}}

A systematic uncertainty in $B$ shifts the overall reconstructed energy scale, so uncertainties in $B$ are expected to propagate to $E_0$. By contrast, the $m_\beta$ determination is unaffected  by the $B$ systematic, since $m_\beta$ is altered by the second moment of the energy PSF (spectral broadening), not the first moment (overall energy scale) \cite{robertson:1988aa}. 

The best estimate for $B$ is determined by fitting $^{\mathrm{83m}}$Kr pre-tritium quad trap data.
We choose this data set because its event features most closely resemble those of tritium data. In particular, as shown in \autoref{tab:mml_track_length_fit_results}, the pre-tritium mean track duration differs from that of tritium data by 10\% (compared with 29\% for post-tritium data), and the pre-tritium mean number of tracks differs by 2\% (compared with 43\% for post-tritium data). This choice minimizes differences between parameters fitted from $^{\mathrm{83m}}$Kr data and tritium parameters. Pre- and post-tritium estimates of $B$ differ by $1.6\sigma$, 
indicating that the impact of discrepancies in data-taking conditions on $B$ is relatively small. The best estimate for $B$ is shifted downward by $5\times10^{-7}\,$T to correct for a 14\,kHz (0.3\,eV) mean error in start frequencies. This error is caused by the reconstruction algorithm identifying electron tracks with a small time delay, on average, during which the electron loses energy to radiation. The uncertainty on this shift contributes negligibly to the $B$ uncertainty.

The uncertainty on $B$ includes three contributions from the $^{\mathrm{83m}}$Kr quad trap fitting process:
the statistical uncertainty outputted by the $^{\mathrm{83m}}$Kr maximum likelihood fit ($\pm 3\times10^{-7}\,$T), the uncertainty in the gas composition of pre-tritium $^{\mathrm{83m}}$Kr data ($\pm 7\times10^{-7}\,$T), and simulation uncertainties on the instrumental resolution input to the $^{\mathrm{83m}}$Kr fit ($\pm 5\times10^{-7}\,$T). 
Combining the three uncertainties in quadrature, we measure a mean magnetic field of $B=0.9578099(9)\,$T. Separately, there is a magnetic field uncertainty from a 0.5\,eV uncertainty on the $^{\mathrm{83m}}$Kr K-line energy~\cite{V_nos_2018}. Accounting for the external K-line uncertainty, we find $B=0.9578104(13)\,$T for the tritium analysis.

\subsection{Energy resolution \texorpdfstring{$\sigma$}{} \label{sec:sigma_errors}}

The simulated resolution $\mathcal{I}$ for the tritium data analysis differs from the resolutions used for quad trap $^{\mathrm{83m}}$Kr analyses due to slight differences in mean track duration and mean number of tracks per event.
The standard deviation $\sigma$ of the tritium resolution is estimated by fitting the tritium-specific simulated resolution with the model in \autoref{eq:T2_ins_res_model}. The best fit result is $\sigma=15.10\,$eV.
In the tritium data fits, we include two types of uncertainties on the resolution parameter $\sigma$: (a) uncertainties from the $\mathcal{I}$ simulation process, and (b) an uncertainty on the optimized SNR$_{\rm max}$ value. The procedures for determining (a) and (b) are described below.

There are three contributions to the simulation uncertainty (a): first, Poisson errors on the number of simulated events; second, uncertainties in the efficiency filter matrix; and third, the uncertainty in the number of events contributed from each magnetic trap in the quad trap. The resulting bin errors are propagated to each bin of the histogrammed start frequencies that comprise the simulated resolution. Combined, these uncertainties are approximately Gaussian and are included in a $\chi^2$ fit of $\mathcal{I}$ using \autoref{eq:T2_ins_res_model}. Accordingly, the $0.22$-eV uncertainty on $\sigma$ that the fit outputs accounts for simulation uncertainties.

The SNR$_{\rm max}$ in the tritium simulation  (b) also affects $\sigma$. To estimate the SNR$_{\rm max}$ uncertainty, we compared the optimal SNR$_{\rm max}$ from \autoref{sec:max-SNR-optimization} (14.3) with the result from an alternate method, the first-track SNR matching described in \autoref{sec:sim-events}. 
This produced an estimate of SNR$_{\rm max}$=18.0. The result of the method in \autoref{sec:max-SNR-optimization} provides the SNR$_{\rm max}$ best estimate, because that method ensures that the distribution's width is consistent with data---our primary concern when estimating $\sigma$. Still, the discrepancy between the two estimates sets the uncertainty scale, as it quantifies the impact of small imperfections in the resolution simulation (for example, mis-modeling of pitch angle changes from scattering, which could not be directly compared to data). Thus, we take the SNR$_{\rm max}$ uncertainty to be half the difference between the two estimates: 1.85. This is larger than the uncertainty from the procedure in Sec.~\ref{sec:max-SNR-optimization} (0.15) and twice the difference between the optimal SNR$_{\rm max}$ values for pre- and post-tritium $^{\mathrm{83m}}$Kr data, suggesting that the uncertainty of 1.85 may be conservative.

To find the corresponding uncertainty on $\sigma$, we fit \autoref{eq:T2_ins_res_model} to 100 tritium-specific resolutions, each simulated with an input SNR$_{\rm max}$ sampled from a normal distribution with a mean of 14.3 and standard deviation of 1.85.
The $\sigma$ uncertainty contribution from SNR$_{\rm max}$ is then $1.73\,$eV, the standard deviation of the 100 fit results. Combining simulation and SNR$_{\rm max}$ uncertainties in quadrature, we find $\sigma=15.1\pm1.7\,$eV for the tritium instrumental resolution.

\subsection{Scatter peak amplitudes \texorpdfstring{$A_j(p, q)$}{}
\label{sec:scatter_peak_errors}}
\label{subsubsec:extract_scatter_peak_amplitudes}
\label{subsubsec:pq_extrapolation}

We estimate $p$, $q$ for tritium data using $p$, $q$ results from the two $^{\mathrm{83m}}$Kr quad trap fits, adjusting for the difference in mean number of tracks per event ($N_\text{tracks}^{\text{true}}$) between data sets. More true tracks provide more opportunities to detect an event, even after early tracks are missed---increasing the number of events in higher-$j$ scatter peaks. Accordingly, a higher $N_{\mathrm{tracks}}^{\mathrm{true}}$ produces a slower decline in $\mathcal{A}_j$ as a function of $j$. 
Among data sets obtained with the same magnetic trap configuration, such as the tritium and $^{\mathrm{83m}}$Kr quad trap data, we therefore expect that $N_\text{tracks}^{\text{true}}$ is the dominant factor causing differences in $p$ and $q$.  Besides $N_\text{tracks}^{\text{true}}$, other properties affecting missed track probabilities, such as electron pitch angle distributions, are similar among data sets with the same trap configuration. In addition, the mean track duration $\tau$ is determined to be a sub-dominant factor, since fitted $p$ and $q$ values appear to vary randomly with $\tau$ across data sets, suggesting that they are not strongly correlated with $\tau$.

\begin{figure}
\centering
\includegraphics[width=0.45\textwidth]{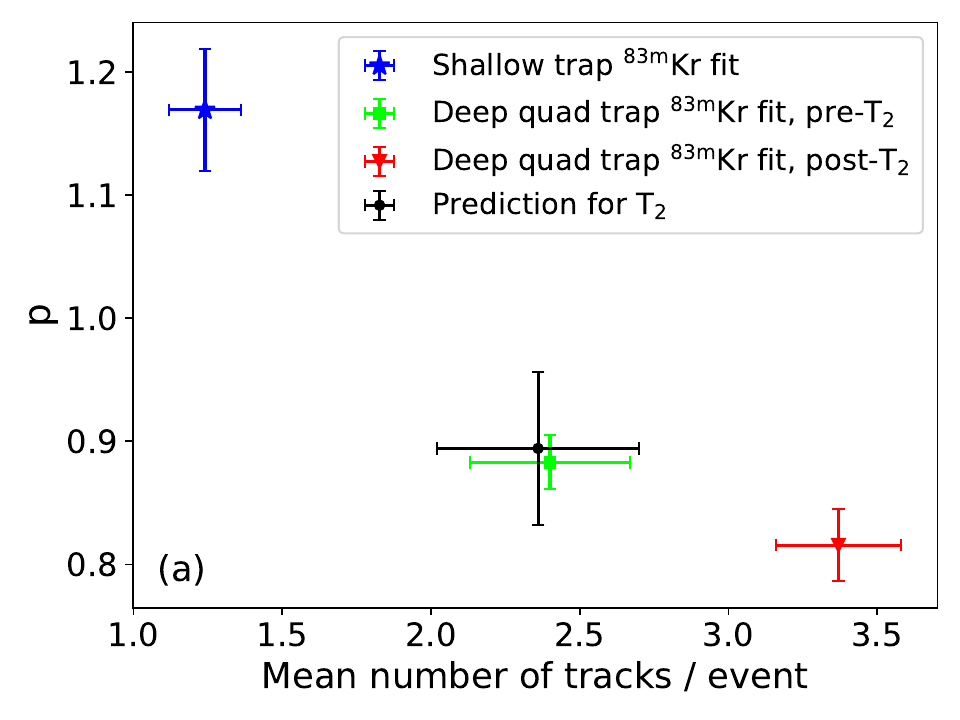}
\includegraphics[width=0.45\textwidth]{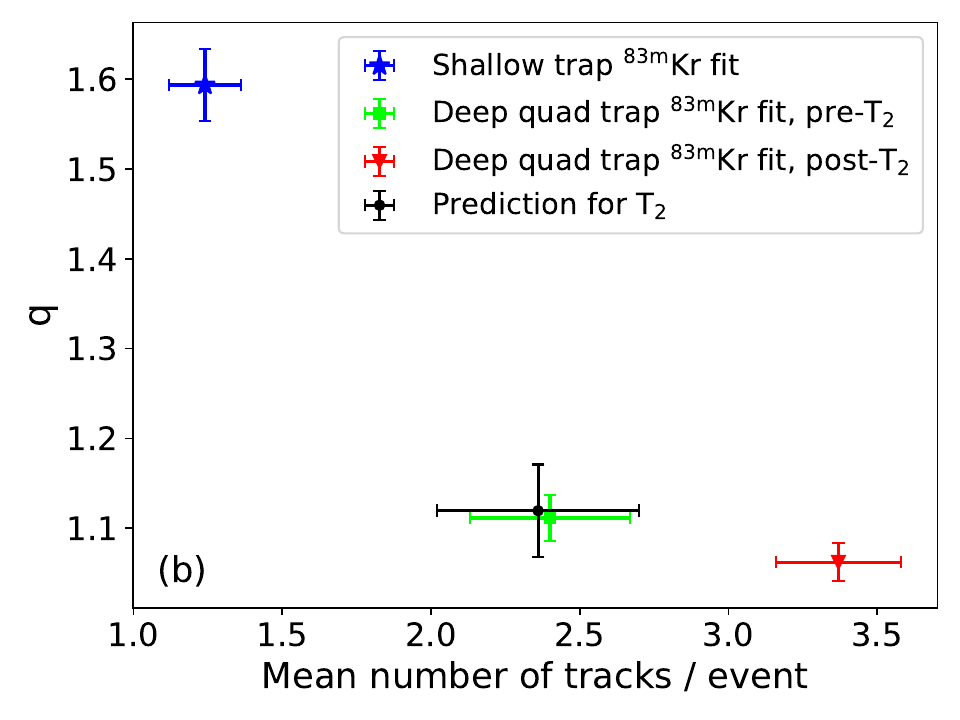}
\caption{Fitted scatter peak amplitude parameters (a) $p$ and (b) $q$ vs.~mean number of tracks per event in each data set. For tritium (black dot), parameters are predicted by extrapolating deep quad trap results to the tritium data's number of tracks per event.}
\label{fig:p-q-vs-ntracks}
\end{figure}

\begin{figure}
\centering   
\includegraphics[width=\linewidth]{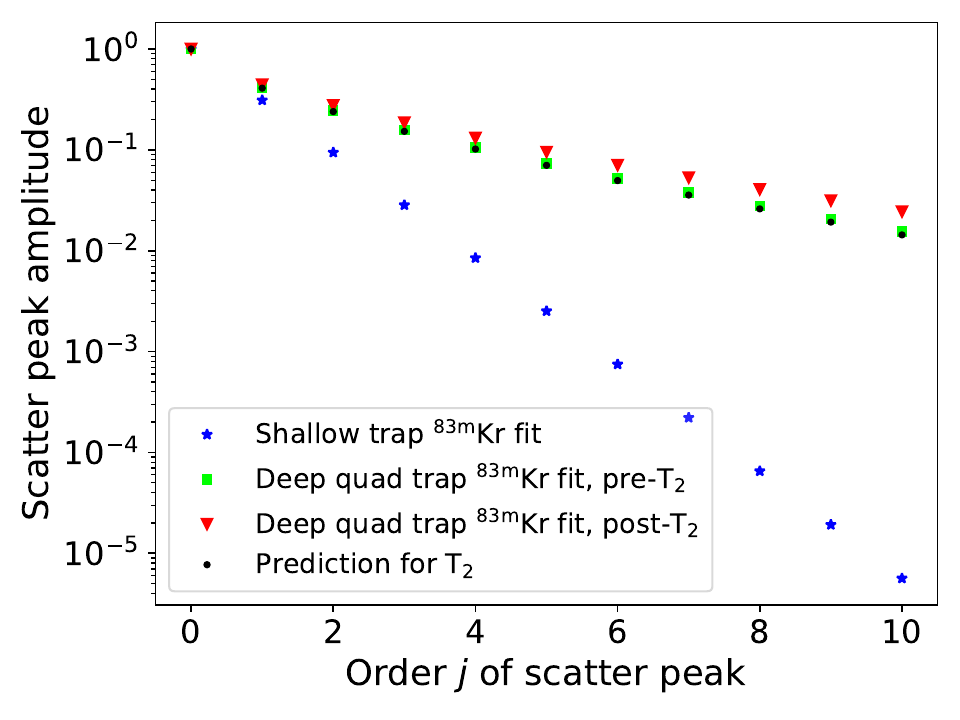}
\caption{Best-estimate scatter peak amplitudes $\mathcal{A}_j$ for $j$ missed tracks. $^{\mathrm{83m}}$Kr estimates are fitted from data; tritium estimates are extrapolated from $^{\mathrm{83m}}$Kr results.}
\label{fig:Aj-results}
\end{figure}

The estimated $N_\text{tracks}^{\text{true}}$ for all data sets are listed in \autoref{tab:data_set_table}. 
Results from pre- and post-tritium quad trap $^{\mathrm{83m}}$Kr fits define functions $p(N_\text{tracks}^{\text{true}})$ and $q(N_\text{tracks}^{\text{true}})$, which predict $p$ and $q$ for tritium data. 
We perform linear extrapolations from these to $p$ and $q$ for the $N_\text{tracks}^{\text{true}}$ in tritium data. \autoref{fig:p-q-vs-ntracks} displays the $p$, $q$ estimates and uncertainties used in this extrapolation. The uncertainty on $N_\text{tracks}^{\text{true}}$ is larger for tritium than for $^{\mathrm{83m}}$Kr data because the tritium data set has fewer events. Uncertainties are propagated via Monte Carlo sampling.  For each of the three quad trap data sets, we sample $N_\text{tracks}^{\text{true}}$ from a normal distribution defined by its mean and uncertainty, given in \autoref{tab:mml_track_length_fit_results}. For each $^{\mathrm{83m}}$Kr data set, we sample $p$ and $q$ from a bivariate normal distribution which accounts for their correlated uncertainty contributions. That set of sampled values predicts one $p$-$q$ pair for tritium. By repeating the sampling process, we construct a bivariate uncertainty distribution of tritium $p$-$q$ values. Means, uncertainties and a correlation are computed from this distribution and reported in the first row of \autoref{table:pq_systematic}.  Note that the mean $p$, $q$ values for tritium are close to the pre-tritium $^{\mathrm{83m}}$Kr values; this is because $N_\text{tracks}^{\text{true}}$  is similar for the two data sets. Best estimates for scatter peak amplitudes as a function of scatter order are shown in \autoref{fig:Aj-results}.

We include three systematic effects in the $p$-$q$ bivariate uncertainty distributions for $^{\mathrm{83m}}$Kr quad trap data:
statistical fit uncertainties, gas composition uncertainties, and resolution ($\mathcal{I}$) simulation uncertainties. These effects produce shifts in different directions for distinct data sets, so they may be propagated in the extrapolation using independent distributions for pre- and post-tritium $^{\mathrm{83m}}$Kr data.
For each $^{\mathrm{83m}}$Kr data set, the covariance matrix for sampling $p$, $q$ is the sum of three covariance matrices associated with the three systematic effects. For statistical fit uncertainties, the matrix is output by the $^{\mathrm{83m}}$Kr maximum likelihood fit. Gas composition uncertainties are determined as described in \autoref{sec:gas_composition}. Specifically, $^{\mathrm{83m}}$Kr data are re-fitted for 300 gas compositions, producing a $p$-$q$ distribution from which a covariance matrix is calculated. Simulation uncertainties in $\mathcal{I}$  are determined using the method in \autoref{sec:res-errors-to-Kr}, which again yields a $p$-$q$ distribution and corresponding covariance matrix.

\begin{table}
\centering
\caption{Estimates and uncertainties of scatter peak amplitude parameters $p$ and $q$ for tritium data.}
\renewcommand{\arraystretch}{1.1}
\begin{tabular}{l c c c}
\hline \hline 
& $p$ & $q$ & Correlation\\ \hline
Best estimate & 0.89 & 1.12 & N/A\\  
Extrapolation uncertainty & 0.06 & 0.05 & 0.60\\  
SNR$_{\rm max}$ uncertainty & 0.09 & 0.01 & 0.56\\
Total uncertainty & 0.11 & 0.05 & 0.38\\
 \hline \hline
 \label{table:pq_systematic}
\end{tabular}
\end{table}   

One additional effect, SNR$_{\rm max}$ uncertainty, is accounted for after the $p$-$q$ extrapolation. This uncertainty is not propagated through the extrapolation because doing so would treat the effect as uncorrelated among data sets. In fact, mis-modeling the SNR$_{\rm max}$ systematically shifts $p$ and $q$ across data sets by altering the width of $\mathcal{I}$, causing $p$ and $q$ to compensate similarly during both quad trap $^{\mathrm{83m}}$Kr fits. We estimate the SNR$_{\rm max}$ uncertainties and correlation for tritium $p$, $q$  using the method in \autoref{sec:res-errors-to-Kr}, with pre-tritium $^{\mathrm{83m}}$Kr data. Covariance matrices for the extrapolation and the SNR$_{\rm max}$ effect are then summed.
Table~\ref{table:pq_systematic} summarizes the results. We find $p=0.89\pm0.11$ and $q=1.12\pm0.05$, with a correlation of 0.38.

\begin{figure*}[tb]
\centering  
\includegraphics[width=\linewidth]{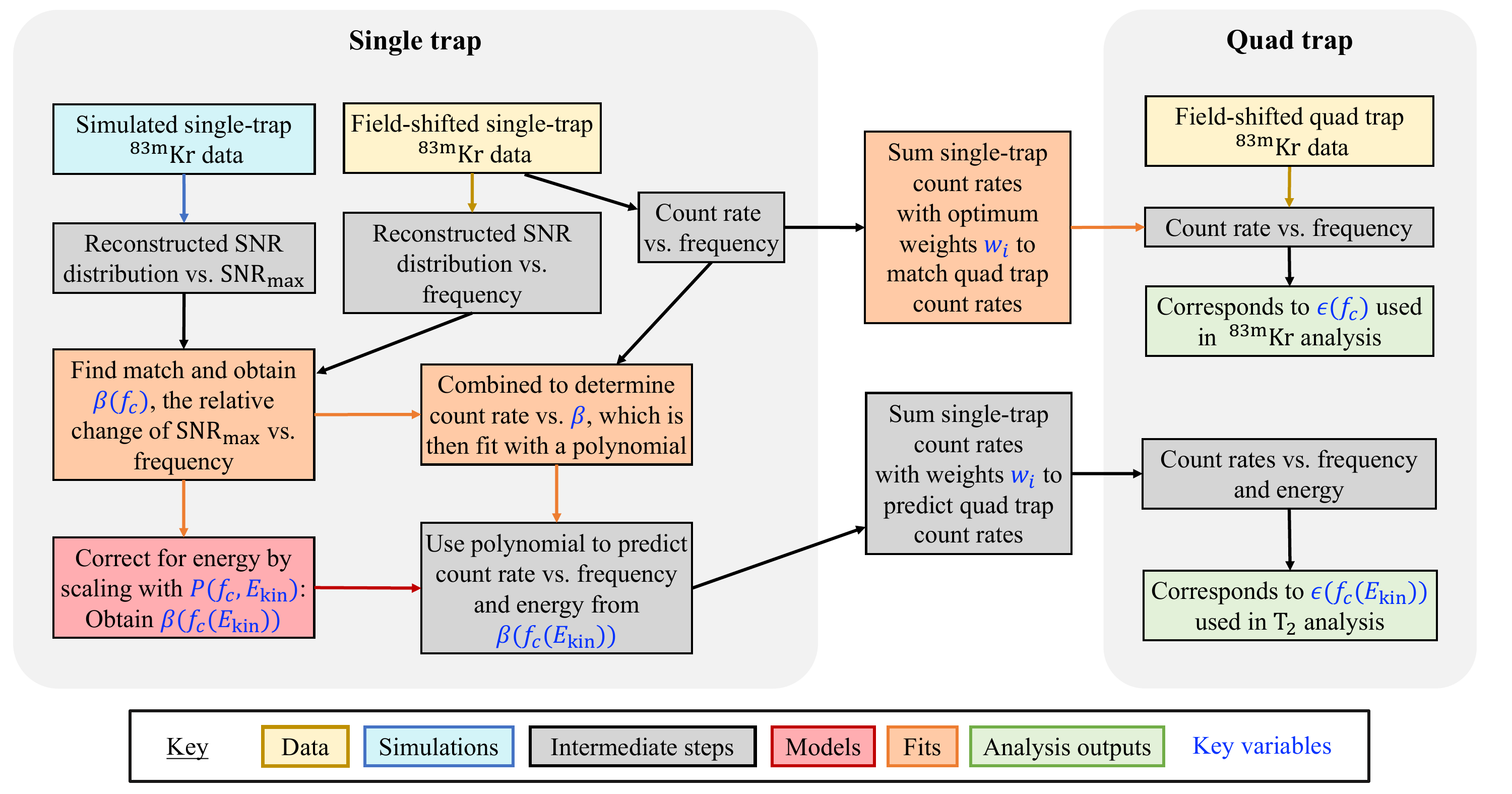}
\caption{Energy correction of the detection efficiency curve: Count rate dependencies on frequency are obtained from field-shifted $\mathrm{^{83m}Kr}$ data in each single-coil trap at fixed electron energy. Matching the data to dedicated simulation sets yields $\beta(f_c)$, the relative change of SNR with frequency. From these, the count rate dependence on SNR is obtained. $\beta(f_c)$ is corrected for energy using the analytic power dependence on energy (\autoref{eq:power_from_pheno}) to obtain $\beta(f_c(E_{\mathrm{kin}}))$. Combining this with the count rate vs.~$\beta$ yields the energy-corrected count rate vs.~frequency. The energy-corrected count rates vs.~frequency for single traps are summed to obtain the quad-trap rates which correspond to the quad-trap  efficiency for tritium: $\epsilon(f_c(E_{\mathrm{kin}}))$.}

\label{fig:efficiency_energy_correction}
\end{figure*}

The SNR$_{\rm max}$ effect also causes $p$ and $q$ to be correlated with $\sigma$, the standard deviation of the tritium $\mathcal{I}$.  We assume that, in the tritium analysis, the correlations of $p$ and $q$ with $\sigma$ match correlations observed in $^{\mathrm{83m}}$Kr fits due to the SNR$_{\rm max}$ effect. Pearson correlations of $p$ and $q$ with $\sigma$ are determined using 100 sets of ($\sigma$, $p$, $q$) values for different SNR$_{\rm max}$ values.
We find that the $p$-$\sigma$ correlation from the SNR$_{\rm max}$ effect is 1.00, but the overall $p$-$\sigma$ correlation is 0.82---accounting for the fact that the parameters' uncertainties include other, uncorrelated contributions. The $q$-$\sigma$ correlation from SNR$_{\rm max}$ is 0.60 and the overall correlation is 0.06.

\subsection{Energy-dependent efficiency \texorpdfstring{$\epsilon$}{}} 
\label{sec:efficiency_for_tritium}

The measured efficiency variation with frequency in \autoref{subsec:efficiency} requires a correction for energy dependence before it can be used for tritium analysis. 

\subsubsection{Efficiency correction for energy dependence}\label{sec:energy_correction}
Determining the detection efficiency vs.~frequency curve for tritium requires an additional step beyond the procedure 
for $\mathrm{^{83m}Kr}$ data. This is because the field-shifted $\mathrm{^{83m}Kr}$ data at different frequencies vary in $B$ but are fixed at an energy of 17.8\,keV, while electrons in the tritium spectrum experience the same $B$ but have varied kinetic energies. This leads to a difference in radiated power and therefore in detection efficiency. The power coupled to the transporting $\mathrm{TE_{11}}$ mode is a function of both frequency and energy, as derived in \cite{Esfahani:2019mpr}:
    \begin{equation}\label{eq:power_from_pheno}
    \begin{split}
        P(f_c, E_{\mathrm{kin}}) \propto  {Z_{11}(f_c) e^2 v_0^2(E_{\mathrm{kin}})}, 
    \end{split}
    \end{equation}
where $Z_{11}$ is the $\mathrm{TE}_{11}$ mode impedance and $v_0$ is the magnitude of the electron's velocity. 

The process of correcting $\epsilon(f_c)$ for the energy-dependence of the transmitted power is depicted in \autoref{fig:efficiency_energy_correction}. 
Since $B$ was constant during tritium data collection and $f_c$ and $E_{\rm{kin}}$ are linked by \autoref{eq:energytofrequency}, the goal of the correction is to obtain the efficiency's simultaneous dependence on $f_c$ and $E_{\rm{kin}}$ which we denote $\epsilon(f_c(E_{\mathrm{kin}}))$. 
The energy dependence of $\epsilon$ is added by combining information from {field-shifted $\mathrm{^{83m}Kr}$ data}, {simulation}, and \autoref{eq:power_from_pheno}.
The energy correction relies on the fact that a relative change of SNR leads to a predictable relative change of count rate. Because of the SNR differences between the individual trap locations in the cell, predicting the relation between SNR and count rates is difficult in the quad-trap but can be found by summing the single-trap count rates.
Hence the energy correction is done for single-trap data and the  statistical weights from \autoref{subsec:trap_weights} are used to sum the results.
 
\begin{figure}
\centering   
\includegraphics[width=\columnwidth]{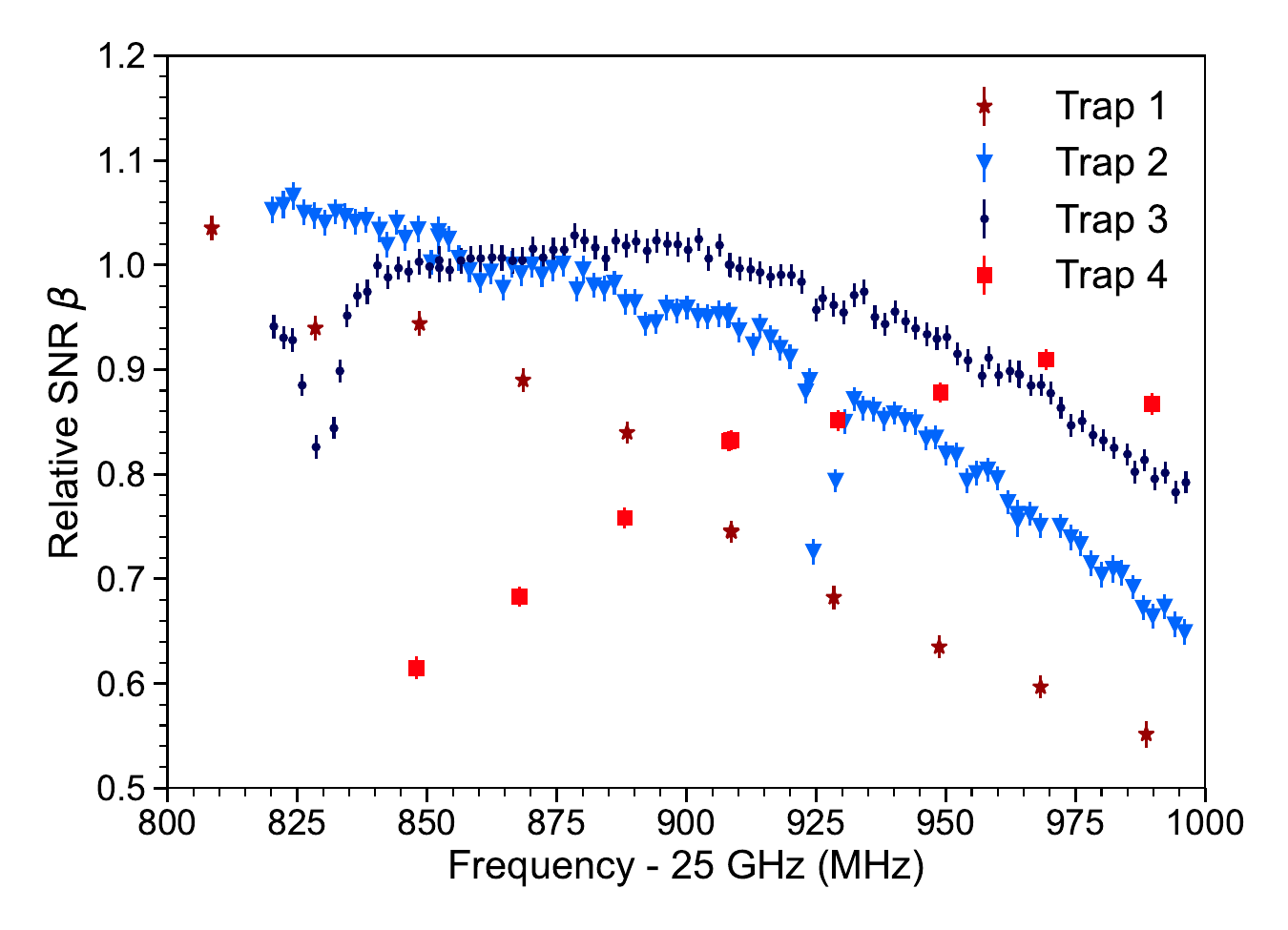}
\caption{Extracted relative SNR change $\beta(f_c)$ during the background field scan in each single-coil trap.}
\label{fig:snr_scaling_vs_frequency}
\end{figure}

\begin{figure}
\centering   
\includegraphics[width=\columnwidth]{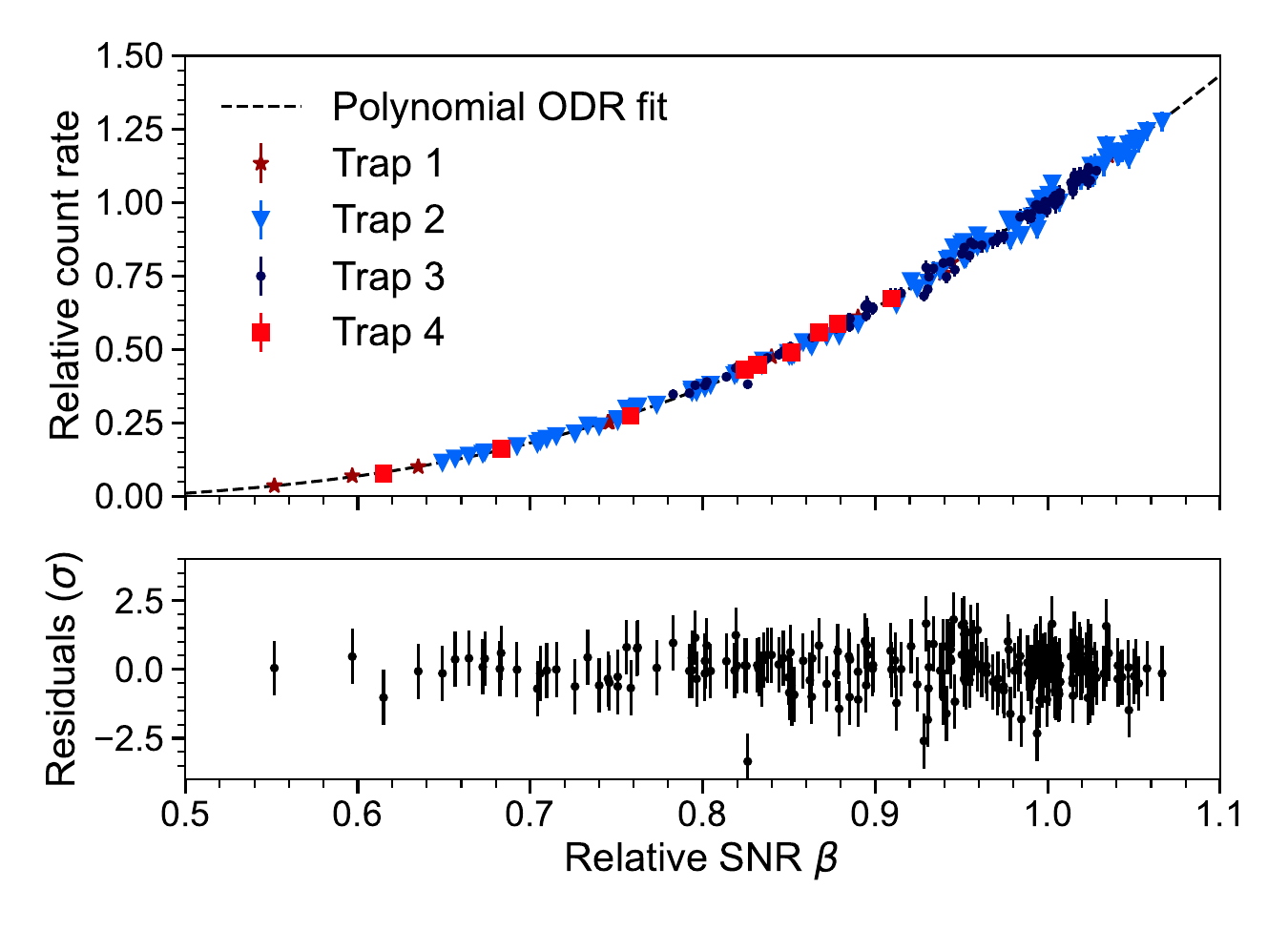}
\caption{Measured relative count rate vs.~extracted relative SNR change during background field scans. The data points originating from ${\sim} 25.828\,\si{GHz}$ and ${\sim} 25.926\,\si{GHz}$ (where slopes in trap 2 and 3 are highest) are excluded from this plot. The count rates in each trap are scaled relative to trap 3 to minimize the orthogonal distance to a common 4\textsuperscript{th}-order polynomial fit. The orthogonal distance regression (ODR) takes the x and y uncertainties into account. 
}
\label{fig:count_rate_vs_snr}
\end{figure}

\begin{figure}[htb]
\centering
\includegraphics[width=0.5\textwidth]{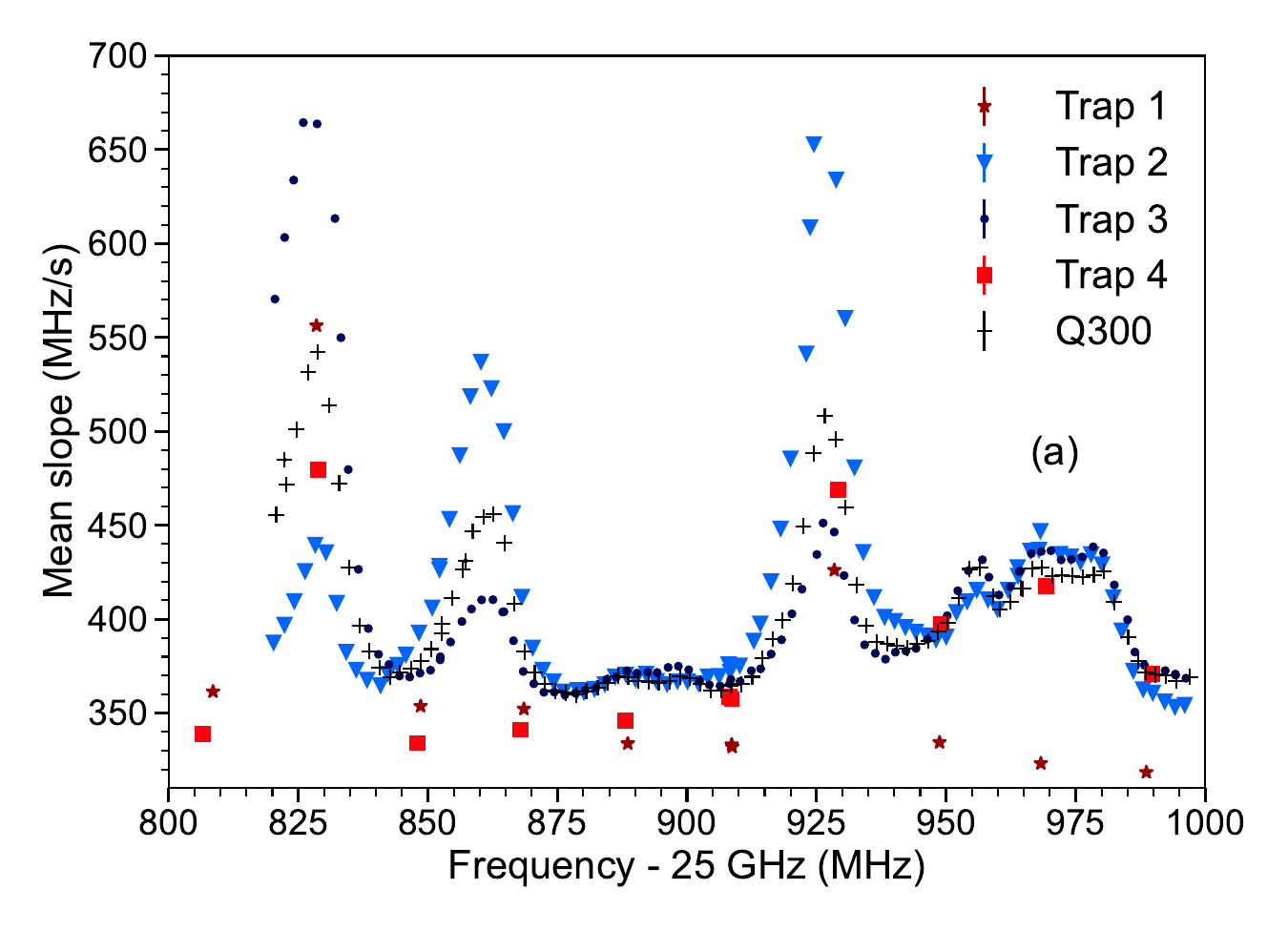}

\includegraphics[width=0.5\textwidth]{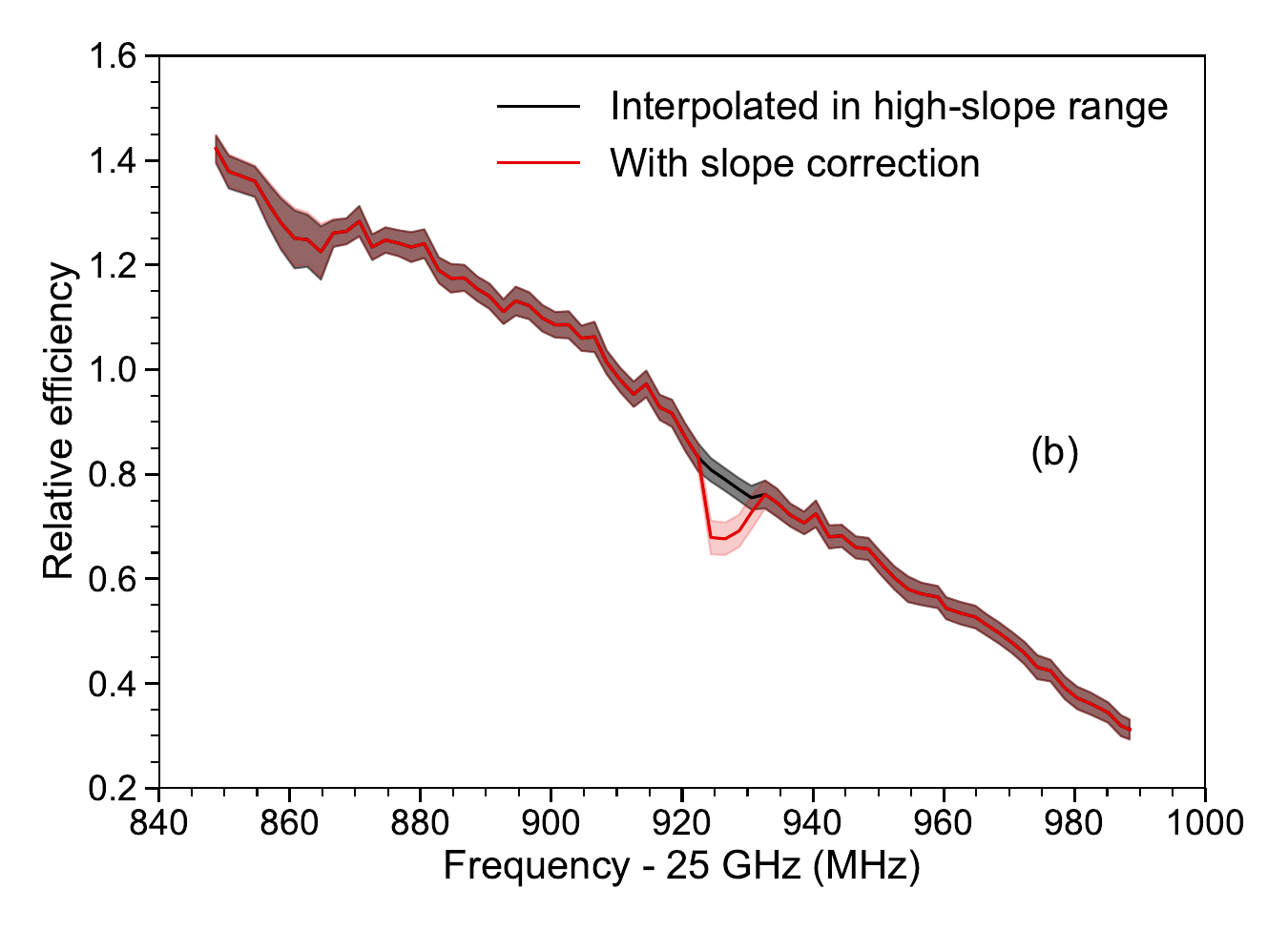}
\caption{(a) Mean recorded first track slope in events that start within $\pm 1 \,\si{MHz}$ of  the fitted $^{\mathrm{83m}}$Kr peak position. (b) Relative detection efficiency predicted for tritium $\epsilon(f_c(E_{\rm{kin}}))$. The tritium efficiency is given relative to the efficiency for $\mathrm{^{83m}Kr}$ K-line events.  As an intermediate step in the construction of $\epsilon(f_c(E_{\rm{kin}}))$, the high-slope region at 25.926\,GHz was excluded from the correction for energy dependence and the efficiency was linearly interpolated in this range (black). The correction for slope variation assumes that the relative decrease of detection efficiency is independent of the kinetic energy, which results in the red curve. The slope peak at 25.828\,GHz lies outside the calibrated efficiency range and the range covered by the tritium spectrum.}
\label{fig:efficiency_slope_correction}
\label{fig:slope_vs_frequency}
\label{fig:final_efficiency}

\end{figure}

We measure the SNR of events in each trap at each frequency in the field-shifting scans relative to the SNR in trap~3 at the un-shifted background field strength. We name this relative SNR $\beta(f_c)$.
In single-trap data, $\beta(f_c)$ (\autoref{fig:snr_scaling_vs_frequency}) is extracted by matching simulations to the SNR distribution in field-shifted $\mathrm{^{83m}Kr}$ data at each step in the field scans (see \autoref{sec:snr_analysis} for more). 
Combining $\beta(f_c)$ with the relative changes of count rate vs. frequency yields the {count rate vs.~$\beta$} (\autoref{fig:count_rate_vs_snr}). With the exception of two narrow frequency ranges at around $25.828\,\si{GHz}$ and $25.926\,\si{GHz}$, the relation between count rate and $\beta$ is bijective and identical in each trap up to a relative scaling factor resulting from trap-depth differences. At $\sim 25.828\,\si{GHz}$ and $\sim 25.926\,\si{GHz}$, the reconstructed event count rate is decreased by the presence of very high track slopes, especially in traps 2 and 3 (see \autoref{fig:slope_vs_frequency}(a)). 
We exclude these frequency ranges (they are treated separately in \autoref{sec:slope_correction}) and fit the remaining count rate vs. SNR from all traps with a 4\textsuperscript{th}-order polynomial. This fit provides a function that enables empirical prediction of a relative  change in detectable tracks from a relative SNR change.

With the relation of count rate vs. SNR in hand, we perform the energy correction for each individual trap by:
\begin{itemize}
    \item Translating relative count rates vs.~frequency in each trap to relative SNR vs.~frequency $\beta(f_c)$.
    \item Multiplying the relative SNR vs.~frequency with the relative dependence of the coupled power on energy: $P(f_c, E_{\mathrm{kin}}) / P(f_c , E_\mathrm{kin}=17.83\,\si{keV})$.
    \item Translating the energy-corrected SNR vs.~frequency $\beta(f_c(E_{\mathrm{kin}}))$ back to relative count rates vs.~frequency using the $\mathrm{4^{th}}$-order polynomial representing count rate vs.SNR.
\end{itemize}

After performing the energy correction in the single traps, we sum the corrected count rates using the trap weights $w_i$ to obtain energy-corrected quad-trap count rates. 
The resulting relative changes of the quad-trap rates with frequency directly correspond to the energy-corrected detection efficiency curve $\epsilon(f_c(E_{\mathrm{kin}}))$, which is shown in \autoref{fig:final_efficiency}(b).  

Each step of the energy correction process contributes an uncertainty to $\epsilon$. All uncertainties are propagated and amount to the width of the grey curve in \autoref{fig:final_efficiency}(b). The main contributions are the statistical uncertainties from the count rates in each constituting trap and the uncertainty from interpolating the coarse field scans in traps 1 and 4 to the dense steps recorded in traps 2, 3, and the quad trap.
Other uncertainties originate from the extraction of $\beta(f_c)$ and the polynomial fit parameter uncertainties of count rates vs.~$\beta$.

\subsubsection{Extraction of SNR dependence on frequency}
\label{sec:snr_analysis}
The relative SNR $\beta(f_c)$ is extracted from the field-shifted $\mathrm{^{83m}Kr}$ data by comparing the data to simulation. We simulate $\mathrm{^{83m}Kr}$ K-line data for a single deep trap and collect the coupled power from each simulated event. Instead of processing the simulated data with the trigger and offline event reconstruction, we apply a response matrix. This matrix maps the true SNR of an event to a distribution of likely reconstructed SNR that integrates to the detection probability for this event. The coupled power of each event is scaled to SNR before being forward folded with the response matrix. The resulting SNR distribution is then compared to the distributions recorded in the field-shifted data for all scan steps. Prior to the comparison, the simulated and recorded data are both reduced to a frequency slice around the peak center of $\pm 1\,\si{MHz}$ (the same cut was applied to the field-shifted $^{\mathrm{83m}}$Kr data). The process of scaling the power to SNR, mapping it to a detected SNR distribution, and comparing it to the recorded data is repeated in a $\chi^2$ minimization with the power-to-SNR scaling factor as the only free parameter. $\beta(f_c)$ is calculated by dividing the power-to-SNR scaling factors by the factor for $\Delta B_{\mathrm{FSS}} = 0$ in trap 3. The results are shown in \autoref{fig:snr_scaling_vs_frequency}. The similarity to \autoref{fig:fss_event_rates} (count rate vs.~frequency from data) gives rise to the direct relation between count rates and $\beta$ in \autoref{fig:count_rate_vs_snr}. 
Note that the lowest frequency point at which $\beta(f_c)$ could be extracted from fits in trap~4 is $\approx 25.85\,\si{GHz}$. At the two scan points below this frequency, the limited statistical power of this trap prevented a successful SNR analysis. As a result, the energy-corrected detection efficiency in the quad trap $\epsilon(f_c(E_{\rm{kin}}))$  is limited to $f_c \gtrsim 25.85 \,\si{GHz}$. 

\subsubsection{Efficiency correction for slope variation}\label{sec:slope_correction}
In \autoref{fig:slope_vs_frequency}(a), it can be seen that the average track slope in the field-shifted data varies strongly across the scanned frequency range. 
The high-slope regions occur at frequencies where there is enhanced coupling of the electrons to trapped resonant modes as described in \autoref{subsec:rf_system}. The largest effects are caused by the TM$_{01}$ mode, which is trapped at the upper end by the quarter-wave plate and at the lower end by the terminator (which were designed to match the mode impedance of TE$_{11}$, not TM$_{01}$).  Other, weaker resonances arise from reflections at the windows, coupling flanges, gas connection, etc.

The detection efficiency mostly depends on frequency and trap location in the waveguide. However, the event reconstruction has a small track-slope dependence, too. Over a large range of slopes ($\Delta s \lesssim 300 \,\si{MHz/s}$), this is simply due to the fact that the electron tracks with larger slopes cross more frequency bins per time and hence the power in a single time-frequency bin is reduced.  The SNR-scaling analysis of the field-shifted data incorporates this effect by returning a decreased optimum $\beta$ in frequency regions of high slopes. 

The largest slopes within the tritium analysis frequency range are found at ${\sim}25.926\,\si{GHz}$ in trap 2. Here the reconstruction algorithm sometimes misses or breaks tracks and the reconstruction efficiency is reduced. 
Therefore, we exclude this frequency region from the count rate vs.~$\beta$ correlation (\autoref{fig:count_rate_vs_snr}), and thus also from the efficiency correction for energy dependence.  Instead, we linearly interpolate $\epsilon$ for this frequency range which results in the black efficiency curve in \autoref{fig:efficiency_slope_correction}(b).
The slope peak at ${\sim}25.828\,\si{MHz}$ is below the efficiency analysis range. The smaller peak at around ${\sim}25.86\,\si{GHz}$ is close to the expected tritium endpoint location but the event reconstruction quality is not further diminished by the increased slopes in this region. 

 To re-introduce the variation in efficiency that is observed in the field-shifted data at ${\sim}25.926\,\si{GHz}$, we multiply the interpolated $\epsilon(f_c(E_{\mathrm{kin}}))$ by the relative efficiency decrease in this region. The uncertainty on $\epsilon(f_c(E_{\mathrm{kin}}))$ increases in the process, since the multiplied ratio comes with an uncertainty that is added in quadrature to the uncertainty from interpolation. The result is shown in red in \autoref{fig:efficiency_slope_correction}(b).

\subsubsection{Efficiency binning}
\label{sec:efficinecy_binning}
Since the tritium analysis is performed with binned data, we obtain an efficiency $\epsilon_k$ for each bin $k$ by integrating the quasi-continuous efficiency $\epsilon(f_c(E_{\mathrm{kin}}))$ over the bin.

The interpolation of the measured count rates in each trap makes it possible to calculate a relative efficiency value at any frequency over the calibrated range covered by the field-shifted $\mathrm{^{83m}Kr}$ data. The correction of $\mathcal{\epsilon}$ for energy dependence produces an efficiency curve that applies for any energy corresponding to a frequency at which the field-shifted $\mathrm{^{83m}Kr}$ data had sufficient statistical power to perform the energy correction. This limits the energy range over which the tritium data can be modeled to 16.2--19.0\,\si{keV}. However, for the background-only energy range above the tritium endpoint, $\epsilon(f_c(E_{\mathrm{kin}}))$ is not used because the background is due only to false tracks from noise and is thus energy-independent and unrelated to signal efficiency.

\subsection{Frequency dependence of energy response function \texorpdfstring{$\mathcal{R}_{\mathrm{PSF}}$}{}}
\label{subsec:frequency_dependent_detector_response}
In $\mathrm{^{83m}Kr}$ data analysis, the response function $\mathcal{R}_{\rm{PSF}}$ is optimized to fit the K-line at its singular frequency position.  However, we know that the SNR varies with frequency and depends on energy, so we can expect that the instrumental resolution $\mathcal{I}$ changes over the frequency and energy ROI. We also expect the shape of the scattering tail to vary with frequency, since the event reconstruction is impacted, for example, by power-coupling changes in the waveguide that manifest in the track slope changes observed in the field-shifted data.

\begin{figure}
\centering
\includegraphics[width=\linewidth]{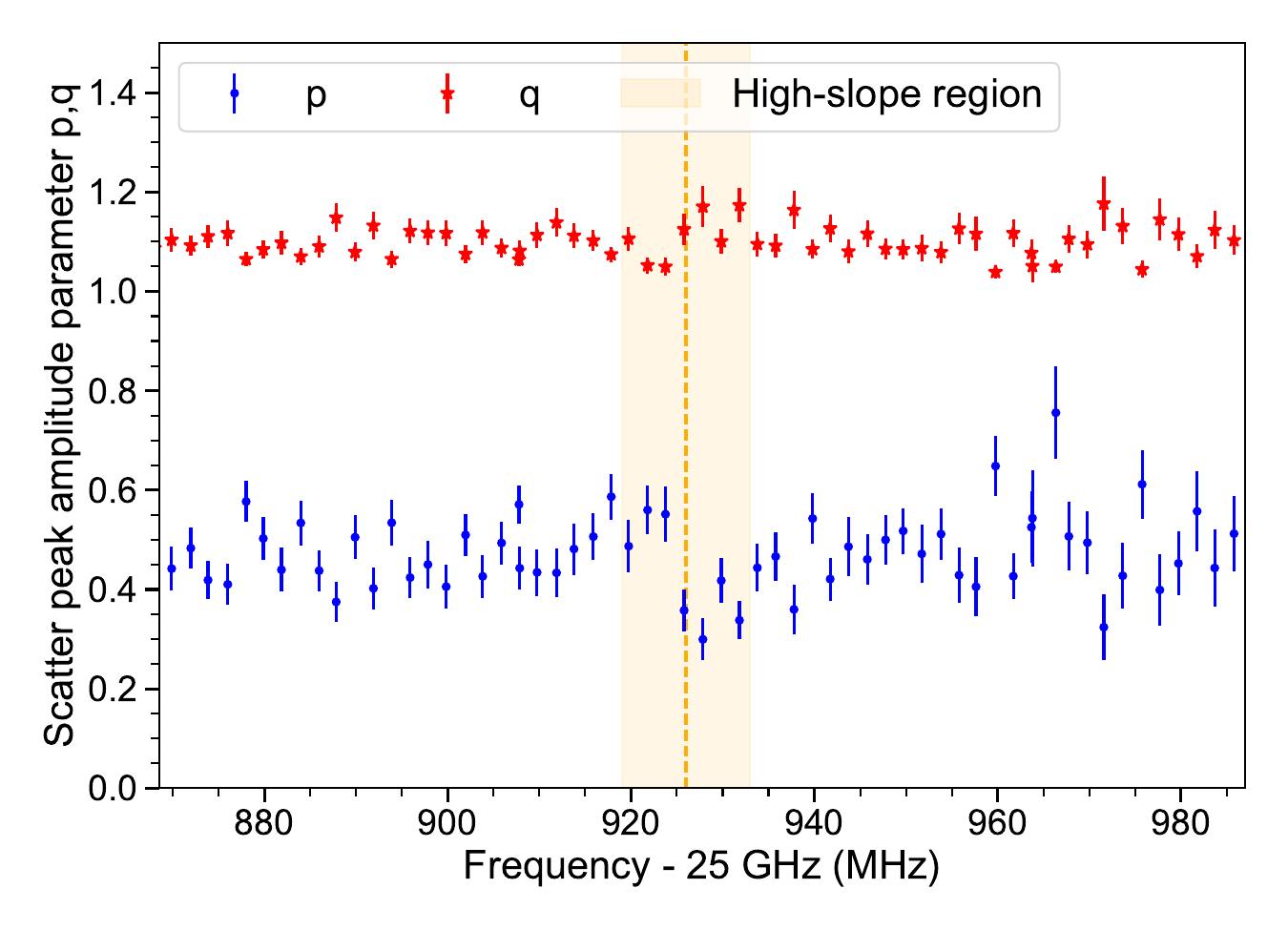}
\caption{The field-shifted quad trap data is analyzed at each magnetic field step with the $^{\mathrm{83m}}$Kr lineshape model. A separate simulated resolution shape was produced in simulations for each step and the response function parameters $p$ and $q$ were extracted from fits. The $p$ and $q$ fit results over the tritium analysis ROI are shown.
The vertical yellow band indicates the high-slope frequency range at $\sim 26.926\,\si{GHz}$.
}
\label{fig:p_and_q_frequency_variation}
\end{figure}
    
To obtain the variation in the shape of $\mathcal{R}_{\mathrm{PSF}}$, we fit the K-line in the field-shifted $\mathrm{^{83m}Kr}$ data at each magnetic field step.  For each frequency position, a simulated instrumental resolution $\mathcal{I}(f_c)$ is created by scaling the optimum $\mathrm{SNR_{max}}$ for the reference field-shifted data from $\Delta B = 0 \,\si{mT}$ by the relative SNR $\beta(f_c)$. The resulting instrumental resolution serves as input to the lineshape fits in which $p$ and $q$ are left free to obtain their variation with frequency. The result is shown in \autoref{fig:p_and_q_frequency_variation}. We verify that the relative variation with frequency is not significantly affected by the gas composition by repeating all fits for the maximum and minimum allowed helium contribution to inelastic scattering. 

The variation of $p$ with frequency is much larger than that of $q$. Monte Carlo studies of tritium data with varying $p$ and $q$ have shown that the shape of $\mathcal{R}_{\mathrm{PSF}}$ and its impact on the tritium analysis results are more sensitive to changes of $q$ than they are to changes of $p$. It is therefore unsurprising that the $^{\mathrm{83m}}$Kr fits constrain $q$ more tightly than $p$.

Note that the mean $p$ and $q$ seen in \autoref{fig:p_and_q_frequency_variation} are  different from their values under tritium conditions due to the differences in gas composition and system noise temperature (heat transfer from the field-shifting solenoid increased the noise temperature by 10\%). However, we expect the tail shape to vary similarly with frequency in the tritium data. By dividing the $p(f_c)$ and $q(f_c)$ values in \autoref{fig:p_and_q_frequency_variation} by their means, we obtain scale factors $s_p(f_c)$ and $s_q(f_c)$. For the tritium analysis, these scale factors multiply the tritium $p$ and $q$ from \autoref{sec:scatter_peak_errors}. 

The expected variation of $\mathcal{I}$ with frequency in the tritium data is obtained similarly to the calculation of $\mathcal{I}(f_c)$ described above. However, the dependence on energy of the efficiency variation $\epsilon(f_c(E_{\rm{kin}}))$ has to be accounted for. We therefore scale the optimum $\rm{SNR_{max}}$ for tritium (identical to the $\rm{SNR_{max}}$ of the pre-tritium $^{\mathrm{83m}}$Kr data) by $\beta(f_c(E_{\rm{kin}}))$.
This allows us to generate the expected $\mathcal{I}(f_c(E_{\rm{kin}}))$ at each frequency in the tritium ROI. From these distributions, we calculate width scale factors $s_\sigma(f_c(E_{\rm{kin}}))$, which multiply $\sigma$, the resolution standard deviation from \autoref{sec:sigma_errors}.

The estimate ranges of $s_p$, $s_q$, and $s_{\sigma}$ are given in \autoref{tab:freq_var_priors}.
The impact of the frequency and energy variation of $\mathcal{R}_{\mathrm{PSF}}$ on the tritium analysis is discussed in \autoref{sec:MCstudies}.

\subsection{Summary of parameter probability distributions\label{sec:summaryofpriors}}

Parameter estimates and uncertainties are summarized in \autoref{tab:priors} and \autoref{tab:freq_var_priors}. Statistical uncertainties contribute more to the endpoint interval than the systematic uncertainties combined. When propagating systematic uncertainties to the endpoint and neutrino mass, we use normal distributions to describe the uncertainties on all parameters in \autoref{tab:priors}. For the energy response ($\mathcal{R}_{\mathrm{PSF}}$) parameters $p$, $q$ and $\sigma$, a multivariate normal distribution is used to account for correlations. Physical parameter bounds (e.g., $\sigma>0$) are enforced during fits to tritium data. While normal distributions do not permit such bounds, the distributions are all localized sufficiently far from bounds, as verified by MC studies (see \autoref{sec:MCstudies}).  

\begin{table}[htbp]
\caption{Estimates with uncertainties for parameters in the model of tritium data, derived from $^{\mathrm{83m}}$Kr calibration data and  simulations.}
\centering
\renewcommand{\arraystretch}{1.15}
\begingroup
\setlength{\tabcolsep}{6pt} 
\begin{tabular}{l c }
\hline \hline
Parameter & Estimate \\ \hline
$\mathrm{H}_2$ inelastic scatter fraction $\gamma_{\mathrm{H}_2}$ & $0.91(5)$  \\
Mean magnetic field $B$ & $0.9578099(13)$\,T \\
Instrumental resolution $\sigma$ & $15(2)$\,eV \\
Scatter peak amplitudes:~$p$ & $0.89(11)$ \\
Scatter peak amplitudes:~$q$ & $1.12(5)$ \\
Number of events & 3770 events after cuts \\
\hline \hline
\label{tab:priors}
\end{tabular}
\endgroup
\end{table}

\begin{table}[htbp]
\caption{Size and uncertainty of variation of $\epsilon$, $\sigma$, $p$, and $q$ with $f_c$ in the tritium model, derived from $^{\mathrm{83m}}$Kr calibration data and  simulations. The uncertainty on $s_\sigma$ does not affect tritium analysis results and was therefore omitted. }
\centering
\renewcommand{\arraystretch}{1.15}
\begingroup
\setlength{\tabcolsep}{6pt} 
\begin{tabular}{l c c }
\hline \hline
\makecell[l]{$f_c$-dependent \\ parameter scaling}& \makecell[c]{Estimate \\ (min--max) }& \makecell[c]{Uncertainty \\ (min--max)} \\ \hline

$\epsilon$ & 0.31--1.42 & 0.02--0.05 \\
$s_\sigma$  & 0.87--1.07 & not included\\
$s_p$  & 0.59--1.53 & 0.07--0.21  \\
$s_q$  & 0.95--1.10 & 0.01--0.06  \\

\hline \hline
\label{tab:freq_var_priors}
\end{tabular}

\endgroup
\end{table}

\section{Tritium data analysis\label{sec:final-analysis}}

\subsection{Tritium analysis procedures}
\label{sec:tritium_analysis_procedures}
We perform Bayesian and frequentist analyses of the tritium data to measure the spectral endpoint $E_0$ and place a limit on the neutrino mass $m_\beta$. 
The analyses are not blind, but they are validated using MC studies.

\subsubsection{Bayesian analysis procedure}\label{sec:Bayesian-analysis}
The likelihood function used for Bayesian inference is the Poisson likelihood for the number of counts per bin according to the tritium model (\autoref{eq:fullT2model}), multiplied by prior distributions for all parameters. For bin $k$, the Poisson rate is $(dN/dE_\mathrm{kin})_k^{\mathrm{ctr}}\cdot\Delta E_{\mathrm{kin}, k}$: the tritium model at bin center multiplied by the bin width.\footnote{Test fits with piecewise integration of the probability density in each bin (10 steps per bin) produced consistent results.}  The bins have a fixed width in frequency, so in energy their widths change by 0.7\,eV between the high and low ends of the ROI.

We evaluate the likelihood function using the Stan statistical software platform, which performs Bayesian inference via the Hamiltonian Monte Carlo (HMC) algorithm~\cite{Carpenter2017,morpho}. This produces numerical posterior distributions for $E_0$ and $m_\beta$, marginalized over nuisance parameters. 
There are two reasons why it is valuable to directly fit $m_\beta$, instead of the unbounded variable $m_\beta^2$ used in frequentist analyses. First, imposing a lower bound at $m_\beta=0$ during the fit is consistent with Bayesian methodology, which relies on the principle that known information about parameters (including physical bounds) should be incorporated during inference, via priors. Second, the Heaviside theta factor in the beta spectrum depends directly on $m_\beta$ (not $m_\beta^2$), so fitting $m_\beta$ is natural and avoids the need to choose an approximate form of the spectrum for $m_\beta^2<0$~\cite{AshtariEsfahani:2021moh,Formaggio:2021nfz}.
Several Stan convergence diagnostics validate the algorithm's performance for our model~\cite{Betancourt2015, PystanWorkflow}. For $E_0$, we report the posterior mean and the 1$\sigma$ quantile credible interval~\cite{PDG:2020}. The 1$\sigma$  highest posterior density (HPD) interval is consistent within $<0.5\,$eV. For the neutrino mass, we report a 90\% credible limit~\cite{PDG:2020}. 

Uncertainties are propagated by incorporating them into prior distributions. The uncertainty distributions described in 
\autoref{sec:summaryofpriors} become normal priors in the model (or multi-normal for $\sigma$, $p$, and $q$, the resolution and scatter peak parameters). The HMC algorithm explores the probability space defined by the priors and the tritium model, propagating prior standard deviations (i.e., uncertainties) to the $E_0$ and $m_\beta$ posteriors~\cite{PDG:2020}.

Bayesian inference requires a choice of prior for every parameter, not only those with measured uncertainties. Thus, the Bayesian model includes priors on the background rate $r_f$, as well as $E_0$ and $m_\beta$. For $r_f$, before analysis, prior knowledge constrained likely background rates within several orders of magnitude. This can be modeled with a lognormal prior (which also has the correct bound $r_f>0$). The prior's median is chosen to be $0.0006$\,events/keV/day, the target 90\% C.L.  used to tune event power cuts. To conservatively scale the prior width, we use the soft upper bound $r_f^{\mathrm{max}}\approx0.2006$\,events/keV/day, the 50\% C.L. from observing no events in $\approx\!0.5$\,keV above the endpoint during an early 7-day data set taken to test the apparatus setup. The prior's standard deviation is set equal to $[0.2006-0.0006]$\,events/keV/day, determining the lognormal prior shape. 
 
Separate fits are performed to measure $E_0$ and place a limit on $m_\beta$. When measuring $E_0$, $m_\beta$ should be constrained to its known range, near zero. Otherwise, the $E_0$ posterior is biased upward due to the large $m_\beta$ probability density above the true value---with little density below, since we require $m_\beta>0$. By contrast, when placing a limit on $m_\beta$, a broad $E_0$ prior may be employed, since there is no bound in the $E_0$ prior which introduces an asymmetry.

For the $E_0$ fit, the $m_\beta$ prior is a gamma distribution constructed so that 5\% of its probability mass falls below 0.0085\,eV (from mass splitting uncertainties~\cite{PDG:2020}) and 10\% falls above 0.8\,eV (KATRIN's 90\% C.L.~\cite{KATRIN:2021uub}). A gamma prior is appropriate because of its lower bound at zero, and because it can take on a wide range of shapes depending on the external conditions provided, making it a generic prior choice for positive parameters. Other positive priors tend to effectively rule out either high values of $m_\beta$ or a region near $m_\beta=0$~\cite{Gelman2014, PriorChoiceRecs}.
We use a weakly-informative normal prior on $E_0$ with a standard deviation of 300\,eV. \footnote{We follow established conventions regarding weakly-informative and `non-informative' priors, see~\cite{Gelman2014, PriorChoiceRecs, Betancourt2017, DAgostini2003}.} MC-spectrum tests confirm that the $E_0$ result is robust to the specific choice of prior: when we analyze MC data, the $E_0$ posterior interval width remains constant as the prior standard deviation is scanned from 100 to 400\,eV, within $<1\,$eV computational uncertainty. That uncertainty is calculated by re-analyzing the same MC spectrum 25 times and computing the standard deviation of resulting interval widths.

For the $m_\beta$ fit, we use the same normal $E_0$ prior, and for the $m_\beta$ prior we use a uniform distribution multiplied by an error function with $\mu_{\mathrm{erf}}=1\,$keV and $\sigma_{\mathrm{erf}}=0.2\,$keV. The $m_\beta$ prior for this fit reflects assumptions made before data-taking: if $m_\beta\gtrsim1\,$keV, the apparatus's frequency bandwidth would be too small, as this would shift $\gtrsim2/3$ of the events out of the ROI.   On the lower edge, another error function tapers the prior at 0.0085\,eV ($\sigma_{\mathrm{erf}}=[1/3$ of that value]). MC tests demonstrate that this prior is weakly informative: when we increase/decrease the high-energy error function's parameters by $\approx50$\%, the $m_\beta$ limit is unaffected.

Frequency data are histogrammed in 71 bins, each about 50\,eV in width. We test several bin widths and observe that the $m_\beta$ limit results are slightly higher for both narrower and wider bins. For example, with 65 bins ($\approx55\,$eV in width), the Bayesian 90\% credible limit on $m_\beta$ increases by 5\,eV, relative to the 71-bin analysis. 

\subsubsection{Frequentist analysis procedure}

The frequentist analysis uses the tritium model of \autoref{eq:betaspectrumfull}.  There are small differences from the Bayesian analysis model. 
Most notably, the analysis uses $m_\mathrm{\beta} ^2$ instead of $m_\mathrm{\beta}$ as the mass fit parameter and does not limit $m_\mathrm{\beta} ^2$ to be positive. This allows the fit parameter to range on either side of zero to allow for statistical fluctuations near the endpoint. A functional form should be chosen that yields parabolic likelihood profiles.  We tried two forms for $m_\beta ^2 < 0$, the Mainz implementation~\cite{Kraus:2004zw}: 
\begin{equation}\label{eq:mainz_method}
\begin{split}
    \left[ \epsilon + 0.66 k \exp \left(-1-\frac{\epsilon}{0.66k} \right) \right] \sqrt{\epsilon^2 + k^2} \cdot \Theta \left(\epsilon+0.66k\right),
    \end{split}
\end{equation}
 and the Livermore implementation~\cite{Formaggio:2021nfz}:
\begin{equation}\label{eq:livermore_method}
 |\epsilon^2 + \frac{k^2\epsilon}{2|\epsilon|}|\  \Theta(\epsilon + k),  
\end{equation}
with  $k^2=-m_\beta ^2$.  These functions replace the phase-space factor $\epsilon\big[\epsilon^2 -  m_\beta^2\big]^{1/2} \Theta(\epsilon-m_\beta)$ of \autoref{eq:betaspectrumfull} for $m_\beta ^2 < 0$.  They bracket the desired parabolic behavior, the Mainz version likelihood decreasing too slowly below 0, and the Livermore version likelihood decreasing too quickly.  In the Feldman-Cousins construction used to find an upper limit on the mass with our data, these functions give the same result and refinements were not pursued.

The frequentist analysis uses uniform 50-eV-wide energy bins and employs extended binned maximum likelihood fits, assuming that the number of events in each bin are Poisson-distributed with an expectation value given by the  numerical integral of the tritium spectrum (\autoref{eq:fullT2model}) over each bin. For minimizing the log likelihood we use the MIGRAD algorithm from the Minuit2 library \cite{iminuit,James:1975dr}. 

All systematic parameter uncertainties listed in \autoref{tab:priors} are modeled as Gaussian. The uncertainty of the $\mathrm{H}_2$ inelastic scatter fraction $\gamma_{\mathrm{H}_2}$ is propagated by adding a normal constraint to the log-likelihood function. A multivariate normal constraint is added for $\sigma$, $p$ and $q$. Only the magnetic field and detection efficiency are not included as floating model parameters. Instead, they are propagated by MC sampling. Hard lower limits of 0 are imposed for the number of background and signal events, the resolution, and for the bin efficiencies.

The endpoint $E_0$ and $m_\beta ^2$ are measured in separate analyses. For the endpoint measurement, $m_\beta ^2$ is fixed at zero. For determining a limit on  $m_\beta ^2$, both parameters are free. 
The analysis proceeds in 3 steps. In the first step, the data is fit with a model in which $\gamma_{\mathrm{H}_2}$, $p$, $q$, and $\sigma$ are constrained nuisance parameters and the signal rate $r_s$, the false event rate $r_f$, and $E_0$ (and $m_\beta^2$) are free. The magnetic field and the detection efficiency in each bin are fixed to their calibrated best values. Uncertainty intervals on $E_0$ (and  $m_\beta ^2$) are extracted by the MINOS profile method \cite{minos}. This way, the uncertainties of $\gamma_{\mathrm{H}_2}$, $p$, $q$, and $\sigma$ are propagated to the interval widths of $E_0$ (and $m_{\beta}^2$).  
In the next step, $\gamma_{\mathrm{H}_2}$, $p$, $q$, and $\sigma$ are fixed to the best-fit values while the magnetic field and efficiency uncertainties are propagated by MC sampling. An Asimov data set \cite{Cowan:2010js} from the first-pass best-fit model is generated and then repeatedly analyzed with a model that uses a sampled magnetic field value $B_j$ and efficiency values $\epsilon_{j,i}$, for energy bin $i$. This produces a distribution of fit results for $E_0$ (and $m_\beta ^2$). The difference between the borders of the highest-density intervals and the best-fit values are added in quadrature to the profile interval widths from the first-pass analysis. In the third analysis step, the frequency variation of $p$ and $q$ is propagated (see \autoref{sec:pq_freq_var_prop}).

The Feldman-Cousins construction~\cite{Feldman:1997qc,Workman:2022ynf} is used to translate the best-fit point of actual data to a $90\%$ C.L.~upper limit on $m_\beta^2$, and consequently on $m_\beta$.
For a range of true masses, pseudo experiments are conducted and the measured $m_\beta ^2$ values are included in an interval following the likelihood-ratio ordering rule. Since systematic parameters are sampled in the generation of pseudo data, their uncertainties are included in these intervals.
We find that the ordering process is impacted by a small number of pseudo experiments with higher than usual likelihood-ratio for a given fitted mass which results in over-coverage of the obtained limits. We determined these outliers to be primarily caused by the proximity of the false event rate $r_f$ to the hard limit of 0, which is implemented in Minuit by a non-linear variable transformation. To address these outliers, we take the median of the likelihood-ratio over small ranges of fitted $m_{\rm{\beta}}^2$ from pseudo experiments for each simulated true mass and find the likelihood-ratio level above which 90\% of experiments are contained.

\subsubsection{Frequency variation of scatter peak amplitudes}\label{sec:pq_freq_var_prop}
The tritium analysis model does not include the frequency variation of $\mathcal{R}_\mathrm{PSF}$, controlled by parameters $\sigma$, $p$ and $q$. In both frequentist and Bayesian analyses, the $\sigma$ variation was found not to affect the reported results, but the $p$ and $q$ variation does have an effect. This variation is captured by the scale factors $s_p(f_c)$ and $s_q(f_c)$, as described in \autoref{subsec:frequency_dependent_detector_response}.

In the frequentist case, pseudo experiments revealed that a neglected frequency dependence of $p$ and $q$ shifts the endpoint result  by $-1.03(90)\,\si{eV}$ on average. No bias was found for $m_\beta ^2$. We therefore include an uncertainty of $E_0$ and $m_\beta ^2$ and a correction of $E_0$ resulting from ignoring this dependence in the frequentist analysis. For any given analysis (of pseudo data or real data), this is done by generating many Asimov data sets from the initial best-fit model modified with a randomized frequency-dependent scaling of $p$ and $q$ by $s_p(f_c)$ and $s_q(f_c)$ and analyzing them with the model that does not have a frequency dependence included. The average shift of the $E_0$ fit result is subtracted from the original fit result and the standard deviations of $E_0$ and $m_\beta ^2$ are added in quadrature to the uncertainties of the experiment's final results. For the $m_\beta$ limit construction with the Feldman-Cousins method, $s_p(f_c)$ and $s_q(f_c)$ are sampled within uncertainties during the pseudo-data generation and hence propagated to the distribution of $m_\beta ^2$ fit results and the confidence limit. 

In the Bayesian analysis, MC studies show that neglecting the frequency dependence of the scatter peak amplitudes shifts the endpoint posteriors by $-3.1 (12)\,\si{eV}$ on average. We correct for this shift by adding  $3.1\,\si{eV}$ to the calculated $E_0$ posterior mean. As in the frequentist case, the $s_p$ and $s_q$ values inputted to the MC data generator for each frequency region are sampled from distributions to account for uncertainty in the frequency variation. That sampling produces a fluctuation in the endpoint results, corresponding to an uncertainty of $[{+4.1},{-4.3}]\,$eV. This contribution is added in quadrature to the $1\sigma$ $E_0$ bounds from the fit to data, increasing the bounds by 0.5\,eV. For $m_\beta$, MC studies show that neglecting the frequency variation of $p$ and $q$ reduces the coverage of 90\% credible upper limits on $m_\beta$ by 3\%. To correct for this effect, MC 90\% limits on $m_\beta$ must increase by 8\,eV.  This correction is added to the $m_\beta$ limit result.

\subsection{Monte Carlo studies}\label{sec:MCstudies}

The Bayesian and frequentist analyses are both validated with Monte Carlo (MC) studies.  Random samples of tritium decay pseudo-data are generated and fitted by the analysis model (used to fit experimental data). For ensembles of MC spectra, we confirm that $E_0$ and $m_\beta$ (or $m_\beta ^2$) intervals have no under-coverage, and that $E_0$ best-fit values do not exhibit large biases.
We follow the procedure in~\cite{AshtariEsfahani:2021moh} for MC studies.

\subsubsection{Generating ensembles of MC spectra}
To generate data, electron energies are sampled from the detailed tritium model described in \autoref{sec:detailed_gen_model} and \autoref{sec:event_rate}. Events are converted to cyclotron frequencies and then binned.
Before generating each spectrum, input parameters are sampled from their uncertainty distributions. This ensures that the ensemble of spectra reflects the relative likelihoods of outcomes, validating inferences~\cite{AshtariEsfahani:2021moh, Cook2006}. The background rate is sampled from the prior distribution in \autoref{sec:Bayesian-analysis}.
The efficiency vs. frequency curve is translated to energy using the sampled $B$ before being randomized within its uncertainty. Other parameters are sampled from distributions similar to those in \autoref{tab:priors}, with minor differences to avoid making simplifications.

Instead of sampling $\sigma$ (not present in the generation model), we separately sample each contribution to the uncertainty on $\mathcal{I}$.  First, counts in each bin are sampled to account for Poisson and efficiency matrix uncertainties. Second, statistical trap weights are sampled from normal distributions before combining resolutions for the four individual traps. Third, the SNR$_{\rm max}$ uncertainty is accounted for by scaling the width of the simulated $\mathcal{I}$ by a factor $s$, with $s$ sampled from a normal distribution. 

Inelastic scatter fractions for H$_2$, $^3$He and CO are sampled from normal distributions defined by the means and uncertainties in \autoref{tab:scattering_fraction_results}. This sampling is repeated until the sum of fractions is within [0.99, 1.01], then fractions are normalized to make the sum exactly 1.
We account for a covariance between the $^3$He fraction (or equivalently, the H$_2$ fraction) and the gas-composition contribution to $p$ and $q$ uncertainties. To do this, we sample $p$ and $q$ from a bivariate normal distribution excluding gas composition uncertainty, then shift $p$ and $q$ based on the sampled $^3$He fraction, using pre-tritium $^{\mathrm{83m}}$Kr fit results to estimate the expected shift.

MC studies are performed both with and without frequency variation of  $p$, $q$, and $\sigma$ in the data generation model.
This is done by convolving a different $\mathcal{R}_{\mathrm{PSF}}$ with the beta spectrum for each frequency step in the field-shifted data, then concatenating the resulting partial-spectra. 
 At each step, the scale factors $s_p(f_c)$ and $s_q(f_c)$ are sampled from normal distributions that account for frequency-variation uncertainties, then multiplied by $p$ and $q$. Uncertainties on $s_\sigma(f_c(E_\mathrm{kin}))$ are negligible in comparison and are neglected.

\subsubsection{Coverage and bias tests}
\label{sec:mc_results}
For $E_0$ fit MC studies, ``true" $m_\beta$ values inputted to the generator are sampled from the narrow prior described in \autoref{sec:Bayesian-analysis} (defined by the KATRIN limit). For $m_\beta$ limit MC studies, $m_\beta$ values are sampled from a uniform distribution over [0, 300]\,eV, to observe how the model performs across a large range of ``true" values. For both studies, $E_0$ values are sampled from a normal distribution with $\sigma=0.07\,$eV (see \autoref{sec:Bayesian-analysis}).

\begin{table}[t]
\caption{Interval coverage (covg.) and best-fit bias from analyzing ensembles of Monte Carlo tritium data sets. The Bayesian $m_\beta$ coverage is computed for trials with MC true $m_\beta > 200$\,eV, as the limit coverage approaches $100\%$ for small masses.}
\label{tab:MCresults}
\centering
\renewcommand{\arraystretch}{1.5}
\begin{tabular}{l  c | c | c}
\hline\hline
 & $E_0 \, 1\sigma$ covg.~& $E_0$\,best-fit bias & $m_\beta$ 90\% C.L.~covg.~\\
\hline
Bayesian & $68\pm2\%$  & $-2.9\pm 0.8\,$eV & $90\pm2\%$ \\
Frequentist &  $69\pm2\%$ &  $-2.5 \pm 0.7 $\,eV& {$90 \pm 1\%$}\\
\hline\hline
\end{tabular}
\end{table}

\begin{figure}[b]
  \centering
  \includegraphics[width=1.0\columnwidth]{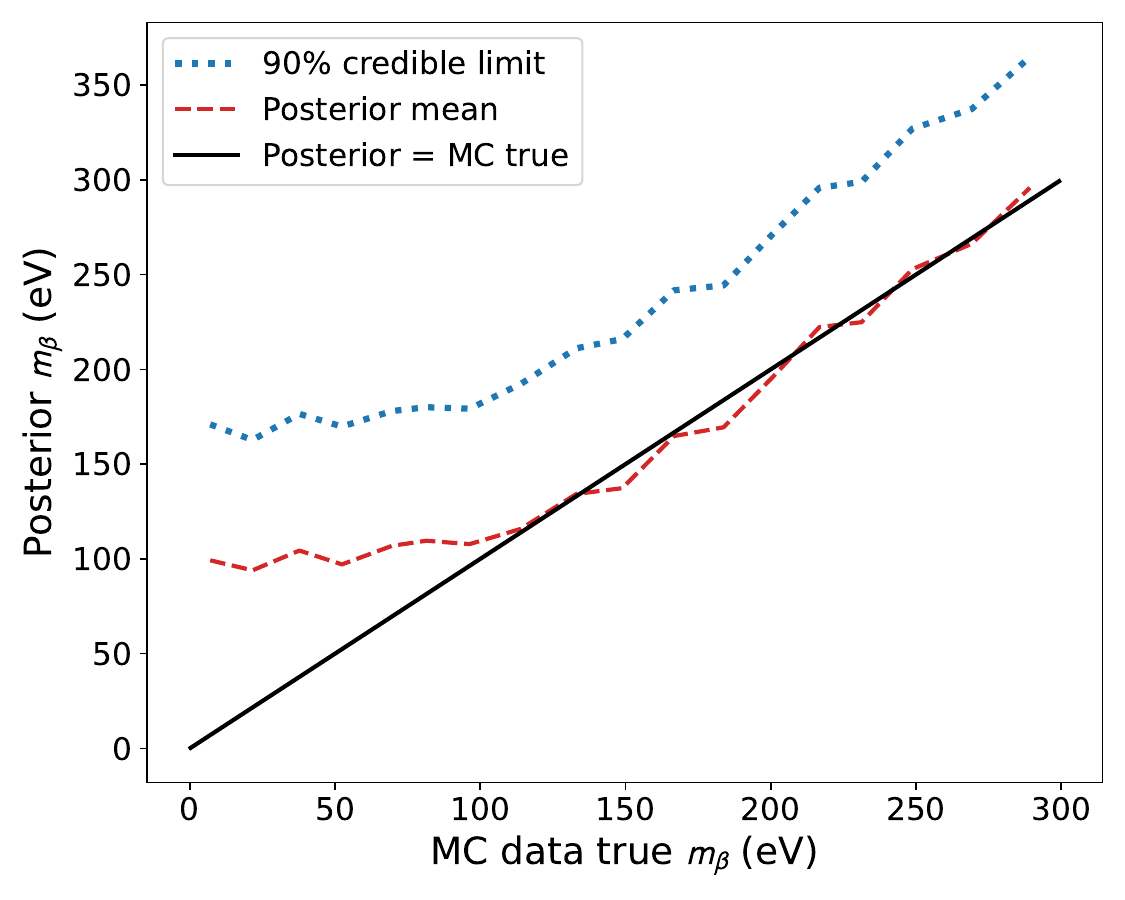}
  \caption{
    {Neutrino mass credible limits and posterior means from Bayesian Monte Carlo studies. As expected, $m_\beta$ is overestimated for low MC true $m_\beta$. Credible limits in this region are still robust. For high true $m_\beta$, posterior means approach the expectation (solid black line) and limit coverages approach the credibility (90\%). In this plot, results are averaged over $\sim\!20\,$eV regions with 33 points per region.}
  }
  \label{fig:mbeta_MC_results}
\end{figure}

MC study results are summarized in \autoref{tab:MCresults}. In the Bayesian case, the values are taken from an MC study with no frequency variation of $\mathcal{R}_{\mathrm{PSF}}$ in the data generator or analysis models. The values for the case with frequency variation were corrected to achieve the same coverage and bias, as explained in \autoref{sec:pq_freq_var_prop}. In the frequentist case, the numbers in the table are taken from an MC study with frequency variation in the data generator; for each spectrum, intervals and best-fits are corrected for this frequency variation using the Asimov sampling method described in \autoref{sec:pq_freq_var_prop}. Accordingly, for both Bayesian and frequentist analyses, \autoref{tab:MCresults} reflects the expected coverage and bias numbers for $E_0$ and $m_\beta$ results that include corrections for the effect of $\mathcal{R}_{\mathrm{PSF}}$ varying with frequency.

We observe no under-coverage of  $E_0$ 1$\sigma$ intervals. $E_0$ best-fit values have a small negative bias. This is likely caused by modeling the correlations between $p$, $q$ and $\sigma$ as linear as a function of frequency, while deviations from linearity were observed for extreme values of these parameters in field-shifted data. The $E_0$ best-fit bias disappears when we omit either the correlations between $p$, $q$ and $\sigma$ or their frequency variation, during both MC data generation and analysis.
For comparison, when different ``best-fit" statistics (e.g. median and mode) are chosen to summarize the Bayesian $E_0$ posterior distributions, the average best-fit values change by a similar amount to the biases in \autoref{tab:MCresults}. 

For the neutrino mass, the Bayesian coverage of 90\% credible limits is $93.8(7)\%$ for ``true" values of $m_\beta \in [0, 300]\,$eV. As expected, this coverage is inflated by trials with true $m_\beta \sim 0\,$eV, for which credible limits necessarily include $\approx\!\!100\%$ of true values.  This effect may be seen in 
\autoref{fig:mbeta_MC_results}, showing the credible upper limits on $m_\beta$ as a function of the true $m_\beta$. When we consider only trials with true $m_\beta > 200\,$eV, which are negligibly affected by the physical bound at 0, the 90\% limit coverage drops to $89.4(17)\%$.\footnote{In analogy with frequentist coverage, we may also consider the coverage for trials with posterior medians near the best-estimate obtained for real data (57\,eV---see \autoref{sec:tritiumresults}). For MC trials with $m_\beta$ medians within 5\,eV of the result from data, the 90\% limit coverage is $91.3(34)$\%.} For $1\sigma$ intervals on $m_\beta$ (like the one reported for tritium data in \autoref{tab:results}), the coverage is $68.4(14)$\%, including all trials in the MC study.  Moreover, as expected, the bias in the $m_\beta$ posterior mean approaches zero for high values of the MC true $m_\beta$.

The frequentist analysis is not limited to $m_\beta ^2 > 0$ and gives no over-coverage for 90\% confidence limits even for small $m_\beta^2$. Neither \autoref{eq:mainz_method} nor \autoref{eq:livermore_method} yields a parabolic $m_\mathrm{\beta} ^2$ likelihood profile, but since our best-fit value for $m_\beta ^2$ is greater than zero, the upper limit in the Feldman-Cousins construction is independent of these descriptions.  The coverage of the 90\% confidence limits is found to be $89.6(10)\%$.
These results validate the model of tritium data used here, as well as the Bayesian and frequentist analysis procedures.

\begin{table}[htbp]
\caption{Best-fit values with 1$\sigma$ uncertainties for the tritium endpoint and neutrino mass or mass squared,  and 90$\%$ credible/confidence limits on the neutrino mass. The literature endpoint value is 18574\,eV~\cite{Bodine:2015sma,Myers:2015}. The reported Bayesian $m_\beta$ result is a posterior median with 1$\sigma$ quantiles.}
\label{tab:results}
\centering
\renewcommand{\arraystretch}{1.5}
\begin{tabular}{lccl}
\hline\hline
 & Bayesian & \phantom{aa}Frequentist\phantom{aa} & Unit \\
\hline
Endpoint\phantom{aa} & $18553^{+18}_{-19}$ & $18548^{+19}_{-19}$ & eV \\ 
$m_\beta $ & $57_{-39}^{+61}$ &  &  eV \\
$m_\beta^2$ &  & $2440_{-13354}^{+10175}$  & eV$^2$ \\
90\% C.L. & $m_\beta < 155 $ & $m_\beta < 152 $ & eV \\
\hline\hline
\end{tabular}
\end{table}

\begin{table}[htb]
\caption{Endpoint uncertainty $\sigma(E_0)$ in the frequentist analysis resulting from systematic effects and statistical precision. Here, individual uncertainties are propagated by sampling each parameter from a PDF and fitting the data with a maximum likelihood fit. The total systematic uncertainty including correlations is 2\,eV smaller than adding individual systematic contributions in quadrature.} 
\centering
\renewcommand{\arraystretch}{1.1}
\begin{tabular}{p{0.40\columnwidth}>{\centering}p{0.36\columnwidth}>{\centering\arraybackslash}p{0.08\textwidth}}
\hline\hline
Effect & Parameters & $\sigma(E_0)$~(eV) \\
\hline
\multicolumn{2}{l}{Systematic w/ correlations}   & $+9, -9$ \\
Systematic quad. sum & & $+11, -11$  \\
\hspace{3mm}Mean magnetic field & $B$ & $+4, -4$ \\
\hspace{3mm}Instr.~resolution std.~& $\sigma$ & $+4, -4$ \\
\hspace{3mm}Scattering energy loss & $\gamma_{\mathrm{H}_2}$, $\mathcal{A} (p,q)$ & $+6, -6$ \\
\hspace{3mm}Bin signal efficiencies & $\epsilon_k$ & $+4, -4$ \\
\hspace{3mm}Frequency dependence & $s_\sigma, s_p, s_q$ vs.~$f_c$ & $+6, -6$ \\

Statistical &  & $+17, -17$ \\
\hline\hline
\label{tab:E0uncertainties}
\end{tabular}
\end{table}

\section{Results}\label{sec:results}

\subsection{Tritium endpoint and neutrino mass limit}\label{sec:tritiumresults}

Table~\ref{tab:results} displays the tritium analysis results for $E_0$, $m_\beta $ or $m_\beta^2$, and 90\% confidence/credible limits on $m_\beta$. Bayesian and frequentist results are similar to one another. In particular, we report a neutrino mass limit of 155 eV (152 eV) from the Bayesian (frequentist) analysis.  In addition, the Bayesian (frequentist) $E_0$ results differ by $1.2\sigma$ $(1.3\sigma)$ from the literature value. The Bayesian $1\sigma$ quantile bounds on $m_\beta$ do not contain zero because all of the posterior probability mass is above zero, since $m_\beta$ is a physical mass parameter. For $m_\beta^2$ in the frequentist analysis, negative values are permitted, and the $1\sigma$ bounds on $m_\beta^2$ are consistent with zero.

Table~\ref{tab:E0uncertainties} reports frequentist systematic and statistical contributions to the total uncertainty on $E_0$. The systematic contributions result from the uncertainties on tritium model parameters, summarized previously in \autoref{tab:priors} and \autoref{tab:freq_var_priors}. The statistical uncertainty on $E_0$ ($\pm 17\,$eV) dominates over systematics ($\pm 9\,$eV), though the latter does have a small effect on the total $E_0$ uncertainty ($\pm 19\,$eV in the frequentist case). All systematic contributions are of a similar size. 
 
\begin{figure}[htb]
  \centering
  \includegraphics[width=1.0\columnwidth]{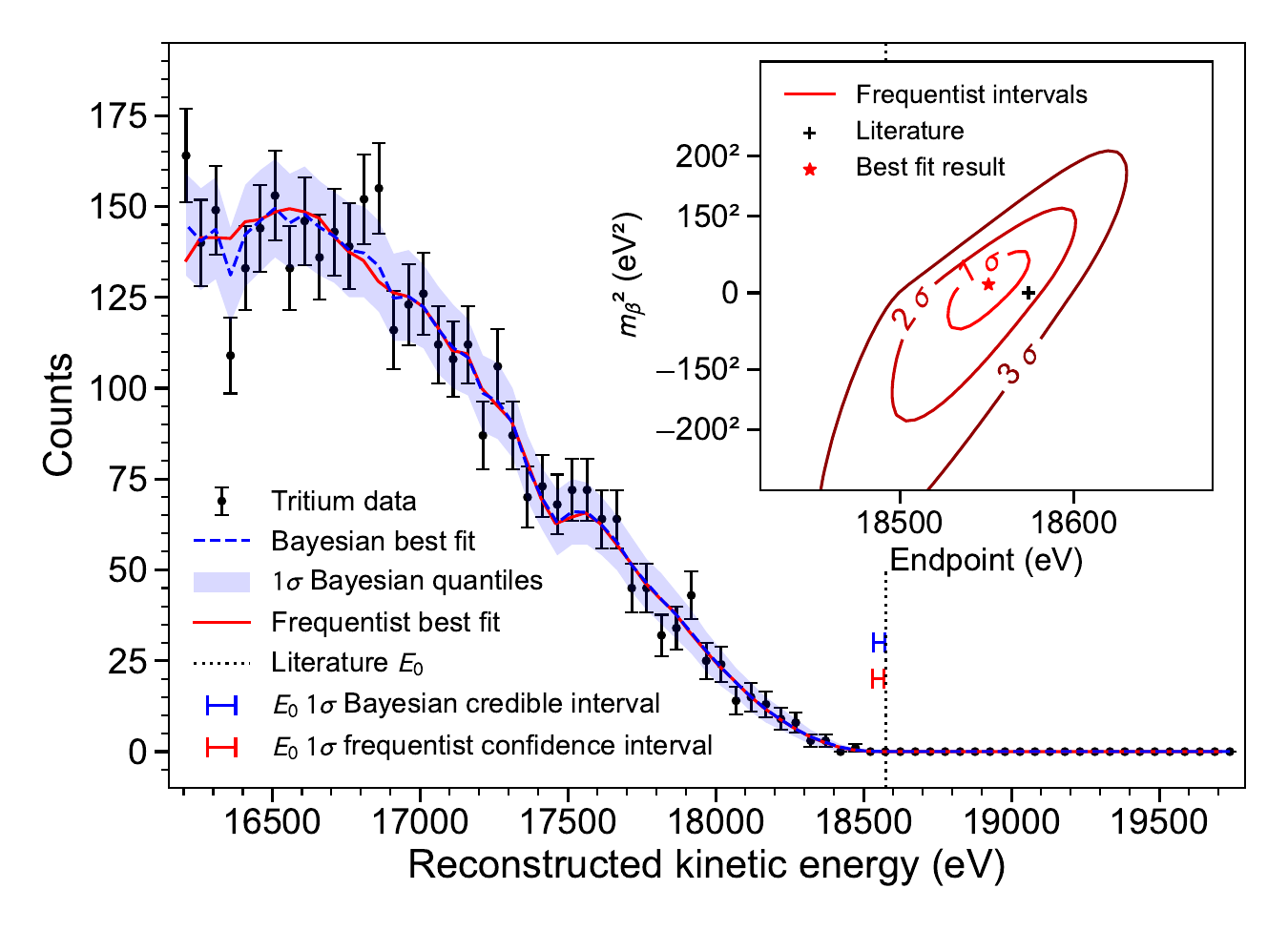}
  \caption{
    Results: Measured tritium beta-decay spectrum with Bayesian and frequentist fits, as well as endpoint intervals from each analysis. (Inset) Frequentist neutrino mass and endpoint contours. The literature sources for the endpoint energy are~\cite{Bodine:2015sma,Myers:2015} and for the  mass the source is~\cite{KATRIN:2021uub}.
  }
  \label{fig:tritium_spectrum}
\end{figure}

 In \autoref{fig:tritium_spectrum}, the measured tritium spectrum is shown with both  Bayesian and frequentist best-fit curves, which agree well.  The small remaining model differences that can be seen result from the different treatment of detection efficiency in parts of the spectrum where the efficiency is changing rapidly with respect to energy. In the frequentist analysis, the best-fit bin-efficiencies $\epsilon_k$ are equal to the values extracted from the field-shifted $\mathrm{^{83m}Kr}$ data. In the Bayesian analysis, $\epsilon_k$ are fitted from the tritium data while constrained by priors that account for uncertainties (like all other nuisance parameters). Accordingly, the Bayesian best-fit efficiency for each bin is the mean of the $\epsilon_k$ posterior. For most bins, the efficiency posteriors are similar to the priors, due to the limited statistical power of tritium data. The exception is the low-energy bins in \autoref{fig:tritium_spectrum}, where the tritium data pulls the $\epsilon_k$  distributions away from the best estimates from field-shifted  $\mathrm{^{83m}Kr}$ data.

The frequentist confidence intervals for the endpoint $E_0$ and neutrino mass squared $m_\beta^2$ are shown in the inset, with the literature values for the endpoint energy~\cite{Bodine:2015sma,Myers:2015} and neutrino mass~\cite{KATRIN:2021uub} close to the 1$\sigma$ contour. The shape of the contours for $m_\beta^2<0$ and the lower limit  given in the third row of \autoref{tab:results} depend on the choice of function for the non-physical regime, here chosen to be \autoref{eq:mainz_method}.

\subsection{Background}
\label{subsec:background_limit}

During the 82-day tritium data taking period, zero counts were observed in the spectrum in the \SI{1.2}{\keV} (23.5 bins) region of the detection window that was above the tritium endpoint energy. From this a background limit of \SI{3e-10}{\per\eV\per\second} at 90\% C.L. is obtained. If the background is energy-independent and Poisson distributed with a mean number $\mu$ of events in the 50-eV wide analysis window $\Delta E$ defined below in \autoref{eq:analysiswindow}, then the probability of there being a single background event in the window and zero events in the window above the endpoint is
\begin{eqnarray}
P&=&\frac{\mu^1 e^{-\mu}}{1!}\frac{\mu^0 e^{-21\mu}}{0!}.
\end{eqnarray}
The probability maximizes at 1.7\% for $\mu=1/22=0.045$.  The bulk of the probability is for zero events in the analysis window;  numbers $>1$ are highly improbable.

\subsection{Neutrino mass sensitivity \label{sec:sensitivity}}

Project 8 has developed and made extensive use of a simple analytic model for predicting the sensitivity  of differential spectrometers to neutrino mass.  In the model, the size of the neutrino mass $m_{\beta}$ is deduced from the count rate in a part of the spectrum of width $\Delta E$ contiguous with the endpoint, and it is assumed that the relative endpoint energy and the background are very well determined from high-statistics data in other parts of the spectrum.  The width of this ``analysis window'' is optimized with respect to the background rate $b$, energy resolution $\Delta E_{\rm res}$, and any other contributions to line broadening such as the final-state distribution in molecular T$_2$ beta decay, gas scattering, field inhomogeneity, etc.  Systematic contributions from imperfect knowledge of the resolution contributions are added in quadrature with the statistical contribution.  Reduced to the bare essentials, the optimum analysis window and the uncertainty in $m_\beta^2$ are~\cite{Formaggio:2021nfz}
\begin{eqnarray}
\Delta E &\simeq& \sqrt{\frac{b}{r}+(\Delta E_{\rm res})^2}, \label{eq:analysiswindow}\\
\sigma_{m_\beta^2} &\simeq & \frac{2}{3}\sqrt{\frac{1}{rt}\left(\Delta E + \frac{b}{r\Delta E}\right)},  \label{eq:sig} 
\end{eqnarray}
where $t$ is the run time, $\Delta E_{\rm res}$ is the FWHM of the combined resolution contributions and \begin{eqnarray}
     r&=&\frac{d^2N}{dt~d(\epsilon^3)}
 \end{eqnarray}
 is the count rate in an energy interval $\epsilon$ contiguous with the endpoint.   The analytic model assumes Gaussian statistics. 
A complete description may be found in \cite{Formaggio:2021nfz}.

The analytic sensitivity prediction model can now be confronted with data (for the first time) in Phase II. 
Table~\ref{tab:ph2sens} lists relevant parameters.  
\begin{table*}[htb]
    \centering
    \caption{Calculation of Phase II sensitivity, and comparison to analytic sensitivity.}
    \renewcommand{\arraystretch}{1.15}
    \begingroup
    \setlength{\tabcolsep}{6pt} 
    \begin{tabular}{lrrrl}
    \hline\hline
         Parameter & Value & Fractional  & Absolute  &  Unit \\
         && Uncertainty & Uncertainty & \\
    \hline
    Total effective volume & 14.6 &  & 1.2 &  mm$^3$\\
    Trigger efficiency &0.229&& 0.003 &  \\
    T\&ER efficiency &0.101&& 0.002 &  \\
    Combined trigger, T\&ER & 0.082 & 0.024 & 0.002 & \\
    Mean track duration &156 &0.07 & & $\mu$s \\
    Total cross section &$3.9\times 10^{-22}$ & 0.05 &  & m$^2$ \\
    Electron speed &0.2625c&  &  & \\
    Density as if all H, D, and T &$2.09\times 10^{17}$& 0.09  & & m$^{-3}$ \\
    Fraction of atoms that are H, D, or T &0.918 & 0.046 & & \\
    Activity fraction T/(H+D+T) &0.389 & 0.1 & & \\
    T$_2$ density &$7.45\times 10^{16}$& 0.14 & & m$^{-3}$ \\
    Background (Poissonian) &$<3\times 10^{-10}$&  &  & eV$^{-1}$ s$^{-1}$ \\
    Detectable source activity & 0.32 & & & Bq \\
    Detectable rate $r$ &$9.5\times 10^{-14}$&  & & eV$^{-3}$ s$^{-1}$ \\
    \hline
    Volume $\times$ Efficiency  & 1.20 & 0.09 & & mm$^3$  \\
    \hline
    Run time & 7185228 &0.008& & s \\
    $m_\beta^2$ ($E_0$ free,  no sys.) &$2473$& & ${}^{+9822}_{-13233}$ & eV$^2$  \\
    $\sigma(m_\beta^2)$  &$9822$ &  & $1520$ & eV$^2$  \\
    \hline\hline
    \end{tabular}
    \endgroup
    \label{tab:ph2sens}
\end{table*}
We first calculate the effective volumes $V_{\rm eff}$ from the magnetic field $B$ and the range of pitch angles that can be accommodated without exceeding a modulation index of 1, or equivalently an axial amplitude of \SI{2.8}{mm} (see \autoref{tab:quadtrapcoils}). 

\begin{figure}[htb]
    \centering
    \hfill
    \includegraphics[width=\columnwidth]{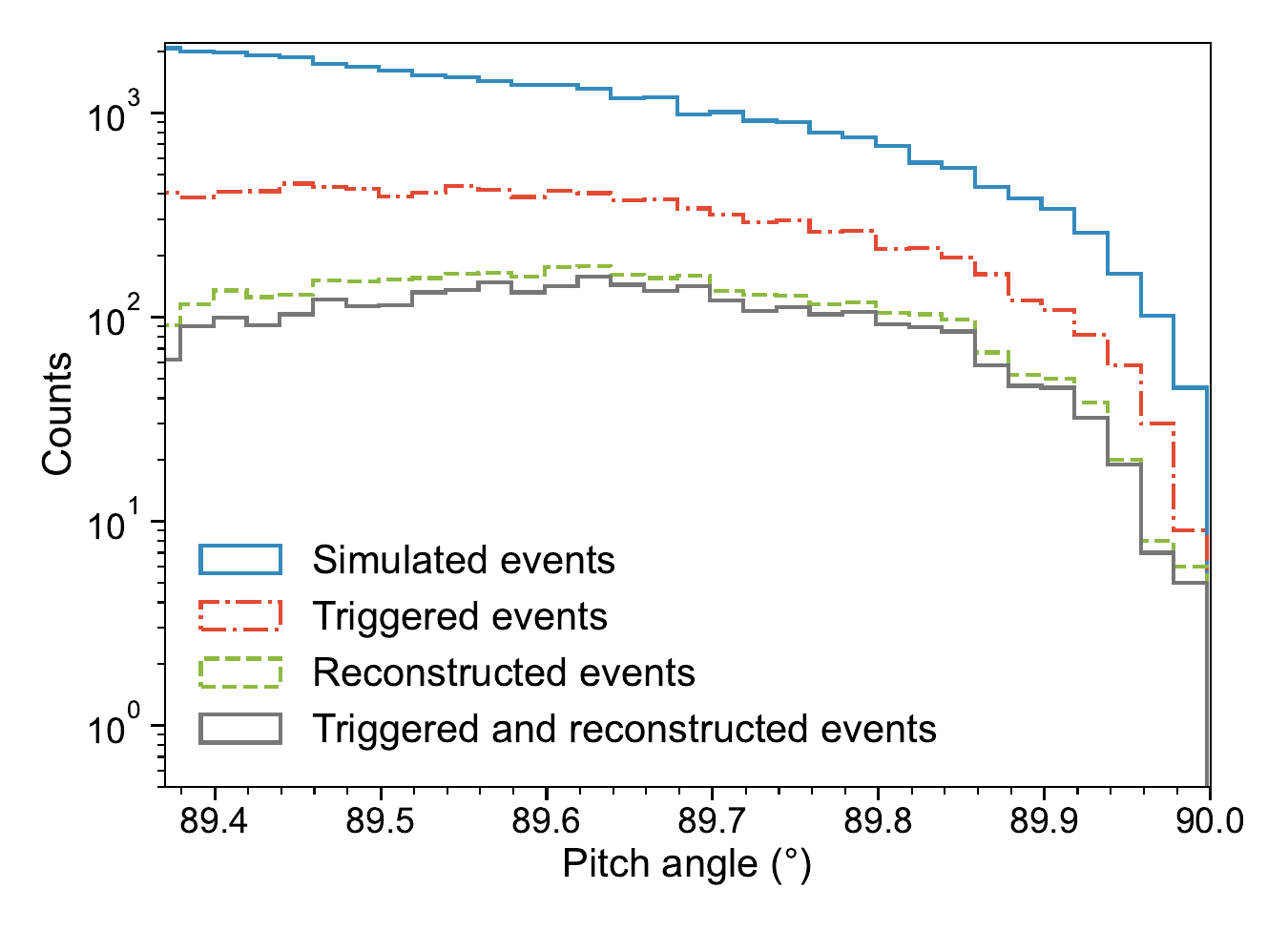}
    \caption{Number of simulated trapped electrons vs. pitch angle (blue) in the quad trap. The grey histogram shows the subset of events that were detected by the DAQ trigger (red) and the reconstruction methods (green). From the fraction of detected events the efficiency vs. pitch angle is determined. 
    }
    \label{fig:faketrackefficiencies_vs_pitchangle}
\end{figure}

Only a subset of electrons trapped within this axial range is detectable above the noise threshold, mainly because electrons at larger radii couple less strongly to the propagating TE$_{11}$ waveguide mode. We compute detection efficiencies by generating simulated tracks in all 4 traps, selecting the subset with average minimum pitch angles of 89.37 degrees (see \autoref{tab:quadtrapcoils}), and passing those events through triggering and T\&ER (track and event reconstruction) processors.  These efficiencies are correlated and the net detection efficiency is not simply the product of the two.   The results are in \autoref{fig:faketrackefficiencies_vs_pitchangle}.

The density (hydrogen equivalent) found from track length data is \SI{2.09e17}{\per\cubic\metre}, with an effective T$_2$ density of \SI{7.45e16}{\per\cubic\metre}  from mass spectrometer data.  The balance is inactive hydrogen, with a small amount of helium that is treated as hydrogen for these purposes.  The standard deviation in the neutrino mass is taken from the frequentist analysis of the Phase II data.  It is derived with the endpoint energy $E_0$ and background $b$ floating, which is the assumption made in deriving the sensitivity curves.    We use only the positive side of the 1-standard-deviation uncertainty range of the data point, because the best-fit value is in the positive,  physical regime.  Negative fit values require a functional form to be chosen for the negative regime~\cite{Formaggio:2021nfz}.

\begin{figure}[htb!]
    \centering
    \includegraphics[width=0.46\textwidth]{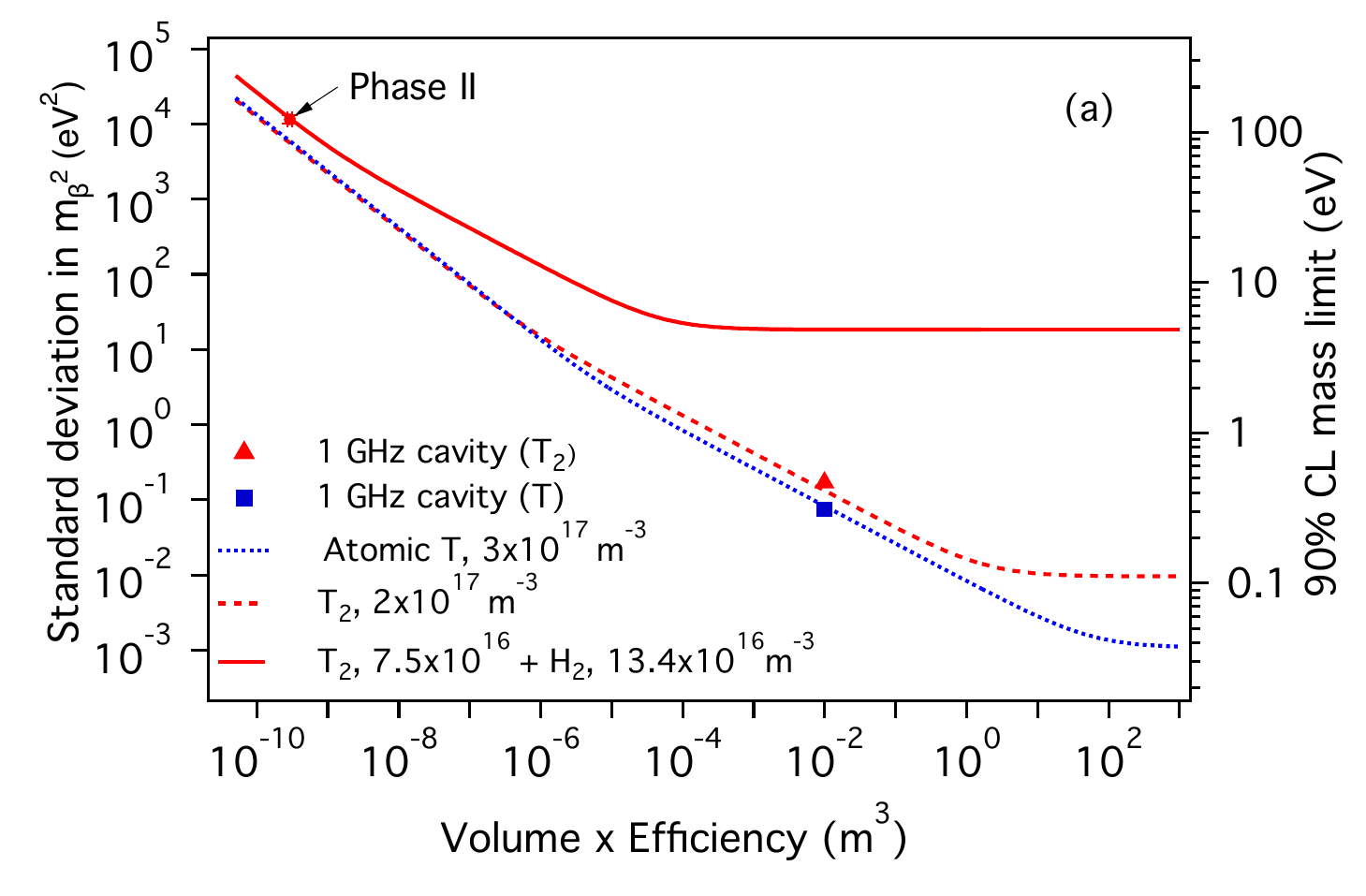}
    \vspace{-0.1in}
    \includegraphics[width=0.48\textwidth]{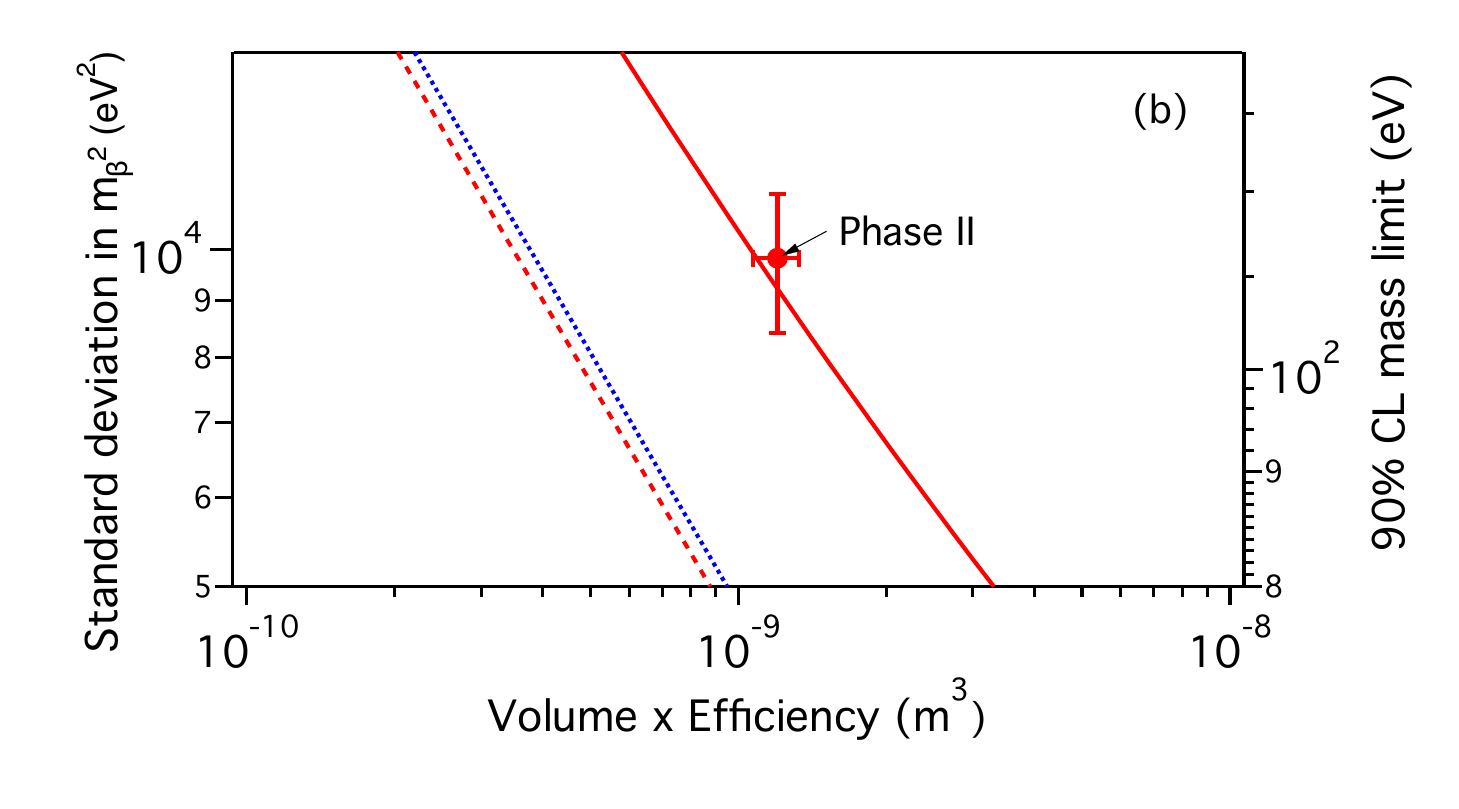}
    \vspace{-0.1in}
    \includegraphics[width=0.47\textwidth]{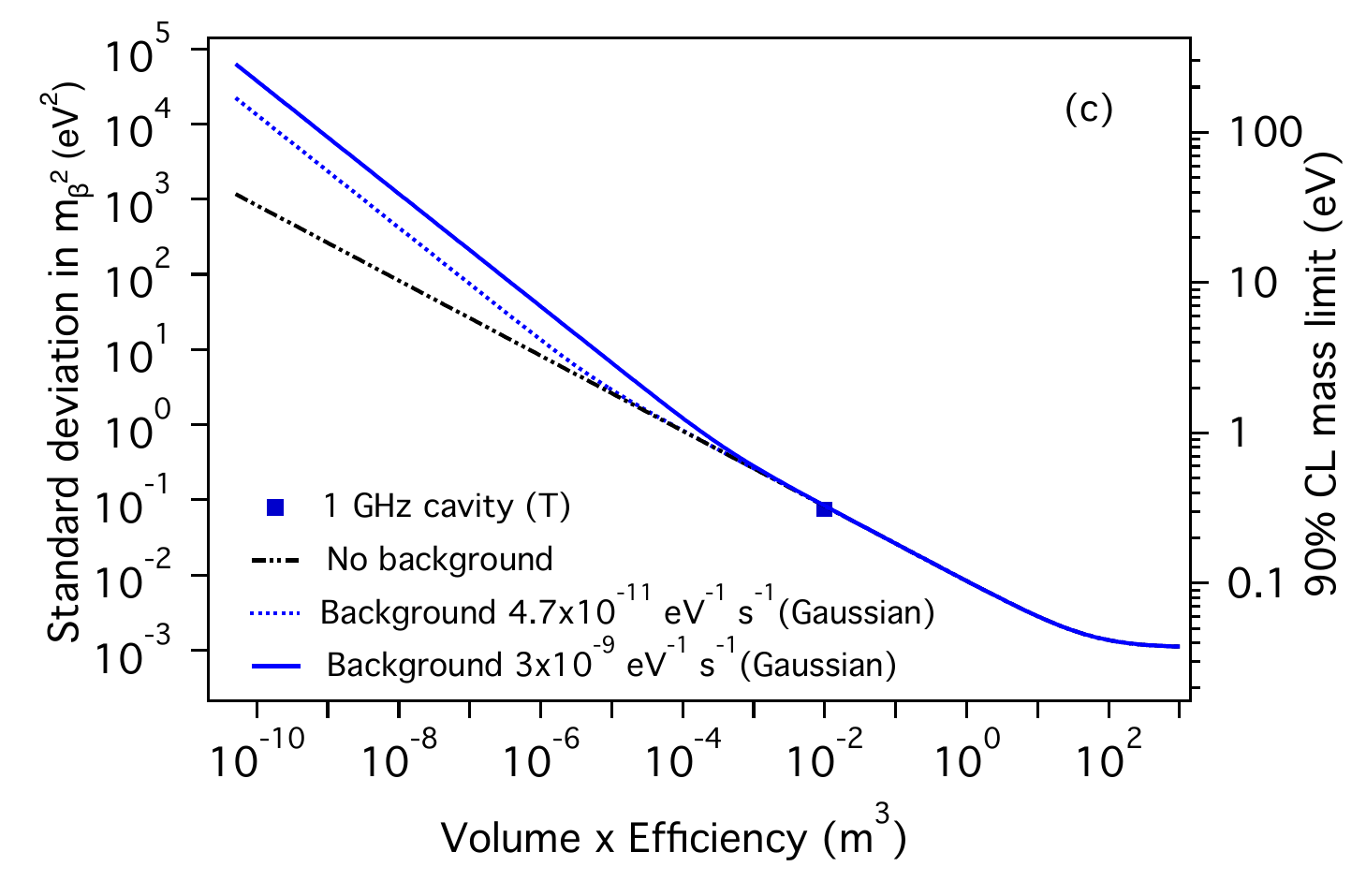}
    \caption{(a) 
    Statistical uncertainty in neutrino mass for \SI{3e7}{\second} runs for the Phase II gas mixture, a pure molecular tritium source, and an atomic tritium source. The Phase II data point is scaled to this run time, as described in the text. (b) Expanded view of the region with the Phase II data point for the actual Phase II run time of \SI{7.2e6}{\second}, with all curves redone for this run time. (c) The atomic T calculation (run time \SI{3e7}{\second}) with different background levels. 
    }
    \label{fig:ph2sensitivitydata}
\end{figure}

In  \autoref{fig:ph2sensitivitydata}(a), the statistical uncertainty in neutrino mass from Phase II data is compared to the predicted sensitivity for a T$_2$ density of \num{7.5e16}  molecules m$^{-3}$, resolution of 50 eV FWHM,  and a background limit of \SI{3e-10}{\per\eV\per\second}  (90\% C.L.).  A mean Gaussian rate of $b=4.$\SI{7e-11}{\per\eV\per\second} for the background reproduces the probability in the analysis window for the sensitivity calculation.  
    The theoretical curves show the evolution with volume and efficiency for experiments run for \SI{3e7}{\second}: with the Phase II gas mixture, resolution and background (red solid line);  with improved resolution (\SI{1.5}{\eV}  FWHM), and pure T$_2$ (red dashed line); and with atomic T  for \SI{0.5}{\eV} FWHM resolution and a density that optimizes the experimental reach in neutrino mass (blue dotted line). Systematic uncertainties have not been fully quantified, however.
  Figure \ref{fig:ph2sensitivitydata}(b) shows an expanded view of the region with the Phase II data point and the actual run time of \SI{7.2e6}{\second}. Uncertainty in the gas density is included with the data point rather than by broadening the theoretical curve.

In regimes where $\Delta E$ is dominated by $\Delta E_{\rm res}$, the curves fall with a logarithmic slope of $-1/2$.  Where $\Delta E$ is dominated by $b/r$, the slope steepens to $-3/4$.  The curves flatten out when systematic uncertainties become important, in this case taken to be a \SI{1}{\percent} uncertainty in the resolution.  The curves shown in Figs.~\ref{fig:ph2sensitivitydata}(a) and ~\ref{fig:ph2sensitivitydata}(c) are calculated for a run time of \SI{3 e7}{\second}  and the effective volume for the Phase II data point is scaled to the actual run time.  Rescaling in that way is only accurate for an experiment for which the sensitivity is dominated by $\Delta E_{\rm res}$ rather than background and signal rate, because the signal scales with volume and time while the background scales only with time. The expanded view in panel (b) is therefore calculated with the actual running time of 82 days. The sensitivity determined for Phase II agrees well with the sensitivity curve.

The role of background is illustrated in Fig.~\ref{fig:ph2sensitivitydata}(c). Project 8 plans to scale up in future experiments by moving to a resonant-cavity-based detector geometry, with one or several large cavities each producing the equivalent of a single spectrogram. It is a design criterion that the SNR remain high enough to hold false-event background to the necessary level, a criterion to be met through choice of frequency, unloaded cavity Q, temperature, amplifier technology, and bandwidth~\cite{Project8:2022wqh}. As a result, the background rate can be maintained at a near-constant level as cavity volume increases, in contrast to the intuitively expected increasing background rate with volume. The background scales with the number of cavity apparatuses. For the future experiments envisioned in Fig. 31(a), the number of apparatuses is assumed to be 1 (as in the present Phase II experiment). On the other hand, the three curves in Fig. 31(c) assume zero background, the Gaussian background estimated for Phase II, and a higher background level corresponding to 10 apparatuses with similar SNR conditions to Phase II if its background is at the \SI{90}{\percent} C.L. limit. For large experiments, the effect of background on the neutrino mass limit remains negligible.

\section{Conclusions and outlook}\label{sec:outlook}

A major motivation in Phase II of Project 8 was to obtain information of value for the future scaling up of the CRES technology for neutrino mass measurement.  With the waveguide-based CRES apparatus described in this work, we have carried out the first measurement of the tritium beta spectrum by this new method. We report an upper limit on neutrino mass at 90\% C.L.~of 155 (152) eV in a Bayesian (frequentist) analysis.  In both analyses, the extrapolated endpoint energy for molecular T$_2$ decay and the neutrino mass limit are found to be consistent with literature values obtained by traditional methods.  No background events were observed in the 82-day running period.  These results from a small-scale apparatus are the first steps in a phased approach to a CRES-based experiment with high sensitivity to neutrino mass. 

The waveguide CRES cell is quite efficient in coupling the electron to the propagating electromagnetic field.  However, because the electron's guiding center in the trap moves along the direction of propagation, the Doppler effect is maximal.  This limits the pitch angle range that can be used, and hence the efficiency with which useful signals can be obtained from the gas in the trap region.  Two approaches to  circumventing this limitation in a large volume are to use an antenna array that is directed at signals emitted by the electron perpendicular to its axial motion, and to use a resonant cavity wherein the phase velocity is much larger and the Doppler shift correspondingly smaller.  For reduced complexity and better signal-to-noise ratio, the Project 8 collaboration is focused on the cavity approach.  The sensitivities of large-volume cavity experiments are marked in \autoref{fig:ph2sensitivitydata}. 

The CRES method is inherently capable of high resolution, as may be seen simply from the widths of tracks in \autoref{fig:spectrogram}.  These tracks are roughly a single bin wide in frequency space, about 1 part per million, which corresponds to an energy width at the tritium endpoint of \SI{0.5}{\eV}.  That measure of the intrinsic resolution depends on the signal-to-noise ratio, and the system noise temperature in Phase II was about \SI{132}{\kelvin} when the CRES cell was at \SI{85}{\kelvin}.  Substantially lower physical temperatures can be realized, especially for atomic tritium in a magnetic trap.  The total experimental resolution has other contributions, primarily the varying magnetic field experienced by electrons moving in a trap that may itself not be ideal, with spatial inhomogeneities and temporal instabilities.  
Scattering of electrons from gas molecules can lead to energy loss prior to the detection of an event. 

The analysis is complex, a consequence of the experiment's exploratory nature---and in particular, of certain unanticipated aspects of design and analysis. For example, the gas composition stability and measurement was more important than expected, given the large role of scattering in the data.
In future experiments, scattering effects will be less significant and better controlled, due to higher energy resolution, purer source gas, and improved composition monitoring. In addition, in Phase~IV (Project 8's final planned phase), the energy loss from scattering ($\gtrsim 12\,\si{eV}$) will  fall outside the likely energy region of interest.
Another unexpected complication was the extent to which the efficiency as a function of frequency and energy was modulated by parasitic reflections in the waveguide, demanding a detailed analysis methodology. In the future, the energy region will be smaller and the mode structure will be a key part of the experimental design, avoiding large efficiency variations and further simplifying the analysis.

The zero background observed is both encouraging and expected. In the CRES method there are no physical detectors for electrons (for example, silicon detectors) that
can be background sources. Electrons or gammas from external radioactivity and cosmic rays do not produce trapped electrons unless an interaction with a gas atom occurs, because electrons produced at the wall can make at most one cyclotron orbit before striking the wall again, while those produced at the ends can make at most a single axial cycle before leaving. Cosmic ray interactions with the dilute source gas itself are negligible even for very large-scale experiments.   In addition, in a differential spectrometer, an electron must be in the right energy range to create a background event.  Electrons in a CRES system are not slowed nearly to rest and reaccelerated as in a retarding-field analyzer, and the counting of ubiquitous slow electrons is thereby avoided.  The primary background that remains is false tracks from random aggregations of noisy pixels.   This background was the target of a detailed study before data-taking began in Phase II, so that a power threshold could be set to limit false events to less than 1 in 100 days at \SI{90}{\percent} C.L.  The success of that strategy is one of the most important conclusions of this study. For a fixed signal-to-noise ratio, the background rate is independent of volume.  A good signal-to-noise ratio is key to eliminating the RF background at minimal cost in efficiency, and in future CRES designs it is a basic requirement. 
\vspace{0.2in}

\section*{Acknowledgments \label{sec:ack}}

The authors wish to thank two anonymous referees for their valuable suggestions reflected in the final version of this paper.  This material is based upon work supported by the following sources: the U.S. Department of Energy Office of Science, Office of Nuclear Physics, under Award No.~DE-SC0020433 to Case Western Reserve University (CWRU), under Award No.~DE-SC0011091 to the Massachusetts Institute of Technology (MIT), under Field Work Proposal Number 73006 at the Pacific Northwest National Laboratory (PNNL), a multiprogram national laboratory operated by Battelle for the U.S. Department of Energy under Contract No.~DE-AC05-76RL01830, under Early Career Award No.~DE-SC0019088 to Pennsylvania State University, under Award No.~DE-FG02-97ER41020 to the University of Washington, and under Award No.~DE-SC0012654 to Yale University; the National Science Foundation under Award No.~PHY-2209530 to Indiana University, and under Award No.~PHY-2110569 to MIT; the Cluster of Excellence “Precision Physics, Fundamental Interactions, and Structure of Matter” (PRISMA+ EXC 2118/1) funded by the German Research Foundation (DFG) within the German Excellence Strategy (Project ID 39083149); the Karlsruhe Institute of Technology (KIT) Center Elementary Particle and Astroparticle Physics (KCETA); Laboratory Directed Research and Development (LDRD) 18-ERD-028 and 20-LW-056 at Lawrence Livermore National Laboratory (LLNL), prepared by LLNL under Contract DE-AC52-07NA27344, LLNL-JRNL-845409; the LDRD Program at PNNL; Indiana University; and Yale University.  Portions of the research were performed using the Core Facility for Advanced Research Computing at CWRU, the Engaging cluster at the MGHPCC facility, Research Computing at PNNL, and the HPC cluster at the Yale Center for Research Computing.  The $^{83}$Rb/$^{83{\rm m}}$Kr isotope used in this research was supplied by the United States Department of Energy Office of Science through the Isotope Program in the Office of Nuclear Physics.

\bibliographystyle{apsrev4-2}
\bibliography{phase2-prc}

\appendix

\begin{figure*}[tb]
\centering
\includegraphics[width=1\textwidth]{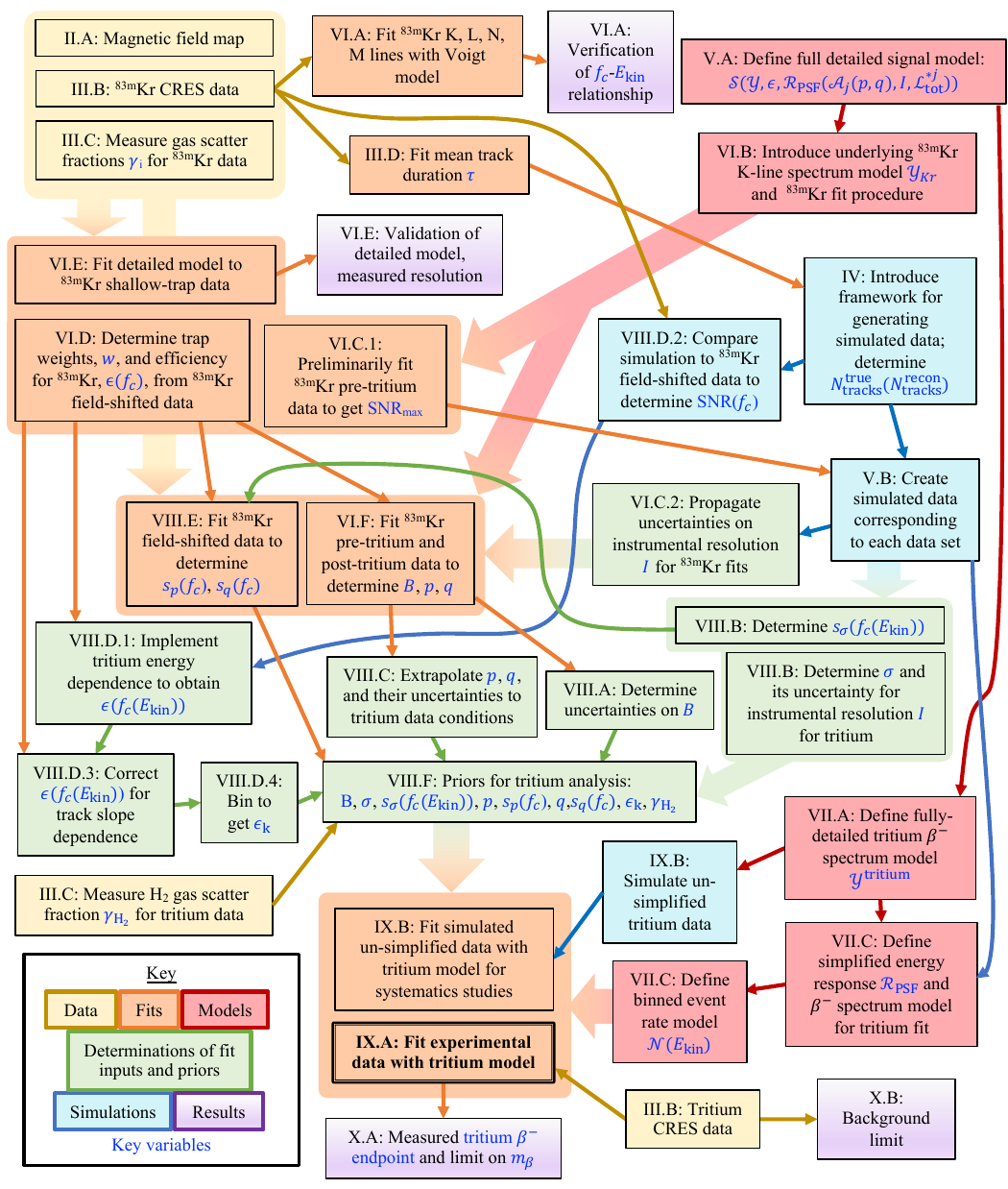}
\caption{Flow chart of the analysis procedure. Thin arrows represent relationships between individual boxes, while thick arrows describe shared inputs from and/or outputs to a group of boxes in the same bubble. }
\label{fig:flowchart}
\end{figure*}

\section{The effect of track duration distribution}
\label{appendix:track_distribution_response_function_etc}
The dependence of efficiency and response function on the track duration distribution comes about as follows. In the track reconstruction process, track duration is a powerful tool for discriminating between noise and signal. The offline event reconstruction \cite{TERpaper:2022} removes events of which the first track's signal-to-noise ratio is below certain thresholds. These thresholds are specific to the combination of first track duration and number of tracks in the event. Hence, the gas pressure in the CRES cell impacts the overall event detection efficiency as well as the first track detection efficiency, thereby affecting the electron energy point-spread function $\mathcal{R}_{\mathrm{PSF}}$ (see \autoref{sec:Kr_detector_response}). Track duration also impacts the reconstructed mean number of tracks per event, which in turn affects the shape of the tail from missed tracks in $\mathcal{R}_{\mathrm{PSF}}$, as discussed in \autoref{sec:scatter_peak_errors}.

\section{Illustration of full analysis procedure}
\label{sec:flowchart}

\autoref{fig:flowchart} shows the analysis flow in greater detail than \autoref{fig:simple_flowchart}. It also displays interdependencies of $^{\mathrm{83m}}$Kr and tritium analyses.

\section{Validating \texorpdfstring{$^{\mathrm{83m}}$Kr}{} scatter peak amplitudes with simulation}\label{appendix:scatter_peak_amplitude_simulation}

In the energy response function model, the amplitude $\mathcal{A}_j(p, q)$ of scatter peak $j$ is proportional to the probability of missing $j$ tracks before detecting an event. Simulations validate the model for $\mathcal{A}_j$ (\autoref{eq:scatter_peak_amplitude}) as well as the fitted values of $p$ and $q$ for quad trap $^{\mathrm{83m}}$Kr data.

We perform a set of toy model simulations (not using Locust) in which inelastic and elastic scattering are modeled separately, with pitch angle changes included. The event detection process is approximated by power and track length cuts. In the Locust simulations in this paper, pitch angle changes are ignored and assumed to be zero. When one is concerned with the properties of first tracks (e.g.,~the start frequency spread captured by $\mathcal{I}$), that approach is sufficient. However, this simplification does not allow the prediction of accurate $\mathcal{A}_j$ values.

For the simulations, it is assumed that inelastic scattering leads to energy loss and small pitch angle changes, while elastic scattering removes electrons from the trap before the next inelastic scattering event~\cite{DavidJoy:ElectronScattering}. 
The inelastic scattering angle $\theta_s$ follows the distribution in \autoref{eq:inelastic scatter angle}~\cite{Rudd:1991differential}.
The elastic scattering is modeled by assuming a fixed fraction $\kappa$ of electrons leave the trap between inelastic scatters due to elastic scatters. The track duration  follows the exponential distribution in \autoref{eqn:track_length_distribution}. The coupled power from a radiating electron is calculated given its instantaneous pitch angle and axial position.

The detection status of an electron is determined by whether the electron power and track length are both above the corresponding preset thresholds. The SNR threshold chosen matches the threshold applied to tracks in real data by the Phase~II track detection algorithm, prior to combining single tracks into full events (see \cite{TERpaper:2022}). Similarly to in \autoref{sec:sim-events}, it is assumed that the highest simulated power corresponds to the estimated maximum SNR observed in data. The power of all simulated events is translated to SNR accordingly. The track length detection threshold  is set to the minimum recorded track length in data. Tuning $\alpha$ and $\kappa$,  the $\mathcal{A}_j$ curve and the mean event length can be simultaneously matched to values extracted from $\mathrm{^{83m}Kr}$ data (see \autoref{fig:scatter-peak_amplitude_simulation}).

\begin{table}[ht]
\caption{Values of the parameters $\alpha$ and $\kappa$ found by tuning the scatter amplitude ($\mathcal{A}_j$) curves from simulations to match those from fits to $^{\mathrm{83m}}$Kr data.}
    \label{tab:scatter_peak_amplitude_alpha_and_kappa}
    \begin{center}
\begin{tabular}{c|c|c|c}
\hline
Data set & \(\alpha\) & \(\kappa\) & Mean of sampled scattering angles\\
\hline
\hline
Pre-tritium & 0.0018 & 0.185 & 0.48$^\circ$ \\
Post-tritium & 0.0025 & 0.150 & 0.64$^\circ$ \\
Shallow trap & 0.0025 & 0.400 & 0.64$^\circ$ \\
\hline
\end{tabular}
\end{center}
\end{table}

\begin{figure}[ht]
  \includegraphics[width=1.0\columnwidth]{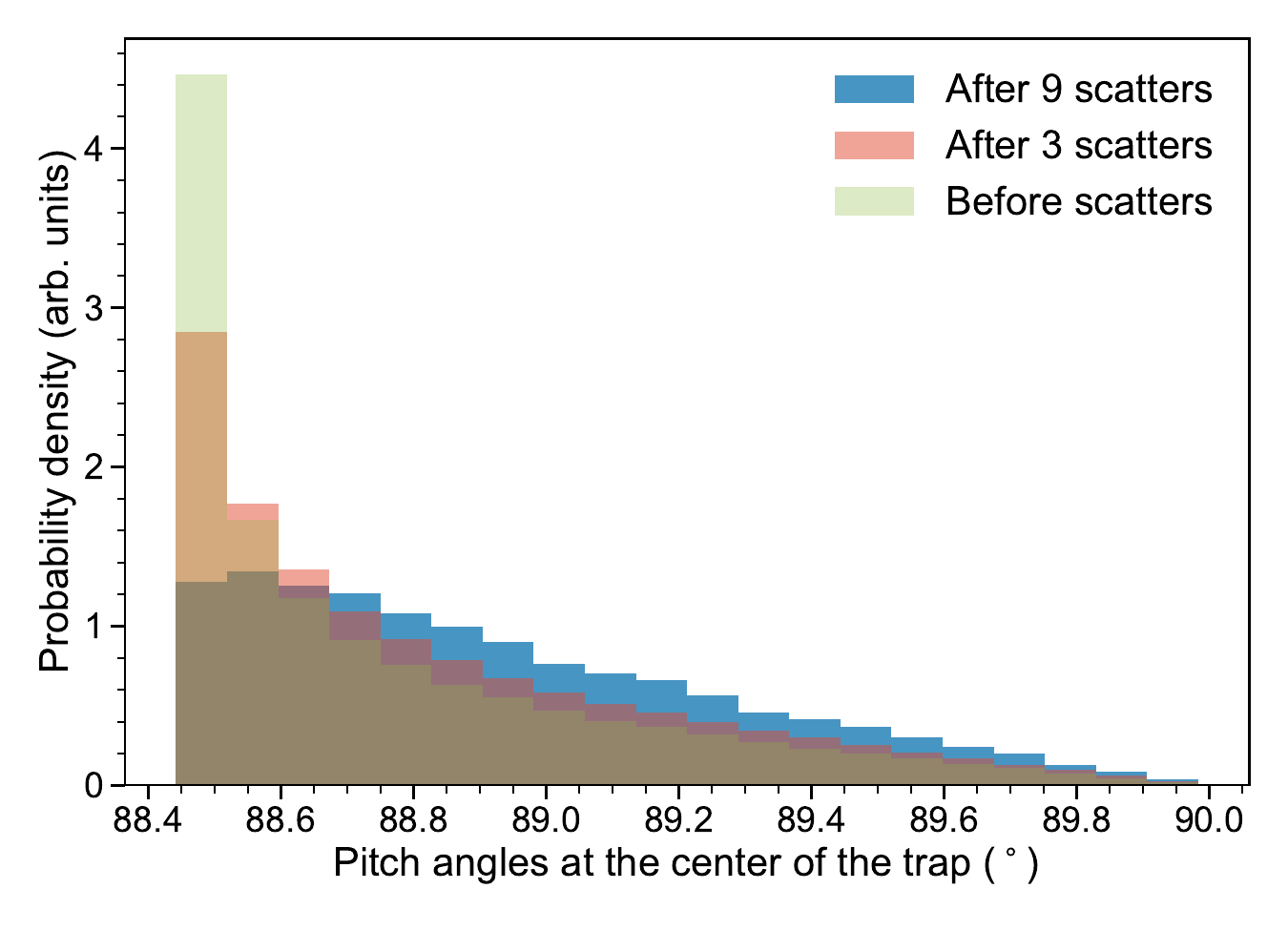}
  \caption{
   Evolution of the distribution of pitch angles of trapped electrons at the center of the trap from simulation.
   After more scatters, a larger fraction of electrons assume larger pitch angles, which corresponds to higher SNRs---thus explaining the deviation in the $\mathcal{A}_j$ curve from a pure exponential, with an excess in higher order scatter peaks.
   }
  \label{fig:evolution_pitch_angle_distribution}
\end{figure}

\autoref{tab:scatter_peak_amplitude_alpha_and_kappa} compiles the values of  $\alpha$ and $\kappa$ found by tuning the $\mathcal{A}_j$ curves to match those from fits for the three $^\mathrm{83m}$Kr data sets.  This simulation reveals that the fraction of trapped electrons with larger pitch angles increases from scatter to scatter as shown in \autoref{fig:evolution_pitch_angle_distribution}. This explains why the scatter peak amplitude curve deviates from an exponential function in the direction of more electrons in the higher order scatter peaks. However, this simulation cannot be directly used to predict $\mathcal{A}_j$ in tritium data, since $\alpha$ and $\kappa$ are related to the gas composition and differ among data sets. Instead, the values of  $p$ and $q$ for tritium data are found by extrapolating from the $p$ and $q$ results of the pre-tritium and post-tritium $\mathrm{^{83m}Kr}$ fits, as described in \autoref{subsubsec:pq_extrapolation}.

\section{Determining parameters for the reduced resolution model}
\label{sec:simplified_res_sims}
The tritium analysis model (\autoref{subsec:tritium_analysis_approximations}) approximates the instrumental resolution $\mathcal{I}$ with a weighted sum of two Gaussians (\autoref{eq:T2_ins_res_model}). We perform fits to simulated resolutions with different $\sigma$ values to determine how $\sigma_0^{[1]}$ and $\sigma_0^{[2]}$ depend on $\sigma$.
To vary $\sigma$ in the simulations, the maximum detectable track SNR (SNR$_{\rm max}$) is varied. As discussed in \autoref{sec:max-SNR-optimization}, $\sigma$ is determined primarily by SNR$_{\rm max}$.

To determine the expressions for $\sigma_0^{[1]}$ and $\sigma_0^{[2]}$ in \autoref{sec:T2-det-response}, we use three pieces of information:
\begin{enumerate}
\item Simulations show that $\sigma_0^{[2]} = 0.8\sigma_0^{[1]} - 5.2\,$eV for our instrumental resolution. We observe this linear relation by generating 100 simulated resolutions with different SNR$_{\rm max}$ values, then fitting each resolution with \autoref{eq:T2_ins_res_model}. In these fits, the Gaussian means, Gaussian standard deviations and $\eta$ are all fitted.
\item We observe that $\sigma \approx \eta\sigma_0^{[1]} + (1-\eta)\sigma_0^{[2]}$ for the same 100 simulated resolutions. The relation is approximate because an exact expression for $\sigma$ must depend on $\mu_0^{[1]}$ and $\mu_0^{[2]}$. In that relation, the right hand side is computed from fit results, and $\sigma$ is calculated directly from the simulated resolutions. 
\item We observe that $\eta=0.66$ is constant within fit errors, for the SNR$_{\rm max}$ uncertainty range in tritium data (quantified in \autoref{sec:sigma_errors}).
\end{enumerate}
Combining the first and second relations, and fixing $\eta$, we find $\sigma_0^{[1]}(\sigma)$ and $\sigma_0^{[2]}(\sigma)$---expressions that depend only on $\sigma$ and constants.

\end{document}